\documentclass[a4paper,12pt]{article}

\usepackage[english]{babel}
\usepackage[utf8x]{inputenc}
\usepackage[T1]{fontenc}

\usepackage[a4paper,top=2.5cm,bottom=2cm,left=2.5cm,right=2.5cm,marginparwidth=1.75cm]{geometry}
\usepackage[a4paper,top=2.5cm,bottom=2cm,left=2.5cm,right=2.5cm,marginparwidth=1.75cm]{geometry}

\usepackage{amsmath}
\usepackage{graphicx}
\usepackage[colorlinks=true, allcolors=blue]{hyperref}
\usepackage{setspace}
\usepackage{apacite}
\usepackage{dsfont}
\usepackage{booktabs}
\usepackage{threeparttablex} 
\usepackage{threeparttable}
\usepackage{chngcntr}		
\usepackage{microtype}     
\usepackage{caption}        
\usepackage{rotating}
\usepackage{pdflscape}    
\usepackage{afterpage}
\usepackage{geometry}
\usepackage[titletoc]{appendix}
\usepackage{amssymb}
\usepackage{subcaption}
\usepackage{ragged2e}
\usepackage{tabularx}
\usepackage{bbm}

\usepackage{tikz}
\usetikzlibrary{shapes,decorations,arrows,calc,arrows.meta,fit,positioning}
\tikzset{
	-Latex,auto,node distance =1 cm and 1 cm,semithick,
	state/.style ={ellipse, draw, minimum width = 0.7 cm},
	point/.style = {circle, draw, inner sep=0.04cm,fill,node contents={}},
	bidirected/.style={Latex-Latex,dashed},
	el/.style = {inner sep=2pt, align=left, sloped}
}
\tikzset{
	vertex/.style = {
		circle,
		fill            = black,
		outer sep = 2pt,
		inner sep = 1pt,
	}
}
\tikzstyle{line} = [draw, -latex']
\usetikzlibrary{arrows,decorations.markings,patterns,calc}

\usepackage{changes}

\usepackage[hang,flushmargin]{footmisc}

\newcommand{\bigCI}{\mathrel{\text{\scalebox{1.07}{$\perp\mkern-10mu\perp$}}}} 

\newcommand\blfootnote[1]{%
  \begingroup
  \renewcommand\thefootnote{}\footnote{#1}%
  \endgroup
}

\newtheorem{assump}{Assumption}
\newtheorem{thm}{Theorem}[section]
\newtheorem{defin}{Definition}[section]
\usepackage{enumitem}

\title{Effect or Treatment Heterogeneity? \\ Policy Evaluation with Aggregated and Disaggregated Treatments
\blfootnote{Michael Knaus gratefully acknowledges financial support from the Swiss National Science Foundation (SNSF) (grant number SNSF 407740\_187301). The paper was circulated and presented previously under different titles. We would like to thank Martin Huber, Michael Lechner, Jana Mareckova, Julian Sch\"ussler, Anthony Strittmatter, and the participants of sessions at IAAE2021, ESEM2021, COMPIE2021, and the 
FFHKT Econometrics Seminar for valuable comments and discussions. All remaining errors are ours.}
} 

\author{Phillip Heiler\thanks{Aarhus University, Department of Economics and Business Economics, CREATES, TrygFonden's Centre for Child Research, Fuglesangs All\'e 4, 8210 Aarhus V, Denmark, \href{mailto:pheiler@econ.au.dk}{pheiler@econ.au.dk}.} \and Michael C. Knaus\thanks{University of T\"ubingen, Mohlstra{\ss}e 36, 72074 T\"ubingen, Germany. Michael C. Knaus is also affiliated with IZA, Bonn, \href{mailto:michael.knaus@unisg.ch}{michael.knaus@uni-tuebingen.de}. }}

\date{\today}

\begin{document}
\maketitle

\doublespacing

\begin{abstract}
\singlespacing
Binary treatments are often ex-post aggregates of multiple treatments or can be disaggregated into multiple treatment versions. Thus, effects can be heterogeneous due to either effect or treatment heterogeneity. We propose a decomposition method that uncovers masked heterogeneity, avoids spurious discoveries, and evaluates treatment assignment quality. The estimation and inference procedure based on double/debiased machine learning allows for high-dimensional confounding, many treatments and extreme propensity scores. Our applications suggest that heterogeneous effects of smoking on birthweight are partially due to different smoking intensities and that gender gaps in Job Corps effectiveness are largely explained by differential selection into vocational training.
\\[4ex]

\end{abstract}

\newpage
\section{Introduction} 
The analysis of causal effects is at the heart of empirical research in economics, political science, the biomedical sciences, and beyond.  To evaluate and design policies, interventions, or programs for units with different background characteristics, it is crucial to develop a thorough understanding of the heterogeneity present in causal relationships. There is now a large literature that develops and applies identification and estimation strategies for causal or treatment parameters that explicitly take into account such heterogeneity, see \citeA{Athey2017} or \citeA{Abadie2018EconometricEvaluation} for recent overviews. 
	
Most attention is on \textit{effect heterogeneity} of binary treatments, while less is given to \textit{treatment heterogeneity}. However, many binary treatments in applications can be conceived as heterogeneous in the sense that they summarize (many) underlying \textit{effective treatments} that impact the outcome of interest. In such cases it is not clear whether effect heterogeneity as defined in the canonical binary treatment setting reflects heterogeneous effects or heterogeneity in the effective treatments. This paper proposes new estimands to disentangle these sources of heterogeneity in a general setting where the analyzed binary indicator does not coincide with the effective treatments. The distinction between sources of heterogeneity is crucial for evaluating and improving assignment mechanisms. 
Consider the following two scenarios:
\footnote{See also Supplementary Appendix \ref{sec:app-ex} for a numerical example.}

\textit{Scenario 1} (binarized treatments): Multiple or continuous treatments are \textit{ex-post} subsumed into a binary indicator (e.g.~different smoking intensities become ``smoking yes/no''). Such aggregations are often motivated by simplicity or data availability, but can have unintended consequences: First, discovered effect heterogeneity can be a spurious byproduct of aggregation and thus falsely be attributed to unit background characteristics. Second, actual effect heterogeneity could be masked as a consequence of the aggregation.

\textit{Scenario 2} (multiple treatment versions): A binary treatment takes different versions after assignment, e.g.~access to a training program with multiple specializations. Here, effect heterogeneity could result from different version targeting and not from different effectiveness of the versions themselves. This distinction is crucial for policy makers to assess the quality of the version assignment mechanism.

In this paper we propose a novel method for decomposing canonical effect heterogeneity into new estimands that are representative of (i) heterogeneous effects and (ii) heterogeneity from different underlying treatment compositions. These decomposition parameters serve as summary measures to evaluate the consequences of (dis)aggregating treatment variables for the causal analysis. 
Furthermore they provide a simple framework for comparing the quality of treatment version assignments and their heterogeneity across units or groups.
		
We develop a simple but flexible nonparametric method for estimation and statistical inference for the decomposition parameters. Our framework allows for the use of machine learning techniques such as random forests, deep neural networks, or high-dimensional sparse regression models in the estimation of the nuisance parameters. We provide high-level conditions regarding the required rates for machine learners, their interaction with the nonparametric decomposition step, and the number of effective treatments $J$. We also provide sufficient conditions for explicit example estimators.

The decomposition can be used to conduct hypothesis tests that consider all effective treatments simultaneously. This allows to test selection and effect heterogeneity without the need for multiple testing procedures. 
It compares favorably to conventional multi-valued treatment effect analysis under many effective treatments $J\rightarrow\infty$, expanding sets of nuisance parameters, and extreme propensity scores. In particular, regular inference is still achievable even if propensity scores are arbitrarily close to zero (limited overlap). This result is obtained by leveraging local superefficiency properties of probability estimators.
The large sample theory extends to other parameters that combine unbiased signals with machine learning inputs and estimated weights. Monte Carlo simulations suggest that coverage rates are close to nominal in finite samples. 

We provide two applications of our decomposition method, one for each leading scenario: First, we show that parts of the finding that the detrimental effect of smoking on birth weight is largest for white mothers can be explained by white mothers smoking more heavily conditional on being smokers. Similarly, different effects for different age groups are partly due to teenage mothers smoking less intensively compared to older mothers. Second, we investigate the lower effectiveness of access to the Job Corps training program for women compared to men. We find evidence that the well-documented gender gap is largely explained by the vocational training curriculum, which focuses more on lower paying service jobs for women and more on higher paying craft jobs for men. Imposing the same mix of vocational training as part of our decomposition removes 73\% of the total gender differences in the effect on earnings.

The paper is structured as follows:	Section \ref{sec:lit} discusses the related literature. Section \ref{sec:decomp} outlines the decomposition of the causal effect parameters and discusses their identification. Section \ref{sec_Estimation1} contains the estimation and inference method. Section \ref{sec_LargeSample1} introduces the technical assumptions and discusses the large sample properties. Section \ref{sec_MC1} provides the Monte Carlo study. Section \ref{sec:app} contains the application. Section \ref{sec:conc} concludes. We also provide an \href{https://github.com/MCKnaus/causalDML/blob/master/R/HK_decomposition.R}{implementation in R} and replication notebooks.\footnote{For Section \ref{sec:app-sc1} see \href{https://mcknaus.github.io/assets/code/Replication_NB_smoking.nb.html}{mcknaus.github.io/assets/code/Replication\_NB\_smoking.nb.html} and for Section \ref{sec:app-sc2} \href{https://mcknaus.github.io/assets/code/Replication_NB_JC.nb.html}{mcknaus.github.io/assets/code/Replication\_NB\_JC.nb.html} on GitHub.}
	
\section{Related Literature} \label{sec:lit}

The proposed decomposition 
complements the literature that considers (dis)aggregated binary treatments. 
\citeA{Lechner2002ProgramPolicies} discusses how to aggregate average effects of multiple treatments into composite treatment effects. \citeA{Hotz2005PredictingLocations} and \citeA{Hotz2006EvaluatingProgram}
investigate the consequences of summarizing different training components in one binary indicator and emphasize the potential lack of external validity under latent treatment heterogeneity. 
\citeA{McCall2016Government-SponsoredAdults} discuss the challenges to determine the optimal degree of coarsening of multi-valued treatments in applications.
Similarly, a recent stream of papers formalizes structural causal models and interpretations of compound treatments \cite{Cole2009TheInference,VanderWeele2009ConcerningInference,Hernan2011CompoundInference,Petersen2011CompoundGraphs}. \citeA{VanderWeele2013CausalTreatment} note that non-homogeneous treatments violate the second component of the “Stable Unit Treatment Value Assumption”  \cite<SUTVA,>{Rubin1980RandomizationComment}: 
no-multiple-versions-of-treatment, which requires a homogeneous treatment or at least the treatment variation irrelevance assumption of \citeA{VanderWeele2009ConcerningInference}. 
\citeA{VanderWeele2013CausalTreatment} formalize a setting where this assumption is violated and provide several new identification results and estimands. 
Aggregating heterogeneous treatments has also been discussed in the context of instrumental variables \cite{Angrist1995Two-stageIntensity,Marshall2016CoarseningEstimates,Andresen2021Instrument-basedRestriction,Harris2022InterpretingEducation}, regression discontinuity designs \cite{Cattaneo2016InterpretingCutoffs}, and models with spillovers and interactions \cite{Manski2013IdentificationInteractions,Vazquez-Bare2022IdentificationExperiments}. These papers mostly discuss the consequences of (dis)aggregation of treatments on unconditional estimands and their connection to (weighted) causal effects. Our paper focuses on the consequences of (dis)aggregation on effect heterogeneity. 

The focus on effect heterogeneity is motivated by the surging literature that develops \cite<e.g.>{Athey2016,Athey2017a,Kunzel2017,Knaus2021} and applies \cite<e.g.>{Davis2020RethinkingJobs,Knaus2022HeterogeneousApproach,Buhl-Wiggers2022SomeInterventionb} flexible machine learning methods to the estimation of heterogeneous causal effects. We build on the double/debiased machine learning framework by \citeA{Chernozhukov2018}. They use Neyman-orthogonal score functions and sample splitting in conjunction with machine learning methods for estimation of low-dimensional parameters that depend on nuisance quantities. 

Regarding heterogeneity analysis, there is now a series of papers that obtain (functional) parameters by localizing these score functions using (nonparametric) regression or machine learning methods \cite{Lee2017,Zimmert2019NonparametricConfounding,Colangelo2020DoubleTreatments,Kennedy2020OptimalEffects,Semenova2021DebiasedFunctions,Fan2022EstimationData,Knaus2022DoubleUnconfoundedness,Heiler2022HeterogeneousPolarization}. 
Our theoretical contribution builds on the structural function approach by \citeA{Semenova2021DebiasedFunctions} with least squares series estimation \cite{Newey1997ConvergenceEstimators,Belloni2015SomeResults,Cattaneo2020LargeEstimators}. We extend some of the inferential results by \citeA{Semenova2021DebiasedFunctions} to settings where pseudo-outcomes are constructed as a weighted average of Neyman-orthogonal scores with (estimated) weights and potentially many treatments. 

The paper is also related to the literature regarding inference on effect parameters under extreme propensity scores or ``limited overlap'' \cite{Khan2010IrregularEstimation,Rothe2017RobustOverlap,Ma2020RobustWeighting,Hong2020InferenceOverlap,Heiler2021ValidScores}. Limited overlap occurs by construction when allowing for ``many treatments'' $J\rightarrow\infty$. In this case, the set of nuisance parameters is expanding and classic multi-valued treatment effect parameters \cite<e.g.>{Cattaneo2010EfficientIgnorability} are irregularly identified which complicates inference. 
The decomposition method, however, always yields three aggregate (functional) parameters independently of $J$. As a consequence, regular estimation and inference regarding heterogeneity is still feasible 
as long as $J$ does not grow too fast. 
In finite samples, determining what constitutes a many treatments setup is difficult as $J$ is always a finite number and a small lower bound for propensities are hard to distinguish from a zero lower bound \cite{Rothe2017RobustOverlap}. Thus, a method that is robust to a potentially large number of treatments provides safeguard for empirical practice.

\section{Decomposition and Identification} \label{sec:decomp}

\subsection{The Setting}\label{sec:setting}

Assume we observe independent data $(Y_i,D_i,T_i,X_i)$ for $i=1,\dots,n$. $Y_i$ denotes the outcome of interest, $D_i \in \{0,1\}$ is the analyzed binary indicator, $T_i \in  \mathcal{T} =  \{0,1,\dots,J\}$ indicates the effective treatment\footnote{Note that \citeA{Manski2013IdentificationInteractions} also uses the term ``effective treatments'' in the context of interference. Like in our setting, it describes the treatments that create variation in potential outcomes. In the following the term ``treatment'' refers to effective treatment if not stated differently.}, and $X_i$ contains confounding variables. We consider settings that are characterized by two features: (i) Not $D_i$, but the effective treatment $T_i$ has a direct influence on the outcome creating potential outcomes $Y_i(t)$ for each $t \in \mathcal{T}$. Thus, we assume SUTVA with respect to the effective treatment such that $Y_i = \sum_t \mathbbm{1}(T_i =t) Y_i(t)$. (ii) Conditional on $T_i$, the binary indicator $D_i$ is deterministic, i.e.~it perfectly separates the support $\mathcal{T}$. 
We use directed acyclic graphs \cite<DAGs, see e.g.>{Pearl1995CausalResearch} to outline our main scenarios:

\begin{figure}[!h] \centering \caption{Analyzed indicator is ex-post aggregate of confounded multiple treatment:} \label{fig_DAG_expost1}
	\begin{tikzpicture}
		\node (1) at (0,0) {Treatment $T$};
		\node (2) [right = of 1] {Outcome $Y$};
		\node (3) [above = of 2] {Binarized Treatment Indicator $D$};
		\node (4) [below = of 1] {Confounders $X$};
		
		\path (1) edge (2);
		\path (1) edge (3);
		
		\path (4) edge (2);
		\path (4) edge (1);		
	\end{tikzpicture}
\end{figure}
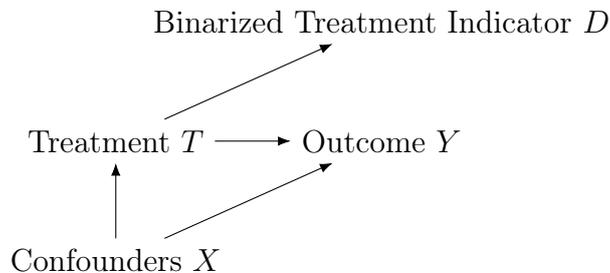

Figure \ref{fig_DAG_expost1} outlines the causal structure of \textit{Scenario 1} where the binary indicator variable $D_i$ is the result of an ex-post aggregation and not directly related to the outcome in a structural sense. In practice, this aggregation is often conducted after the outcome realizes, which makes it unlikely for $D_i$ to affect $Y_i$ directly. This is indicated by a missing arrow from $D_i$ to $Y_i$. 

\begin{figure}[!h] \centering \caption{Randomized binary treatment precedes confounded treatment versions:}  \label{fig_DAG_exante1}
	\begin{tikzpicture}
		\node (1) at (0,0) {Binary Treatment $D$};
		\node (2) [right = of 1] {Treatment Version $T$};
		\node (3) [right = of 2] {Outcome $Y$};
		\node (4) [below = of 2] {Confounders $X$};
		
		\path (1) edge (2);
		\path (2) edge (3);
		
		\path (4) edge (2);
		\path (4) edge (3);		
	\end{tikzpicture}
\end{figure}
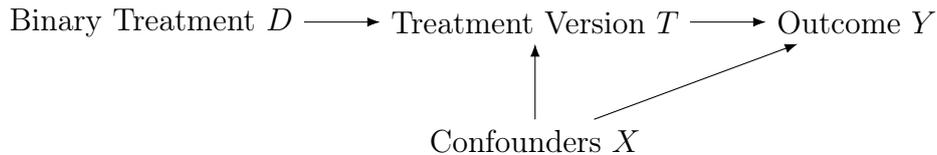

The DAG in Figure \ref{fig_DAG_exante1} depicts the causal structure of \textit{Scenario 2} where a randomized binary treatment $D_i$ precedes the confounded allocation of treatment versions $T_i$. Here, $D_i$ is not an ex-post variable with regards to $Y_i$. $Y_i$ and $D_i$ are associated as the latter determines which treatment versions are available, but has no direct effect beyond that. Its effect is completely mediated through the treatment versions $T_i$. 



We denote $D_{t,i} = \mathbbm{1}(T_i =t)$ to indicate that unit $i$ is observed in treatment $t$ and define $e_t(x) = P(D_{t,i} = 1 | X_i = x)$ as corresponding propensity score. Without loss of generality, we assume throughout that $T_i=0$ denotes a homogeneous control condition.
Thus, the binary indicator is defined as $D_i=\sum_{t\neq 0}D_{t,i}$ and $D_{0,i} = 1-D_i$ in what follows.

\subsection{Dissecting Aggregate Effect Heterogeneity}
We are interested in cases with causal structures as described in Section \ref{sec:setting} but analysis limited to binary $D_i$. 
Here, typical quantities of interest are conditional average treatment effects ($CATE$) or aggregations thereof like the average treatment effect ($ATE$).
Canonical strong ignorability assumptions for $D_i$ are then imposed to exploit quantity $\tau(x) = E[Y_i | D_i = 1, X_i = x] - E[Y_i | D_i = 0, X_i = x]$ for identification of the $CATE$. However, when $D_i = 1$, the potential outcome is not uniquely defined unless $J=1$. 
Therefore, the question is what does this $\tau(x)$ actually identify?
Given the setting outlined in Section \ref{sec:setting}, we can backwards engineer the actually identified estimand in terms of potential outcomes of the effective treatment: 

\noindent
\begin{align}
	\tau(x) 
	&= \sum_{t\neq 0} E\left[D_{t,i} Y_i(t) | D_i = 1, X_i = x\right] - E[Y_i(0) | D_i = 0, X_i = x] \nonumber \\
	&=\sum_{t\neq 0}E[Y_i(t) | D_{t,i}=1,D_i=1,X_i=x]\frac{e_t(x)}{\sum_{t \neq 0}e_t(x)} - E[Y_i(0) | D_i = 0,  X_i=x] \nonumber \\
	&= \sum_{t\neq 0}  E[Y_i(t) | D_{t,i}=1,X_i=x] \frac{e_t(x)}{\sum_{t \neq 0}e_t(x)} - E[Y_i(0) | D_i = 0,  X_i = x] \nonumber \\ 
	&= \sum_{t\neq 0}  \underbrace{E[Y_i(t) - Y_i(0) | X_i=x]}_{t-\text{specific CATE}} \frac{e_t(x)}{\sum_{t \neq 0}e_t(x)}  \nonumber \\
	&\quad\quad\quad + \sum_{t\neq 0}  \underbrace{\{E[Y_i(t) | D_{t,i}=1,X_i=x] - E[Y_i(t) | X_i=x]\}}_{\text{selection effects $D_i = 1$}} \frac{e_t(x)}{\sum_{t \neq 0}e_t(x)} \nonumber \\
    &\quad\quad\quad  - \underbrace{\{E[Y_i(0) | D_{i}=0,X_i=x] - E[Y_i(0) | X_i=x]\}}_{\text{selection effect $D_i = 0$}}      \label{eq:decomp1} 
\end{align}

Equation \eqref{eq:decomp1} shows that the estimand consists of three components: First, a weighted average of $CATEs$ of the effective treatments, $\tau_t(x) = E[Y_i(t) - Y_i(0) | X_i = x]$, with weights depending on the conditional probability of the respective effective treatment. Second, a weighted average of effective treatment specific selection effects. Third, a selection effect into the control group. The selection effects are positive if units with characteristics $x$ that are actually observed in treatment $t$ show higher potential outcomes than the general population described by $x$, or negative if vice versa. The second and third term is relevant if there is selection into the effective treatments even after conditioning on observed confounders. This can e.g.~occur in the case of a randomized binary treatment in Scenario 2 where the selected heterogeneity variables $X_i$ might not include all confounders for the treatment versions.

The decomposition in \eqref{eq:decomp1} highlights that the interpretation of the underlying estimand becomes more nuanced in the presence of heterogeneous treatments. What is supposed to be an easily interpretable $CATE$ depends now on the potentially unknown distribution of effective treatments and selection into those treatments.  Thus, without further assumptions, heterogeneous effects attributed to the binary indicator can be driven by different $CATEs$, different compositions of the effective treatments, different selection effects of the effective treatments, or combinations thereof.

The development of an identifiable decomposition for a parameter such as \eqref{eq:decomp1} requires conditional independence or related assumptions. For example, the leading scenarios in Figure \ref{fig_DAG_expost1} and Figure \ref{fig_DAG_exante1} imply the same conditional independence relationship between effective treatment and potential outcomes despite not being Markov equivalent:
\begin{equation}\label{eq:cia-ytx}
  Y_i(t) \bigCI \  T_i\ |\ X_i = x, ~\forall ~t \in \mathcal{T}\text{ and }x \in \mathcal{X}
\end{equation}
Condition \eqref{eq:cia-ytx} implies the more conventional ``weak unconfoundedness'' assumption for multi-valued treatments \cite<see e.g.>{Cattaneo2010EfficientIgnorability,Yang2016PropensityTreatments}. The latter is sufficient for the decomposition proposed in Section 3.3. Therefore we maintain it throughout the paper together with a common support assumption: 
\begin{assump} \label{ass:si-v}
(ignorability of effective treatment) 

(a) Weak unconfoundedness: $Y_i(t) \bigCI D_{t,i} | X_i=x$, $\forall$ $t \in \mathcal{T}$ and $x \in \mathcal{X}$.

(b) Common support: $0 < P[D_{t,i} = 1 | X_i =x] \equiv e_t(x)$, $\forall$ $t \in \mathcal{T}$ and $x \in \mathcal{X}$.
\end{assump}

Assumption \ref{ass:si-v} is a standard assumption in the multiple treatments setting \cite{Imbens2000TheFunctions,Lechner2001}. It imposes that (a) the set of conditioning variables 
contains all confounders
and (b) there are comparable units across all treatments. 
Under Assumption  \ref{ass:si-v} $E[Y_i(t) | D_{t,i}=1,X_i=x] = E[Y_i(t) | X_i=x]~\forall~t \in \mathcal{T}$ and the selection effects in \eqref{eq:decomp1} disappear. The underlying estimand then simplifies to
\begin{align}
	\label{eq:ncate}
	\tau(x) = \sum_{t\neq 0} \frac{e_t(x)}{\sum_{t \neq 0}e_t(x)} \tau_t(x) \equiv nATE(x).
\end{align}

We call this estimand the \textit{natural conditional average treatment effect} $nATE(x)$ because it is the result of the actual or ``natural'' effective treatment composition. It is important to note that, even under Assumption \ref{ass:si-v}, the differences between units characterized by $x$ and $x'$ can result from different treatment shares, different treatment $CATEs$, or both. We thus could detect seemingly heterogeneous effects, even if the treatment $CATEs$ are constant within treatments but not homogeneous between treatments, i.e.~$\tau_t(x) = \tau_t~\forall~ t \in \mathcal{T}, x \in \mathcal{X}$ but $\tau_t \neq const. ~\forall~ t \in \mathcal{T}$, as long as the probabilities to be observed in the different effective treatments are heterogeneous. 
This fundamentally affects the interpretation of heterogeneous effects even if the underlying effective treatments are not observable. If they are observable, however, we can further decompose heterogeneous effects of the binary indicator in what follows.

\subsection{The Decomposition}
In this section we demonstrate how to disentangle actual effect heterogeneity and heterogeneity driven by selection into effective treatments. We propose to decompose the $nATE(x)$ in two parts: 
\begin{equation}
	\label{eq:decomp2}
	\underbrace{\sum_{t\neq 0} \frac{e_t(x)}{\sum_{t \neq 0}e_t(x)} \tau_t(x)}_{nATE(x)} = \underbrace{\sum_{t\neq 0} \frac{\pi_t}{\sum_{t \neq 0}\pi_t} \tau_t(x)}_{rATE(x)} + \underbrace{\sum_{t\neq 0} \left( \frac{e_t(x)}{\sum_{t \neq 0}e_t(x)} - \frac{\pi_t}{\sum_{t \neq 0}\pi_t}\right) \tau_t(x)}_{\Delta(x)}
\end{equation}

where $\pi_t=E[D_{t,i}]$ are the unconditional treatment probabilities.\footnote{
In principle, analogous decompositions could be constructed with alternative weights for the effective treatments, e.g.~equal weighting $1/J$. However, the unconditional effective treatment probabilities ensure that $nATE(x) = rATE(x)$ in the case of completely randomized effective treatments.
This shows resemblance to the comparison between the canonical $ATE$ and $ATET$: $ATE$ is an average effect under hypothetical random assignment, $ATET$ under actual treatment assignment. Both coincide under a completely randomized binary treatment.} The first component on the right hand side fixes the composition of the effective treatments at the population value. It resembles a situation where effective treatments are randomly allocated using the population level selection probabilities. Thus, we refer to it as the \textit{random conditional average treatment effect} $rATE(x)$. All heterogeneity in $rATE(x)$ is driven by ``real'' effect heterogeneity within treatments, $\tau_t(x) \neq \tau_t(x')$ for some $x,x'\in\mathcal{X}$, as the underlying treatment composition is held fixed. In other words, differences in $rATE(x)$ describe effect heterogeneity \textit{compositionis paribus}. Thus, we can exploit potential heterogeneity in $rATE(x)$ to test for classic (or ``within'') effect heterogeneity.

The second component of the decomposition $\Delta(x)$ is the part of $nATE(x)$ stemming from the interaction of non-constant effective treatment probabilities and different effective treatments having different effects (``between'' treatment effect heterogeneity). The decomposition is redundant, i.e.~$\Delta(x) = 0~\forall~x\in \mathcal{X}$, under (i) effective treatment composition homogeneity $\frac{e_t(x)}{\sum_{t \neq 0}e_t(x)} - \frac{\pi_t}{\sum_{t \neq 0}\pi_t} = 0~\forall~t \in \mathcal{T}$ and $x \in \mathcal{X}$, (ii) treatment variation irrelevance $E[Y_i(t) | X_i=x] = E[Y_i(t')| X_i=x]~\forall~x\in \mathcal{X},~ t, t' \in \mathcal{T}$ \cite{VanderWeele2009ConcerningInference}
, or (iii) if positive and negative components net out to zero. Hence, $\Delta(x) \neq 0$ is a necessary condition for unequal treatment probabilities and between treatment effect heterogeneity and thus a violation of SUTVA. Furthermore, heterogeneity in $\Delta(x)$ is a necessary condition for heterogeneous assignment probabilities, within treatment effect heterogeneity, or both. Thus, the decomposition addresses a variety of relevant policy questions. 
The focus on such necessary conditions offers statistical advantages over testing related conditions in the standard multi-valued treatment effect setup when there are \textit{many} effective treatments, see Section \ref{sec_LargeSample1}.

Under Assumption \ref{ass:si-v}, the conditional average potential outcome of treatment $t$ is identified as $\mu_t(X_i) \equiv E[Y_i(t) | X_i] = E[Y_i(t) | D_{t,i} = 1, X_i] =  E[Y_i | D_{t,i} = 1, X_i] $ and accordingly the decomposition terms are identified as:
\begin{align}
	\label{eq:ident}
	nATE(x) &= \sum_{t\neq 0} \frac{e_t(x)}{\sum_{t \neq 0}e_t(x)} (\mu_t(x) - \mu_0(x)) \nonumber \\
	rATE(x) &= \sum_{t\neq 0} \frac{\pi_t}{\sum_{t \neq 0}\pi_t} (\mu_t(x) - \mu_0(x)) \nonumber\\
	\Delta(x) &= \sum_{t\neq 0} \left( \frac{e_t(x)}{\sum_{t \neq 0}e_t(x)} - \frac{\pi_t}{\sum_{t \neq 0}\pi_t}\right) (\mu_t(x) - \mu_0(x))  
\end{align}

Aggregations or projections of the three estimands are thus also identified. In particular, let $Z_i = f(X_i)$ denote a (low dimensional) function (e.g.~subset) of confounders supported on $\mathcal{Z}$ and define
\begin{align}
	\label{eq:ident-z}
	nATE(z) &= E[nATE(X_i) | Z_i = z] \nonumber \\
	rATE(z) &= E[rATE(X_i) | Z_i = z] \nonumber\\
	\Delta(z) &=  E[\Delta(X_i) | Z_i = z].  
\end{align}
Focusing on specific subgroups defined by $Z_i = z$ provides concise, predictive summaries of heterogeneity or allocation differences without compromising on the dimensionality of confounders $X_i$ and is standard in the literature on effect heterogeneity \cite{Chernozhukov2017GenericExperiments,Semenova2021DebiasedFunctions}.
The unconditional decomposition terms $nATE = E[nATE(X_i)]$, $rATE = E[rATE(X_i)]$, and $\Delta = E[\Delta(X_i)]$ are special cases thereof.\footnote{The $rATE$ is a special case of composite treatment effects \cite{Lechner2002ProgramPolicies}. If $J\rightarrow \infty$, it can approximate integrated dose-responses of continuous treatments \cite<e.g.>{Kennedy2017Non-parametricEffects}. 
$\Delta$ is similar to the population average prescriptive effect in the context of policy learning \cite{Imai2021ExperimentalRules}.} Thus, we focus on the former throughout the paper.
Estimation and inference methods are presented in Section \ref{sec_Estimation1}.

 The interpretation of $\Delta(x)$ depends on the scenario: In
    \textit{Scenario 1}, $\Delta(x)$ and its aggregates have descriptive interpretation. They describe how much of $nATE(x)$ is driven by an underlying effective treatment mix that deviates from the population mix. A non-constant $\Delta(x)$ indicates that the binarization has consequences for detected heterogeneous effects. Thus, it helps to understand heterogeneity resulting from the binarization. 
    In \textit{Scenario 2}, $\Delta(x)$ and its aggregates provide information for assignment evaluation. Positive (negative) values indicate that assignment of treatment versions is better (worse) than random assuming that individuals act equivalently under the hypothetical random assignment compared to the observational assignment \cite{Heckman2020Epilogue:Revisited}.
    A non-constant $\Delta(x)$ indicates that the selection quality of versions varies across different groups. Thus, the estimand provides an evaluation of the actual assignment mechanism.

\section{Estimation and Inference} \label{sec_Estimation1}
In this section, we outline a flexible estimation approach for the (conditional) decomposition terms and propose a method for conducting valid statistical inference. The method accommodates the use of modern machine learning and other non- or semiparametric methods in the estimation of the required nuisance parameters. 

We propose to approximate the conditional expectations of the decomposition terms $g(z)$ by a linear combination of transformations $b(z)$ of heterogeneity variables $z$, i.e. 
\begin{align}
		g(z) = b(z)'\beta_0 + r_g(z)
\end{align} 
where $\beta_0$ is the parameter vector of the best linear predictor given as solution to equation $E[b(Z_i)(g(Z_i) - b(Z_i)'\beta_0)] = 0$. $r_g(z)$ is the approximation error and $b(z)$ can be basis transformations of the regressors of interest such as polynomials, splines, wavelets, or other functions. The number of components in $b(\cdot)$ is allowed to grow with the sample size which allows us to be agnostic about the shape of the true $g$-function.

Let in the following $\eta = \eta(x) = (\mu_0(x),\dots,\mu_J(x),e_0(x),\dots,e_J(x))'$ denote the vector of nuisance quantities and write $\eta = \eta_i = \eta(X_i)$ with subscript and argument suppressed whenever it does not cause confusion. Also define the unconditional selection probability vector $\pi = (\pi_0,\dots,\pi_J)$. 

\begin{table}[!h]\centering \caption{Score Functions of the Decomposition Parameters} \label{tab_scores1}
\begin{threeparttable}
	\begin{tabular}{c|l} \hline \\[-1.5ex] 
		Parameter & Score function $\psi_i({\eta},{\pi}) = \psi_i^{[Parameter]}({\eta},{\pi})$ \\[.5ex]  \hline \\
		$nATE$ &  $\Psi_i({\eta}) - \psi_i^{[0]}({\eta})$ \\ \\ 
		$rATE$ &  $\frac{\sum_{t\neq 0} {\pi}_t \psi_i^{[t]}({\eta}) }{\sum_{t \neq 0}{\pi}_t} - \psi_i^{[0]}({\eta})$ \\ & \\
		$\Delta$ & $\Psi_i({\eta}) - \frac{\sum_{t\neq 0} {\pi}_t \psi_i^{[t]}({\eta}) }{\sum_{t \neq 0}{\pi}_t}$ \\[1.5ex] \hline
	\end{tabular}
\begin{justify} \footnotesize The scores $\psi_i^{[t]}({\eta})$ and $\Psi_i({\eta})$ are defined in equations \eqref{eq:score-v} and \eqref{eq:score-agg}, respectively. \end{justify}
\end{threeparttable}		
\end{table}

We follow the general idea of \citeA{Semenova2021DebiasedFunctions} to construct ``Neyman-orthogonal'' scores $\psi_i(\eta,\pi)$ such that $g(z) = E[\psi_i(\eta,\pi)|Z_i=z]$. These scores are defined by having an (approximate) zero Gateaux derivative with respect to the underlying nuisance parameters at the true parameter vector \cite{Chernozhukov2018}. The robust scores for the three decomposition parameters considered here are weighted combinations of the well-known Neyman-orthogonal scores for average potential outcomes \cite{Robins1995SemiparametricData}, also known as augmented inverse probability weighting (AIPW) scores: 
\begin{align}
    \psi_i^{[t]}({\eta}) &= \mu_t(X_i) + \frac{D_{t,i}(Y_i - {\mu}_t(X_i))}{e_t(X_i)} \label{eq:score-v} \\
    \Psi_i({\eta})&= E[Y_i | D_i = 1, X_i] + \frac{D_i(Y_i - E[Y_i | D_i = 1, X_i])}{P(D_i = 1 | X_i)} \nonumber \\ 
    &=\frac{\sum_{t\neq 0}\mu_t(X_i)e_t(X_i)}{\sum_{t\neq 0}e_t(X_i)}  + \frac{D_i\bigg[Y_i -\frac{\sum_{t\neq 0}\mu_t(X_i)e_t(X_i)}{\sum_{t\neq 0}e_t(X_i)}\bigg]}{\sum_{t\neq 0}e_t(X_i)}  \label{eq:score-agg}
\end{align}

where $\psi_i^{[t]}({\eta})$ is the score of the treatment $t$ specific average potential outcome and $\Psi_i({\eta})$ is the score for the group described by the binary indicator. Table \ref{tab_scores1} shows how to combine these scores to form unbiased signals of the decomposition parameters. These combinations retain Neyman-orthogonality with respect to $\eta$, see Appendix \ref{app_neyman}, but inference has to be adjusted for uncertainty in the estimation of $\pi$, see Section \ref{sec_LargeSample1}.

Consider now the projection of the score functions onto the space spanned by the $k$-dimensional transformation of $Z_i$, $b(Z_i)$. This yields the estimator \begin{align}
		\hat{\beta} &= \bigg(\sum_{i=1}^nb(Z_i)b(Z_i)'\bigg)^{-1}\sum_{i=1}^nb(Z_i)\psi_i(\hat{\eta},\hat{\pi}) \label{eq_bhat1}
	\end{align}
where the score of a decomposition term with estimated nuisance quantities $\psi_i(\hat{\eta},\hat{\pi})$ serves as pseudo-outcome in the corresponding least squares regression on $b(Z_i)$. For $\hat{\pi}$ we use simple sample averages, i.e.~$\hat{\pi}_t = n^{-1} \sum_{i=1}^n D_{t,i}$. Estimation of $\hat{\eta}$ can be done via modern machine learning methods or other non- and semiparametric estimation methods with good approximation qualities for the functions at hand. For details regarding the technical assumptions, consider Section \ref{sec_LargeSample1}. We require that all components in $\hat{\eta}$ are obtained via $K$-fold cross-fitting:
	\begin{defin}\textbf{K-fold cross-fitting} (see Definition 3.1 in \citeA{Chernozhukov2018})
		Take a K-fold random partition $(I_f)_{f=1}^K$ of observation indices $[K] = \{1,\dots,n\}$ with each fold size $n_f = n/K$. For each $f \in [K] = \{1,\dots,K\}$, define $I_f^c := \{1,\dots,n\}\backslash I_f$. Then for each $f\in [K]$, the machine learning estimator of the nuisance function are given by \begin{align*}
			\hat{\eta}_{f} = \hat{\eta}((Y_i,X_i,T_i)_{i\in I_f^c}).
		\end{align*}
        Thus for any observation $i \in I_f$ the estimated score only uses the model for $\eta$ learned from the complementary folds $\psi_i(\hat{\eta},\hat{\pi}) =  \psi_i(\hat{\eta}_{f},\hat{\pi}).$ 
	\end{defin}
Cross-fitting controls the potential bias arising from overfitting using flexible machine learning methods without the need to evaluate the complexity of the function class that contains true and estimated nuisance quantities. If finite parametric models such as linear or logistic are assumed for the nuisance quantities, the proposed methodology can be applied without cross-fitting.  
	
Under suitable assumptions, the predictions using estimator \eqref{eq_bhat1} are consistent for $g(z)$. Moreover, it is possible to conduct asymptotically valid inference, i.e.~for any $z_0 = z_{0,n}$ we can construct $(1-\alpha)\%$ confidence intervals for the decomposition parameter as
	\begin{align}
		CI_{1-\alpha}(g(z_0)) = \bigg[b(z_0)'\hat{\beta} \pm q_{1-\alpha/2}\sqrt{\frac{b(z_0)'\hat{\Omega}b(z_0)}{{n}}} \bigg]\label{eq_CI1}
	\end{align}
where $q_{1-\alpha/2}$ denotes the $(1-\alpha/2)$-quantile of the standard normal distribution and $\hat{\Omega}$ is a consistent sample estimator of the asymptotic variance $\Omega$ (see Section \ref{sec_LargeSample1} and Appendix \ref{sec_AsyVar1}). The estimator explicitly takes into account the additional uncertainty from estimating the unconditional treatment probabilities in the decomposition terms. The interval in \eqref{eq_CI1} is also valid for the best linear predictor $b(z_0)'\beta_0$ under misspecification if the approximation error is not too large. It provides asymptotically accurate confidence intervals around the true $g$-function if the approximation error vanishes at a suitable rate as the number of basis functions or transformations increases. For the technical details consider Section \ref{sec_LargeSample1}.

\section{Large Sample Properties} \label{sec_LargeSample1}
\subsection{Assumptions and Main Results} \label{sec_LargeSampleAssMain}
In this section, we present and discuss the large sample properties of the proposed decomposition estimator.
First, we introduce the relevant definitions. We then discuss the assumptions required for (i) all decomposition parameters, (ii) $nATE$, and (iii) $rATE/\Delta$ specifically and their connections to the literature. We contrast (ii) and (iii) as the $nATE$ tends to require less restrictive conditions compared to $rATE/\Delta$. We then present the main Theorem and outline potential extensions. 

In the following, quantities like $\psi_i()$ or $r_g()$ are used in their generic sense, i.e.~for a given choice of decomposition parameter $nATE$, $rATE$ or $\Delta$. 
$a \lesssim b$ means $a/b = O(1)$ and $a\lesssim_P b$ means $a/b = O_p(1)$. For a general matrix $M$ denote its largest (smallest) eigenvalue by $\lambda_{max}(M)$ ($\lambda_{min}(M)$). 
Let $\eta \in T$ where $T$ is a convex subset of some normed vector space. Denote the realization set of the estimated nuisance quantities by $\mathcal{H}_n = \mathcal{E}_n \times \mathcal{M}_n \subset T$ with $
	\mathcal{E}_n = E_{0,n}\times E_{1,n}\times\dots\times E_{J,n}$ and $
	\mathcal{M}_n = M_{0,n}\times M_{1,n}\times\dots\times M_{J,n}$,
where $E_{t,n}$ and $M_{t,n}$ are the realization sets that contain estimates $\hat{e}_t(X_i)$ and $\hat{\mu}_t(X_i)$ with probability $1-u_n$. Define their $L_q$ error rates \begin{align*}
	s_{t,n,q} = \sup_{\hat{e}_t \in E_{t,n}} E[(\hat{e}_t(X_i) - e_t(X_i))^q]^{1/q}, \quad 
	m_{t,n,q} = \sup_{\hat{\mu}_t \in M_{t,n}} E[(\hat{\mu}_t(X_i) - \mu_t(X_i))^q]^{1/q} \end{align*} and the slowest $L_q$ rates over all treatments 
	$s_{n,q}  = \sup_{t \neq 0} s_{t,n,q}$ and 
	$m_{n,q} =  \sup_{t \neq 0} m_{t,n,q}$.
Note that $\psi_i^{[t,0]}(\eta) = \psi_i^{[t]}(\eta) -\psi_i^{[0]}(\eta)$. By definition $\varepsilon_i = \psi_i(\eta,\pi) - E[\psi_i(\eta,\pi)|Z_i]$ where $\psi_i(\eta,\pi)$ corresponds to the score function of the decomposition parameter of choice from Table \ref{tab_scores1}. Denote $g(z)= E[\psi_i(\eta,\pi)|Z_i=z]$ where $g \in \mathcal{G}$ with $\mathcal{G} = \mathcal{G}_n$ being a function class potentially depending on $n$. Thus, $g(z) = b(z)'\beta_0 + r_g(z)$ where $\beta_0$ is the parameter of the best linear predictor defined as the root of equation $E[b_i(g(Z_i) - b_i'\beta_0)] = 0$ with $b_i = b(Z_i)$ being the $k$-dimensional basis functions. Also define the potential outcome mean error $\varepsilon_i(t) = Y_i(t) - E[Y_i(t)|X_i]$ and its conditional variance $\sigma_t^2(X_i) = E[\varepsilon_i(t)^2|X_i]$. For the $nATE$ machine learning bias components, we define 
\begin{align*}
	B_n^{[nATE]} 
	&:= \sqrt{n}\sup_{\hat{\eta} \in H_n} ||E[b_i(\psi_i^{[nATE]}(\hat{\eta},\pi) - \psi_i^{[nATE]}(\eta,\pi))]|| \\
	\Lambda_n^{[nATE]} 
	&:= \sup_{\hat{\eta} \in H_n} (E[||b_i(\psi_i^{[nATE]}(\hat{\eta},\pi) - \psi_i^{[nATE]}(\eta,\pi))||^2])^{1/2}
\end{align*} and equivalently for $rATE/\Delta$ with score functions according to Table \ref{tab_scores1}.
Remainder terms $R_n$ are defined in Appendix \ref{app_Theory1}.
Let $\gamma_t = E[b_i\psi_i^{[t,0]}(\eta)] = E[b_i\tau_t(X_i)]$, $\gamma = (\gamma_1 \dots \gamma_J)$, and define $a_i = (a_i^{[1]} \dots a_i^{[J]})'$ with $a_i^{[t]} = (1-\pi_0)^{-2}(D_{t,i}(1-\pi_0) + D_{0,i}\pi_t - \pi_t)$. Now let $Q = E[b_ib_i']$ and define \begin{align*}
	\Omega &= Q^{-1}E[(b_i(\varepsilon_i + r_i) + \gamma a_i)(b_i(\varepsilon_i + r_i) + \gamma a_i)']Q^{-1} \\
	\Omega_1 &= Q^{-1}E[b_ib_i'(\varepsilon_i + r_i)^2]Q^{-1} \\
	\Omega_2 &= Q^{-1}E[\gamma a_ia_i'\gamma'] Q^{-1}.
\end{align*}  
 We now present the assumptions required for all decomposition parameters. They are meant to hold uniformly over $n$ if not stated otherwise:
\par{\textbf{\underline{Decomposition Assumptions}:}}
\textit{\begin{enumerate}[itemsep=0pt] \singlespacing
		\item[A.1)] (Identification) $Q$ has eigenvalues bounded above and away from zero. 
		\item[A.2)] (Conditional means)  The potential outcomes have bounded conditional means 
		\begin{align*}
			\sup_{t}\sup_{x\in\mathcal{X}}\mu_t(x) &\lesssim 1 
		\end{align*} 
		\item[A.3)] (Control overlap and limited treatment overlap) The control propensities are bounded away from zero and one, i.e.~for some $c \in (0,1/2)$  \begin{align*}
		c <	\inf_{x\in\mathcal{X}}{e_0(x)} \leq \sup_{x\in\mathcal{X}}{e_0(x)} < 1 - c		
		\end{align*}
		and the re-scaled inverse treatment propensity scores are proportional to the number of different treatments  \begin{align*}
			\sup_{t\neq 0} \sup_{x\in\mathcal{X}}\frac{\pi_t}{e_t(x)} \lesssim 1 \ \text{ and} \quad J\pi_t \lesssim 1
		\end{align*}
	for all $t=1,\dots,J$ with $J = o(n)$. 
		\item[A.4)] (Bounded relative prediction error) On the realization set with probability $1 - u_n = 1 - o(1)$, the worst relative prediction error for the cross-fitted treatment propensities are bounded \begin{align*}
			\sup_{t\neq 0}\sup_{\hat{e}_t \in E_{t,n}}\sup_{x\in\mathcal{X}}\frac{e_t(x)}{\hat{e}_t(x)} \lesssim 1
		\end{align*}
\end{enumerate}}

A.1 rules out multicollinearity of the basis functions used for the nonparametric heterogeneity analysis in the last stage. A.2 is a mild heterogeneity restriction on the potential outcomes. A.3 is crucial: It is concerned with the degree of overlap for a general number of treatments $J = J_n$. In particular, it assumes that there is strong overlap for the control group and the aggregate treatment, i.e.~control and aggregate treatment propensities are uniformly bounded away from zero. However, the propensities for treatments $t=1,\dots,J$ are allowed to be arbitrarily close to zero as long as they vanish at most at a rate proportional to their respective unconditional treatment selection probability $\pi_t$. 
This allows for limited overlap at each treatment level which is necessary when their number is allowed to increase with the sample size, i.e.~$J \rightarrow \infty$.
We suggest to assess Assumption A.3 empirically by analyzing the (estimated) distribution of $e_t(x)/\pi_t$ for all $t$: If these re-scaled scores have sufficient density bounded away from zero by the same standard used to assess conventional propensity score distributions \cite{Heiler2021ValidScores}, then the assumption is likely to hold, see Appendix \ref{app:app-sc1} and \ref{app:app-sc2} for examples based on the empirical applications from Section \ref{sec:app}. 
Assuming a homogeneous $1/J$-rate for all $\pi_t$ is without loss of generality: If the product converges to zero for some $t$, it vanishes from relevant first-order approximations and estimation properties are eventually determined by the treatments that obey Assumption A.3. Moreover, if A.3 only applies to a smaller finite subset of treatments, it effectively corresponds to strong overlap for these particular $t$ and thus estimators behave analogously to standard AIPW for a control potential outcome. Note that the growth of $J$ is restricted to rate $o(n)$ such that consistent estimation of unconditional multi-valued treatment effects is still possible, albeit at a slower rate compared to the strong overlap case similar to \citeA{Hong2020InferenceOverlap}. 

A.4 says that the worst relative prediction for the cross-fitted propensities is bounded on the realization set with high probability. This is a non-standard assumption, in particular when $J\rightarrow \infty$. It is likely to hold for frequency based methods, i.e.~estimators that use some form of (weighted) average within the cells defined by $D_{t,i}$ for $t=0,\dots,J$ to construct propensities including advanced machine learning methods. A sufficient, but by no means necessary, condition is uniform consistency of $\hat{e}(x)$ over $\mathcal{X}$ at rate $o(e_t(x)^{-1})$. This can be shown to hold e.g.~for single-index models, see Theorem 2 and Theorem 3 in \citeA{Ma2022TestingOverlap}, and nonparametric kernel regression \cite{Heiler2022NonparametricFrequencies} under weak conditions. A key point is that these estimators inherit a \textit{local superefficiency} property from $\hat{\pi}_t$, i.e.~faster convergence rate $|\hat{\pi}_t - \pi_t| \lesssim_P (nJ)^{-1}$ in regimes with many treatments/vanishing unconditional selection probabilities. A.4 then requires the estimators to have a consistency rate increased by a factor of $\sqrt{J}$ compared to the finite $J$ case. For parametric estimators this holds as long as $J = o(n)$ while for kernel regression, for example, under the usual smoothness assumptions with $\mathcal{X}$ of dimension $d$ and a bandwidth $h$, it requires that $\sqrt{\log(n)nh^d/J} = o(1)$. We provide some more intuition about Assumption A.4 and the links between large $J$ and superefficient nuisance parameter estimation in Section \ref{sec_addNote_SUPER1}. 

We now present the assumptions required for $nATE$ followed by $rATE/\Delta$:

\par{\textbf{\underline{$nATE$ Assumptions:}}}

For some $m > 2$ , we have that: 
\textit{\begin{enumerate}[itemsep=0pt] \singlespacing
		\item[B.1)] (Conditional Moments) The potential outcomes 
		have at least $m$ conditional moments for the treated: 
		$		\sup_{t}\sup_{z\in\mathcal{Z}}E[\varepsilon_i(t)^m|Z_i=z,D_{t,i}=1] \lesssim 1$.
		\item[B.2)] (Approximation) For each $n$ and $k$, there are finite constants $c_k$ and $l_k$ such that for each $g \in \mathcal{G}$ \begin{align*}
			||r_g||_{P,2} &:= \sqrt{\int_{z\in\mathcal{Z}}r_g^2(z)dP(z)} \leq c_k,  \\
			||r_g||_{P,\infty} &:= \sup_{z\in\mathcal{Z}}|r_g(z)|  \leq l_kc_k.
		\end{align*}
		\item[B.3)] (Machine Learning Bias) For some $h_1,h_2 > 0$ with $1/h_1 + 1/h_2 = 1$ we have that \begin{align*}
			B_n^{[nATE]} 
			&\lesssim \sqrt{nk}s_{0,n,2h_2}\bigg(Js_{n,2h_1} + m_{n,2h_1}\bigg) = o(1), \\
			\Lambda_n^{[nATE]} 
			&\lesssim \xi_k \bigg(s_{0,n,2} + \sqrt{J}s_{n,2} + J^{-1}m_{n,2} \bigg) = o(1).
		\end{align*}  
		\item[B.4)] (Basis and Linearization Error) The $k$ basis functions are chosen such that \begin{align*}
			||R_{n,Q}|| &\lesssim_P \sqrt{\frac{\xi_k^2\log k}{n}}\bigg(1 + k^{1/2}l_kc_k\bigg) = o(1).
		\end{align*}
		\item[B.5)] (Basis and Lindeberg Condition) Let $\sqrt{n}/\xi_k - l_kc_k \rightarrow \infty$ such that \begin{align*}
			\frac{J}{[\sqrt{n}/\xi_k - l_kc_k]^{2}}+ \bigg(\frac{(l_kc_k)^{\frac{2}{m}}J^{\frac{1}{m}}}{[\sqrt{n}/\xi_k - l_kc_k]}\bigg)^m = o(1).
		\end{align*}
\end{enumerate}}

B.1 imposes some regularity on the tails of the conditional potential outcomes. B.2 defines the $L_2$ and uniform approximation rates using the basis functions for function class $\mathcal{G}$. If the basis is sufficiently rich to span $\mathcal{G}$, we say it is correctly specified and $c_k \rightarrow 0$ as $k \rightarrow \infty$. However, our results allow for the case of misspecification, i.e.~$c_k \not\rightarrow 0$. This is a standard characterization in the literature on nonparametric series methods, see e.g.~\citeA{Belloni2015SomeResults} for more details and examples. 

B.3 is crucial: It requires high-quality approximation capabilities of the first-stage machine learning methods for the nuisance quantities.
In the case of a finite-dimensional, bounded basis $\sup_{z\in\mathcal{Z}}||b(z)||_{\infty} < C$ and finite $J$, the conditions can be simplified to $\sqrt{nk}s_{0,n,2}(s_{n,2} + m_{n,2}) = o(1)$. This means that the products of the nuisance quantities for the conditional control propensity and treatment propensities/potential outcome means have to converge at least at rate $o((nk)^{-1/2})$ identical to conditional ATE estimation in \citeA{Semenova2021DebiasedFunctions}.
With fixed basis, as in unconditional binary ATE estimation, it reduces to the well-known requirement that nuisance functions attain rate $o(n^{-1/4})$ \cite{Chernozhukov2018}.
For many treatments $J\rightarrow\infty$, flexible $k$, and/or machine learning estimators, the convergence requirements can be more demanding. We discuss these cases and corresponding rate requirements in Section \ref{sec_addNote_SUPER1}.

B.4 controls the approximation error from linearization taking into account the unknown design matrix $Q$ of increasing dimension. The condition is equivalent to linearization in conventional least squares series estimation \cite{Belloni2015SomeResults}.\footnote{This suggests that, for specific series methods such as splines \cite{Huang2003LocalRegression} and local partitioning estimators \cite{Cattaneo2020LargeEstimators}, the rate can be improved to $\sqrt{\xi_k^2 \log k/n}(1+ \sqrt{\log k}l_kc_k)$, see also \citeA{Belloni2015SomeResults}, Section 4 and \citeA{Cattaneo2020LargeEstimators}, Remark SA-4 of their supplemental appendix.} Note that this rate does not depend on $J$ as, for linearization, the treatment dimension enters only through estimation of the expanding set of nuisance parameters. Once the difference between true and estimated nuisance parameters is controlled for via B.3, there is no difference to the standard series estimation/binary ATE case with no or known nuisances.  

B.5 controls the rate of the basis function relative to approximation error such that the Lindeberg condition holds. Note that this rate is required to be faster by a factor of $J$ relative to conventional series estimation. This is due to the fact, that the tails of the summands that determine the first-order asymptotics are selected from a combination of $J$ different potential outcome errors $\varepsilon_i(t)$ for $t=1,\dots,J$. Thus, the conditions for the many treatments case are somewhat stronger then the ones expected for series estimation or (conditional) ATE estimation under a moment assumption such as B.1 and finite $J$.

\par{\textbf{\underline{$rATE/\Delta$ Assumptions}}}

For some $m > 2$ , we have that: 
\textit{\begin{enumerate}[itemsep=0pt] \singlespacing
		\item[C.1)] (Conditional Moments) The potential outcomes 
		have at least $m$ conditional moments for the selected:
		$	\sup_{t}\sup_{z\in\mathcal{Z}}E[\varepsilon_i(t)^m|Z_i=z,D_{t,i}=1] \lesssim 1$.
		\item[C.2)] (Approximation) For each $n$ and $k$, there are finite constants $c_k$ and $l_k$ such that for each $g \in \mathcal{G}$ \begin{align*}
		||r_g||_{P,2} &:= \sqrt{\int_{z\in\mathcal{Z}}r_g^2(z)dP(z)} \leq c_k,  \\
		||r_g||_{P,\infty} &:= \sup_{z\in\mathcal{Z}}|r_g(z)|  \leq l_kc_k. 
		\end{align*}
		\item[C.3)] (Machine Learning Bias) For some $h_1,h_2 > 0$ with $1/h_1 + 1/h_2 = 1$ we have that \begin{align*}
			B_{n}^{[rATE]} &\lesssim \sqrt{nk}Js_{n,2h_1}m_{n,2h_2} &= o(1), \\
			\Lambda_n^{[rATE]} &\lesssim \xi_k (m_{n,2} + Js_{n,2} + \sqrt{Js_{n,2}m_{n,2h_1}m_{n,2h_2}}) &= o(1).
		\end{align*}  
		\item[C.4)] (Basis and Linearization Error) The $k$ basis functions are chosen such that \begin{align*}
			||R_{n,\pi}|| &\lesssim_P J\sqrt{k}(n^{-1/2} + m_{n,2} + Js_{n,2})&= o(1), \\
			||R_{n,Q}|| &\lesssim_P \sqrt{\frac{\xi_k^2\log k}{n}}\bigg(1 + k^{1/2}(J^{1/4} + l_kc_k)\bigg) &= o(1). 
		\end{align*}
		\item[C.5)] (Basis and Lindeberg Condition) Let ${n/kJ^2} \rightarrow \infty$, $\sqrt{n}/\xi_k - l_kc_k \rightarrow \infty$ such that \begin{align*}
			\frac{J^4}{[\sqrt{n}/\xi_k - l_kc_k]^{2}} +  \bigg(\frac{(l_kc_k)^{\frac{2}{m}}J}{[\sqrt{n}/\xi_k - l_kc_k]}\bigg)^{m} = o(1).
		\end{align*}	
		\item[C.6)] (Eigenvalues) $\lambda_{min}(\Omega_2) > 0$ and $\lambda_{max}(\Omega_1)/\lambda_{min}(\Omega) \lesssim 1$.  
\end{enumerate}}

We discuss and contrast Assumptions C.1--C.6 with B.1--B.5: C.1 and C.2 are equivalent to B.1 and B.2 with potentially different $m$, $c_k$, and $l_k$. C.3 controls for the estimation of nuisance parameters. In the case of a bounded basis, the condition for $B_n$ reduces to $\sqrt{nk}Js_{n,2}m_{n,2} = o(1)$ which is expected to be equivalent to the $nATE$ machine learning bias rate when $J$ is finite. However, for large $J$, estimating potential outcome means at $m_{n,2}$ is generally slower than estimating control propensities at $s_{0,n,2}$. In the parametric case, for example, we have that $m_{n,2} = \sqrt{J/n} = \sqrt{J}s_{0,n,2}$, see Section \ref{sec_addNote_SUPER1}. An equivalent argument holds for $\Lambda_n$ leading to an additional $\sqrt{J}$ factor compared to the $nATE$ case. 
Thus, the product rates have to be faster by a factor of $\sqrt{J}$ in this case, i.e.~$rATE/\Delta$ generally require somewhat higher quality first-stage learners in comparison to the $nATE$. 

C.4 provides the error from the linearization. Note that there is an additional term $R_{n,\pi}$ due to the moment functions not being Neyman-orthogonal with respect to the unconditional weights $\pi$. It puts an additional restriction on the growth of the number of treatments. The first condition reduces to $\sqrt{kJ^3/n} = o(1)$ in case of parametric nuisance quantities. $R_{n,Q}$ corresponds to the $nATE$ case plus an additional term of order $\sqrt{\xi_k\log k/n}k^{1/2}J^{1/4}$. This is a result of the interaction between estimation error from estimating the unconditional weights with design matrix $Q$.\footnote{Again, for specific series such as splines or local partitioning, we conjecture that a faster rate of $\sqrt{\xi_k\log k/n}\sqrt{\log k}J^{1/4}$ is attainable.} The additional factors are only of order $J^{1/2}$ and $J^{1/4}$ compared to the $nATE$. This is due to the superefficiency of the unconditional probability estimator whenever $J\rightarrow\infty$. 

C.5 is similar to B.5 but requires more stringent conditions on $J$ and $\xi_k$ compared to the $nATE$. The Lindeberg condition for $nATE/\Delta$ is driven by the tails of a weighted combination of moment functions from many treatment groups which can have high variance when $J$ is large. It is more restrictive compared to the $nATE$, as, for the latter, the weight for each $t$-specific moment function $\psi_i^{[t,0]}(\eta)$ is the actual treatment propensity $e_t(X_i)$, see Table \ref{tab_scores1}. Thus, inverse propensity scores disappear leading to a lower variance for the $nATE$ 
explaining the additional $J$-dependent factors between B.5 and C.5.  

The first condition in  C.6 rules out the degenerate case where $nATE = rATE$.  Naturally, if this applies, the weaker Assumptions B.1--B.5 can be used instead. The second condition excludes the hypothetical case where the sum of noise plus approximation error is perfectly negatively correlated with the ($\gamma$-weighted) error from estimating the unconditional weights $\pi$. Both restrictions are expected to always hold in practice and can also be assessed by looking at the empirical analogues of $\Omega$, $\Omega_1$, and $\Omega_2$.

For the estimation of the asymptotic variance, we also assume that A.V holds. The corresponding high-level conditions and discussion can be found in Appendix \ref{sec_app_AV1}. 
\textit{\begin{enumerate}[itemsep=0pt]
		\item[A.V)] (Asymptotic Variance) The assumptions in Appendix \ref{sec_app_AV1} hold for $nATE$  and $rATE$ or $\Delta$ respectively, i.e.~$||\hat{\Omega} - \Omega|| = o_p(1)$.  
\end{enumerate}} \noindent
A.V can require somewhat stronger moment and growth conditions for basis and/or number of treatments. For example, for the $nATE$, they reduce to the same rates required by \citeA{Semenova2021DebiasedFunctions}, Theorem 3.3, condition (ii) with factor $n^{1/m}$ replaced by $(nJ)^{1/m}$. Under finite $J$, they are again equivalent. We obtain the following Theorem:

\begin{thm} \label{thm_AsyNor}
	Let $\Phi(\cdot)$ denote the Gaussian cumulative distribution function. Suppose Assumptions A.1 -- A.4, A.V, and B.1 -- B.5 (C.1 -- C.6) hold for $nATE$ ($rATE/\Delta$) and $\hat{\beta}$ and $\hat{\Omega}$ are estimated according to \eqref{eq_bhat1} and \eqref{eq_AVest1} respectively. Then, for any $z_0 = z_{0,n}$,  \begin{align*}
		\lim_{n\rightarrow\infty}\sup_{t\in\mathbb{R}}\bigg|P\bigg(\sqrt{n}\frac{b(z_0)'(\hat{\beta}-\beta_0)}{\sqrt{b(z_0)'\hat{\Omega}b(z_0)}} \leq t \bigg) - \Phi(t)\bigg| = 0.
	\end{align*}
	Moreover if the approximation error is small, $\sqrt{n}r_g(z_0)/\sqrt{b(z_0)'{\Omega}b(z_0)} \rightarrow 0$, then
	\begin{align*}
		\lim_{n\rightarrow\infty}\sup_{t\in\mathbb{R}}\bigg|P\bigg(\sqrt{n}\frac{b(z_0)'\hat{\beta}-g(z_0)}{\sqrt{b(z_0)'\hat{\Omega}b(z_0)}} \leq t \bigg) - \Phi(t)\bigg| = 0.
	\end{align*}
\end{thm}

Theorem \ref{thm_AsyNor} demonstrates the asymptotic validity of the confidence intervals proposed in \eqref{eq_CI1}. The result accommodates the case of misspecification often present in applied research. It is most useful under the additional undersmoothing condition that makes any misspecification bias vanish sufficiently fast.\footnote{ In particular, when $\mathcal{G}$ is in a $s$-dimensional ball on $\mathcal{Z}$ of finite diameter (a H\"older class of smoothness order $s$) then the condition simplifies to $n^{1/2}k^{-(\frac{1}{2} + \frac{s}{d})}\log(k) \rightarrow 0$, see also \citeA{Belloni2015SomeResults}, Comment 4.3 for additional details. Note that such undersmoothing does in general not admit IMSE optimal $k$ choices. Alternatively, bias-correction methods could be employed \cite{Cattaneo2020LargeEstimators}.}

Theorem \ref{thm_AsyNor}  extends to alternative combinations of Neyman-orthogonal scores other than $rATE$ or $\Delta$. In particular, the results for the $rATE$ can directly be applied to any convex combination of conditional average treatment effects as long as the weights are either (i) deterministic sequences (relative to $J$) or (ii) can be estimated at the same rate as $\pi_t$. This can be useful when comparing heterogeneity of a given selection mechanism to alternative, hypothetical (estimated or true) allocation policies different from random selection as considered in this paper even when there are many different treatments.

\subsection{Convergence Rates when $J$ is large: Examples} \label{sec_addNote_SUPER1}
In this section we provide some basic examples and intuition about the properties of probability and nuisance function estimation when $J$ is large and how this relates to the machine learning bias Assumptions B.3/C.3. We first discuss the necessity of Assumption A.3 and the consequences for the unconditional probability estimates. We then show how these properties translate into different convergence rates for propensity scores and potential outcome means under simplified parametric assumptions. We then discuss the explicit requirements for the machine learning bias Assumptions B.3/C.3 for the flexible high-dimensional nuisance parameter case using Lasso methodology under approximate sparsity and many treatments. 

\subsubsection{Large $J$ and Assumption A.3}
Consider the second part of Assumption A.3: If $J$ is large, then $J\pi_t \lesssim 1$ is a necessary requirement. Because if the product diverges, $J \pi_t > 1$ causes a contradiction with the constraint $\sum_{t\neq 0}\pi_t = 1 -\pi_0$. In principle, one could allow for some $t\neq 0$ such that $J\pi_t = o(1)$. However, restricting A.3 to hold for all $t\neq 0$ is without loss of generality as otherwise it would be asymptotically equivalent to a regime where only a smaller subset of treatments $J' < J$ obey A.3. Thus, all further assumptions and rates would be equivalent with $J$ being replaced by the new number of asymptotically relevant treatments $J'$. 

\subsubsection{Superefficiency of Unconditional Probability Estimators}
(Local) superefficiency of frequency estimators $\hat{\pi}_t = \frac{1}{n}\sum_{i=1}^nD_{t,i}$ and other nuisance parameters is not a new discovery and has been exploited and discussed in different places in the literature, see e.g.~\citeA{Stoye2009MoreParameters}. For general $P(D_{t,i}=1) = \pi_t$, note that $V[D_{t,i}] = \pi_t(1-\pi_t)$. Hence, 
$	\sqrt{n}(\hat{\pi}_t - \pi_t) \overset{d}{\rightarrow} \mathcal{N}(0,\pi_t(1-\pi_t))$
which implies that $|\hat{\pi}_t - \pi_t| \lesssim_P (n/\pi_t)^{-1/2} \lesssim (nJ)^{-1/2}$ due to Assumption A.3. Thus under the many treatments regime $J \rightarrow \infty$, the frequency estimator is superefficient, i.e.~converges at a quicker rate than $n^{-1/2}$. 

\subsubsection{Superefficiency of Parametric Propensity Scores} \label{sec_superscores1}
Superefficiency of $\hat{\pi}_t$ spills over to frequency-based/parametric estimators of propensity scores. For example, consider the case where $\mathcal{X}$ is discrete (finite-dimensional) with $f_x := P(X_i=x) > 0$ for all $x \in \mathcal{X}$. Consider a simple frequency-based estimator for the treatment propensity $t$ as \begin{align*}
	\hat{e}_t(x) = \bigg[\frac{1}{n}\sum_{i=1}^n\mathbbm{1}(X_i=x) + \mathbbm{1}\bigg(\sum_{i=1}^n\mathbbm{1}(X_i=x) = 0\bigg)\bigg]^{-1}\frac{1}{n}\sum_{i=1}^n\mathbbm{1}(X_i=x)D_{t,i}.
\end{align*}
where the additional indicator in the denominator assures existence. By standard arguments, it follows that, for each $x \in \mathcal{X}$, 
	$\sqrt{n}(\hat{e}_t(x) - e_t(x)) \overset{d}{\rightarrow} \mathcal{N}(0,e_t(x)(1-e_t(x))/f_x)$ 
which implies that $|\hat{e}_t(x) - e_t(x)| \lesssim_P (n/e_t(x))^{-1/2} \lesssim (nJ)^{-1/2}$ due to Assumption A.3. Thus, the frequency-based finite-dimensional/parametric propensity score has the same superefficiency property as the unconditional frequency estimator. 

\subsubsection{Slower Convergence of Parametric Mean Functions}
Parametric estimators of potential outcome means, however, are not superefficient. On the contrary, convergence rates are generally slower under the many treatments regime. For example, consider a frequency-based estimator similar to the one for the propensity score: \begin{align*}
	\hat{\mu}_t(x) = \bigg[\frac{1}{n}\sum_{i=1}^n\mathbbm{1}(X_i=x)D_{t,i} + \mathbbm{1}\bigg(\sum_{i=1}^n\mathbbm{1}(X_i=x)D_{t,i} = 0\bigg)\bigg]^{-1}\frac{1}{n}\sum_{i=1}^n\mathbbm{1}(X_i=x)D_{t,i}Y_i
\end{align*}
where $\mathcal{X}$ is again assumed to be discrete (finite-dimensional) with $f_x := P(X_i=x) > 0$ for all $x \in \mathcal{X}$. Without loss of generality, assume that $E[(Y_i(t) - \mu_t(X_i))^2|X_i=x] = \sigma^2$ (Assumption B.1/C.1 would suffice as well). Again, by standard arguments, we have that, for all $x\in\mathcal{X}$, 
	$\sqrt{n}(\hat{\mu}_t(x) -\mu_t(x)) \overset{d}{\rightarrow} \mathcal{N}(0,\sigma^2/e_t(x)f(x))$
which implies that $|\hat{\mu}_t(x) -\mu_t(x)| \lesssim_P (ne_t(x))^{-1/2} \lesssim (n/J)^{-1/2}$ due to Assumption A.3. Thus, the estimator converges at a slower than parametric rate. In fact, using the rates from Section \ref{sec_superscores1}, yields, for any $t\neq 0$, $s_{n,t,2} = Jm_{t,n,2} = \sqrt{J}s_{0,n,2}$. Thus, corresponding components in the machine learning bias assumptions B.3/C.3 will be of equal rate in the parametric case.   

\subsubsection{Convergence for High-dimensional Sparse Nuisance Functions}
Here we provide some intuition regarding the use of nuisance function estimation using (frequency-based) Lasso in high-dimensional approximately sparse models. In particular, we say potential outcome means are generated by 
	$\mu_t(x) = x'\theta_0 + r_{\mu_t}(x)$
where $x \in \mathcal{X} \subseteq \mathbb{R}^{p_{\mu_t}}$ (and equivalently for $e_t(x)$ with logistic link). $p_{\mu_t}$ denotes the number of available regressors that is allowed to be high-dimensional and grow with $n$. Assume that the typical regularity conditions for Lasso hold as in \citeA{Semenova2021DebiasedFunctions}, Lemma B.1. Denote $s_{e_t}^*$ and $s_{\mu_t}^*$ as the corresponding sparsity indices that obey these assumptions. For simplicity, let the number of available regressors and sparsity indices coincide for propensity score and potential outcome estimation, i.e.~$s_{e_t}^* = s_{\mu_t}^* \equiv s^*$ and $p_{e_t} = p_{\mu_t} \equiv p$. For the machine learning bias for the $nATE$ in Assumption B.3 we then conjecture that \begin{align*}
	B_n^{[nATE]} &\lesssim \sqrt{nk}\sqrt{\frac{s^*\log(p)}{n}}\bigg(J\sqrt{\frac{s^*\log(p)}{nJ}} + \sqrt{\frac{s\log(p)}{nJ}}\bigg) \\
		&= s^*\sqrt{\frac{kJ\log(p)^2}{n}}
\end{align*}
based on the same argument as for the parametric frequency-based estimation above.

\noindent
Thus $B_n^{[nATE]} = o(1)$ requires that sparsity indices have to obey \begin{align*}
	s^* = o\bigg(\sqrt{\frac{n}{kJ\log(p)^2}}\bigg).
\end{align*} 
For $rATE/\Delta$, it is similarly required that \begin{align*}
	s^* = o\bigg(\sqrt{\frac{n}{kJ^2\log(p)^2}}\bigg).
\end{align*}
Thus, sparsity conditions for $rATE/\Delta$ are stronger than for $nATE$ in the many treatments regime. This reflects the different variability due to  different weighting between the decomposition parameters as the $nATE$ weights minimize variance. 
Comparing rate requirements for $nATE$ to the ones \citeA{Semenova2021DebiasedFunctions}, we find that $s^*$ here must be slower by a factor of $\sqrt{J}$ compared to their Lemma B.1. This is the price for the expanding set of nuisance parameters when estimating treatment propensities and potential outcome means for each treatment level separately instead of imposing the binary treatment structure to begin with. Moreover, the nonparametric heterogeneity step adds an additional $\sqrt{k}$ to the sparsity requirements compared to standard double machine learning estimation of the binary ATE in \citeA{Chernozhukov2018}. 
An analogous derivation can be conducted for $\Lambda_n$ as well. Note that the given sparsity assumption here is for each treatment probability separately. In practice, we might want to impose some (group-based) sparsity across treatments to improve estimation when many treatments are available. In this case, rates can be improved depending on the total complexity of the propensity scores \cite{Farrell2015}. We leave extensions along these lines for future work.  


\section{Monte Carlo Study} \label{sec_MC1}
In this section we analyze the finite sample performance of the analytical confidence bounds proposed in Section \ref{sec_Estimation1}. In particular, we evaluate the empirical coverage rates of the corresponding confidence intervals in setups with heterogeneous effective treatment probabilities for all the decomposition parameters. We consider the case of three effective treatment levels and a best linear predictor for the heterogeneity analysis using different sample sizes and total number of confounding variables including high-dimensional designs. In the heterogeneity step, we regress the estimated pseudo outcomes on a single confounder and evaluate the coverage rates for the parameters of the linear predictor. All nuisance parameters are estimated via 2-fold cross-fitting using $\ell_1$-regularized linear regression for the outcome models as well as $\ell_1$-regularized multinomial logistic regression for the propensity scores.\footnote{We have also conducted similar simulations for correctly specified parametric models. Coverage rates are similar or slightly better in small $n$/large $p$ setups. Results are available upon request.} Tuning parameter selection is done via 5-fold cross-validation. The true models satisfy the necessary sparsity assumptions required for high-quality approximation of the machine learning methods \cite{Belloni2013LeastModels,Farrell2015,Belloni2016Post-selectionControls}. We consider two parameterizations: Design A has linear log-odds and potential outcomes in the heterogeneity dimension while Design B also includes nonlinear components (sign, trigonometric, polynomial, and rectified linear functions). For more details on the designs please consider Appendix \ref{app_MCdesign}. 

\begin{table}[!t] \caption{Monte Carlo Simulations: Results} \label{tab_MC2resultsALL} 
	\begin{subtable}{0.5\textwidth}	\centering \caption{Design A, $n = 1000$}
		\begin{tabular}{l|cccc} \hline \hline \\[-1ex]
			&           &          $rATE$ & $nATE$ & $\Delta$ \\ \hline \\[-1ex]
			$p = 10 $ & $\alpha$ & 0.9492 & 0.9448 & 0.9018  \\      
			&          $\beta$ &  0.9526 & 0.9530  & 0.9524  \\      
			&            &                 &        &        \\
			$p = 100 $ & $\alpha$ & 0.9550  & 0.9484 & 0.8540  \\
			&           $\beta$ &  0.9466 & 0.9552 & 0.9532 \\
			&             &                 &        &        \\
			$p = 1000 $ & $\alpha$ & 0.9486 & 0.9444 & 0.7780  \\
			&            $\beta$ &  0.9538 & 0.9494 & 0.9468 \\
			&             &                 &        &        \\
			$p = 5000 $ & $\alpha$ & 0.9548 & 0.9472 & 0.7102 \\
			&            $\beta$ &  0.9510  & 0.9442 & 0.9510  \\ \hline \hline 
		\end{tabular}
	\end{subtable} 	\begin{subtable}{0.5\textwidth}	\centering \caption{Design A, $n = 5000$}
		\begin{tabular}{l|cccc} \hline \hline \\[-1ex]
			&           &          $rATE$ & $nATE$ & $\Delta$ \\ \hline \\[-1ex]
			$p = 10 $ & $\alpha$ & 0.9454 & 0.9438 & 0.9322  \\      
			&          $\beta$ &  0.9506 & 0.9546 & 0.9498  \\      
			&            &                 &        &        \\
			$p = 100 $ & $\alpha$ & 0.9426 & 0.9424 & 0.9198 \\
			&           $\beta$ &  0.9530  & 0.9514 & 0.9530  \\
			&             &                 &        &        \\
			$p = 1000 $ & $\alpha$ & 0.9524 & 0.9516 & 0.9084 \\
			&            $\beta$ &  0.9488 & 0.9482 & 0.9490  \\
			&             &                &        &       \\
			$p = 5000 $ & $\alpha$ & 0.9460 & 0.9424 & 0.8820 \\
			&            $\beta$ &  0.9480 & 0.9462 & 0.9440 \\ \hline \hline 
		\end{tabular}
	\end{subtable} \\[3ex]
	\begin{subtable}{0.5\textwidth}	\centering \caption{Design B, $n = 1000$}
		\begin{tabular}{l|cccc} \hline \hline \\[-1ex]
			&           &          $rATE$ & $nATE$ & $\Delta$ \\ \hline \\[-1ex]
			$p = 10 $ & $\alpha$ & 0.9528 & 0.9474 & 0.9422  \\      
			&          $\beta$ &  0.9354 & 0.9406 & 0.9416  \\      
			&            &                 &        &        \\
			$p = 100 $ & $\alpha$ & 0.9486 & 0.9416 & 0.9320  \\
			&           $\beta$ &  0.9556 & 0.9490  & 0.9426 \\
			&             &                 &        &       \\
			$p = 1000 $ & $\alpha$ & 0.9484 & 0.9440  & 0.9250 \\
			&            $\beta$ &  0.9602 & 0.9512 & 0.9390 \\
			&             &                 &        &        \\
			$p = 5000 $ & $\alpha$ & 0.9506 & 0.9462 & 0.9182 \\
			&            $\beta$ &  0.9582 & 0.9498 & 0.9358 \\ \hline \hline 
		\end{tabular}
	\end{subtable} 	\begin{subtable}{0.5\textwidth}	\centering \caption{Design B, $n = 5000$}
		\begin{tabular}{l|cccc} \hline \hline \\[-1ex]
			&           &          $rATE$ & $nATE$ & $\Delta$ \\ \hline \\[-1ex]
			$p = 10 $ & $\alpha$ & 0.9474 & 0.9530  & 0.9474  \\      
			&          $\beta$ &  0.9448 & 0.9472 & 0.9482  \\      
			&            &                 &        &        \\
			$p = 100 $ & $\alpha$ & 0.9504 & 0.9464 & 0.9482 \\
			&           $\beta$ &  0.9502 & 0.9518 & 0.9504 \\
			&             &                 &        &        \\
			$p = 1000 $ & $\alpha$ & 0.9542 & 0.9466 & 0.9432 \\
			&            $\beta$ &  0.9524 & 0.9492 & 0.9452 \\
			&             &                 &        &        \\
			$p = 5000 $ & $\alpha$ & 0.9528 & 0.9452 & 0.9430  \\
			&            $\beta$ &  0.9530  & 0.9536 & 0.9476 \\ \hline \hline 
		\end{tabular}
	\end{subtable} 
	\begin{justify} \footnotesize The table entries contain the coverage rates under the null hypothesis for the parameters $(\alpha,\beta)$ of the linear predictor for different number of regressors ($p$), sample sizes ($n$), and decomposition parameters $rATE$, $nATE$, and $\Delta$. The nominal coverage rate is 95\%. Results are based on 5000 simulations.   \end{justify}	
\end{table}

Table \ref{tab_MC2resultsALL} contains the coverage rates of the confidence intervals based on \eqref{eq_CI1} using double machine learning at a significance level of $5\%$ for both designs with sample sizes $n=1000, 5000$ and number of confounders $p=10,100,1000,5000$. 
For $rATE(z)$ and $nATE(z)$ results are always very close to the nominal coverage rate in both designs. 
For $\Delta(z)$, there is some undercoverage for the intercept $\alpha$ in design A which increases in the number of parameters and decreases with the sample size as expected. The slope parameter coverage for $\beta$ for $\Delta(z)$ is very close to nominal for any sample size, confounding dimension, or design. Overall the inference based on the asymptotic approximation in \eqref{eq_CI1} seems to be mostly reliable in finite samples.

\section{Applications} \label{sec:app}

\subsection{Smoking and Birth Weight (Scenario 1)} \label{sec:app-sc1}

The detrimental effect of smoking on birth weight and its economic costs are well documented \cite<see e.g.>[and references therein]{Almond2005TheWeight,Abrevaya2006EstimatingApproach}. Beyond the standard average effects it is important to understand the heterogeneous effects to e.g.~identify for which subgroups interventions to reduce smoking during pregnancy would be most beneficial. \citeA{Abrevaya2006EstimatingApproach} documents that the negative effect of smoking is less pronounced for black compared to white mothers in a standard subgroup analysis. A variety of papers analyze heterogeneous effects of smoking as a function of mother's age \cite{Abrevaya2015EstimatingEffects,Lee2017,Zimmert2019NonparametricConfounding,Fan2022EstimationData}. They all document increasingly negative effects with higher age.
The aforementioned studies consider ``smoking yes/no'' as the binary treatment. \citeA{Cattaneo2010EfficientIgnorability} notes that smoking is not a homogeneous treatment, but that the negative effects become more extreme for higher intensities of smoking. Thus, the binary indicator ``smoking'' represents only an aggregation of smoking intensities which directly affect birth weight. This corresponds to \textit{Scenario 1}. We investigate whether the heterogeneous effects documented in the literature can be at least partly explained by different smoking intensities of different groups. 

We analyze the dataset of \citeA{Almond2005TheWeight} used by \citeA{Cattaneo2010EfficientIgnorability} with five intensities of smoked cigarettes per day as the effective treatment $T_i \in \mathcal{T} = \{0,1-5,6-10,11-15,16-20,>20\}$, the binary indicator defined as $D_i = \mathbbm{1}(T_i > 0)$, the outcome $Y_i$ being birth weight in gram, and the confounders $X_i$ including age, education, ethnicity, and marital status of mother and father as well as health indicators and pregnancy history of the mother.\footnote{We thank Matias Cattaneo for sharing the full data. A random subsample is available on his \href{https://github.com/mdcattaneo/replication-C\_2010\_JOE}{GitHub repository}.} The dataset comprises 511,940 observations after removing the 0.1\% of the observations with missing values in relevant variables and 52 confounders. The nuisance parameters are estimated with 2-fold cross-fitting using an ensemble learner of the unconditional mean, random forests, lasso and ridge regression with 2-fold cross-validated weights. For the propensity scores, we use logistic lasso and ridge. 

Smoking behavior differs along the heterogeneity variables ethnicity and age showing that white and older smoking mothers smoke more heavily.\footnote{Appendix \ref{app:app-sc1} and in particular Figure \ref{fig:smoke-dis} provides the smoking distributions by heterogeneity variables.} Combined with the result of \citeA{Cattaneo2010EfficientIgnorability} that different smoking intensities have different effects, this suggests that at least part of the heterogeneity could be explained by different smoking intensities. 

\begin{figure}[t] 
\caption{Heterogeneous effects and decomposition by ethnicity} \label{fig:decomp_race}
\centering
\begin{subfigure}{0.8\textwidth}
\includegraphics[width=\linewidth]{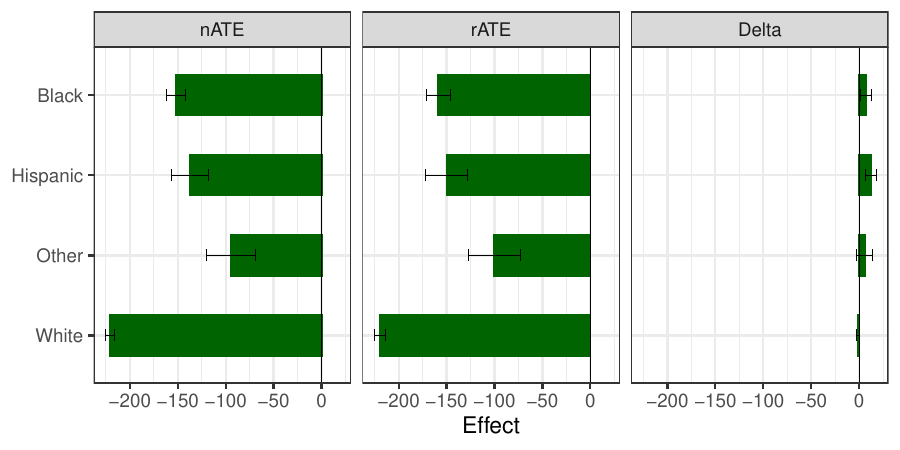}
\caption{Subgroup effects for ethnicity} \label{fig:decomp-race1}
\end{subfigure}

\begin{subfigure}{0.8\textwidth}
\includegraphics[width=\linewidth]{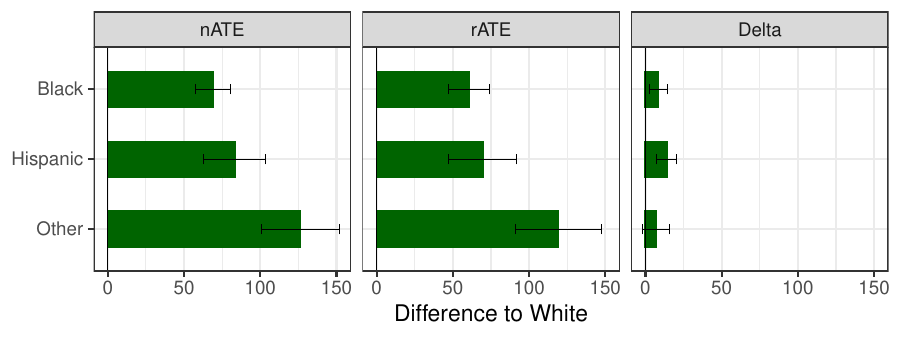}
\caption{Effect heterogeneity with white as benchmark} \label{fig:decomp-race2}
\end{subfigure}

\subcaption*{\textit{Note:} Point estimates of the decomposition parameters with 95\%-confidence interval.}
\end{figure}

Figure \ref{fig:decomp_race} contains the result of the decomposition for the heterogeneity variable ``ethnicity''. The upper panel shows the decomposition for each subgroup. It is obtained by running an OLS regression of the estimand specific pseudo-outcome on a set of four dummy variables indicating ethnicity of the mother without a constant. The $nATE$ in the left part corresponds to standard subgroup analysis. Like previous studies, we find that smoking reduces the birth weight of newborns more for white women than for Blacks, Hispanics and others. Given that smoking is a binarized treatment, it is not clear how much is really effect heterogeneity and how much is driven by the fact that subgroups differ in their smoking intensity. The decomposition term $rATE$ fixes the intensity of smoking for all subgroups at the population level. It provides the subgroup specific effect of smoking if all groups had the same smoking intensity. Under this harmonized smoking intensity the negative average effect of smoking is smaller for white women and larger for the others. $\Delta$ in the right graph quantifies the difference between $nATE$ and $rATE$. It shows relatively small differences suggesting that different smoking intensities are not the main driver of the differences between white mothers and the other groups. However, they are also not negligible as the lower panel of Figure \ref{fig:decomp_race} shows. It quantifies the heterogeneous effects by subtracting the effects for white mothers from the other three groups. We observe that a significant portion of the difference between black/hispanic mothers and white mothers is driven by different smoking intensities. For black vs.~white mothers the difference in the $nATE$ is 69 gram of which 12\% are due to different smoking intensities ($\Delta = 8$). For hispanic vs.~white mothers it explains around 17\% ($\Delta = 14$). 

\begin{figure}[t]
    \centering
    \caption{Effect Heterogeneity by Age}  \label{fig:dec-age}
    \includegraphics[width=1\linewidth]{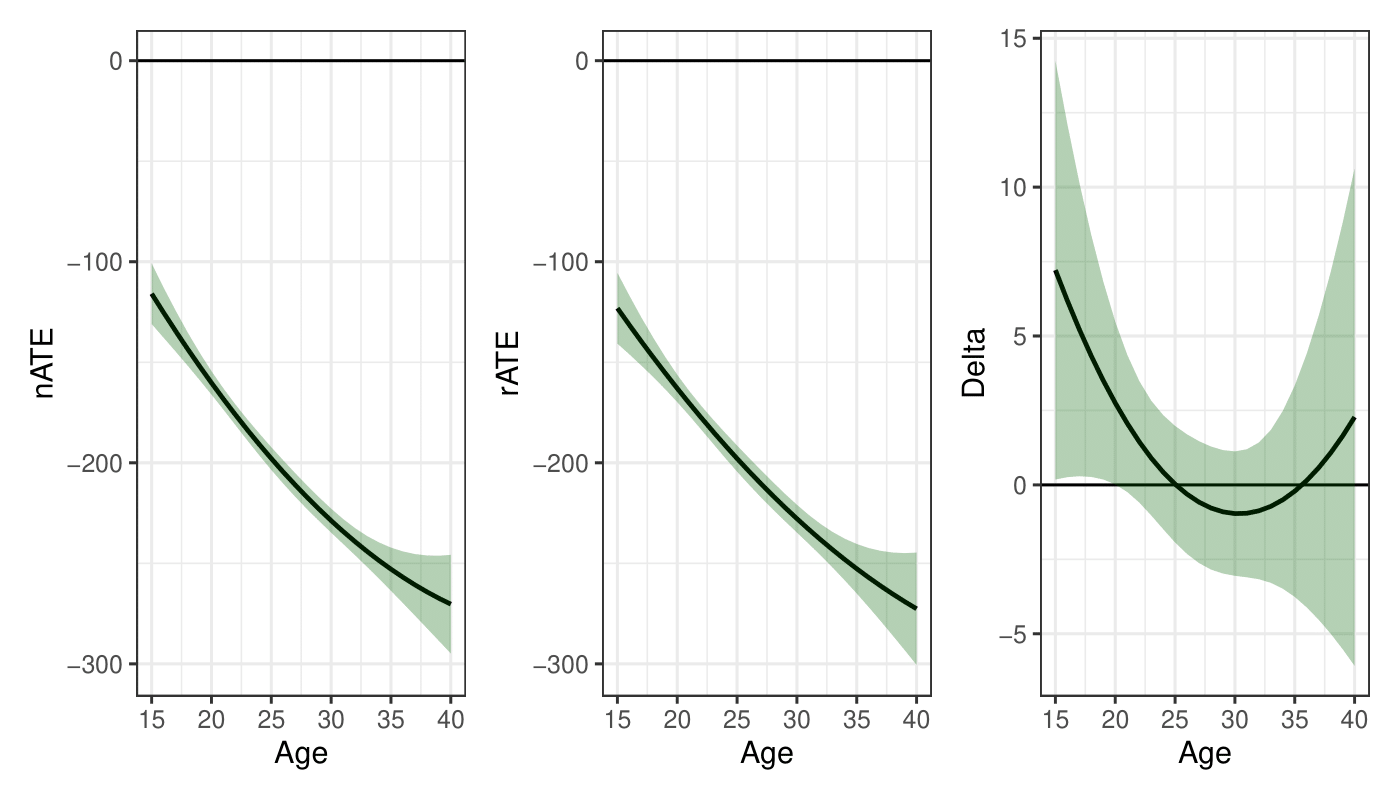}
    
    \subcaption*{\textit{Notes:} B-spline estimated decomposition parameters with 95\%-confidence interval.}
\end{figure}

Figure \ref{fig:dec-age} depicts the heterogeneity analysis along age. We use B-splines as basis functions of age. We select the nodes and order via leave-one-out cross-validation for each parameter and apply the most flexible/low-bias model for all parameters to ensure that the $rATE$ and $\Delta$ curves add up to the $nATE$ curve. The left panel of Figure \ref{fig:dec-age} replicates the well-established findings of previous papers that the $nATE$ is much smaller for younger mothers than for older mothers.
In the extreme case where different smoking intensities would fully explain the heterogeneous $nATE$, we would see a flat $rATE$ curve in the middle graph. However, we only observe that the effect of teenage mothers would be more negative if we harmonize smoking intensity over all age groups.
Overall, only a relatively small part of the heterogeneous effects of the binarized smoking indicator can be attributed to different smoking intensities and the larger part seems to be driven by different age groups actually being affected differently.

\subsection{Job Corps (Scenario 2)} \label{sec:app-sc2}

We illustrate \textit{Scenario 2} with an evaluation of the Job Corps (JC) program. JC operates since 1964 and is the largest training program for disadvantaged youth aged 16-24 in the US \cite<see>[for a detailed description]{Schochet2001NationalOutcomes,Schochet2008DoesStudy}. The roughly 50,000 participants per year receive an intensive treatment as a combination of different components like academic education, vocational training, and job placement assistance. Participants plan their educational and vocational curricula together with counselors. This means that although the variable ``access to JC'' is a binary indicator, different versions of JC participation are conceivable. Heterogeneous effects might thus be driven by different effectiveness of JC for different groups, by different tailoring of the curriculum, or a combination thereof. 

We investigate this based on data from an experiment in 1994-1996 \cite{Schochet2019ReplicationStudy}.\footnote{The data is available as public use file via \href{https://doi.org/10.3886/E113269V1}{https://doi.org/10.3886/E113269V1}.} This experiment is basis of a variety of studies looking at different aspects of JC. Many of them report gender differences in the effectiveness of the programs with women benefiting less than men from access to JC \cite<e.g.>{Schochet2001NationalOutcomes,Schochet2008DoesStudy,Flores2012EstimatingCorps,Eren2014WhoProgram,Strittmatter2019HeterogeneousApproach}. One potential explanation for this finding is that men and women focus on average on different vocational training within JC. In particular men receive more often training for higher paying craft jobs, while women focus more often on training for the service sector \cite{Quadagno1995TheCorps,Inanc2017GenderGap}.\footnote{Appendix \ref{app:app-sc2} and in particular Figure \ref{fig:version-dis} provides the distribution of trainings by gender.} We apply our decomposition method to investigate this potential explanation of the gender gap in program effectiveness.

We analyze the intention to treat effect (ITT) of the binary variable indicating random access to JC ($D_i$) on weekly earnings four years after random assignment ($Y_i$). We consider 11 versions of the effective treatment ($T_i$): (i) \textit{No JC} if eligible individuals did not participate (non-compliers), (ii) \textit{JC without vocational training} if eligible individuals entered JC but did not receive vocational training, (iii-ix) training for jobs in the clerical, health, auto mechanics, welding, electrical/electronics, construction, or food sector, (x) other vocational training, (xi) training for multiple sectors.

Nuisance parameters are estimated with the same ensemble as in Section \ref{sec:app-sc1} using 5-fold cross-fitting. We control for 55 covariates that include pre-treatment information about labor market history, socio-economic characteristics, education, health, crime, and JC related variables. These control variables overlap mostly with those of \citeA{Flores2012EstimatingCorps} who also employ an unconfoundedness strategy.\footnote{Considering second-order interactions results in a total of 1428 variables after screening for nearly empty cells (less than 1\% observations) and nearly perfectly correlated variables (correlation higher than 0.99).} In total we work with a sample of 9,708 observations.

The unconditional $nATE$, corresponding to the ITT of eligibility for JC on monthly earnings, is estimated at \$14.2 (S.E. 3.8), which is an increase of 7\% in line with previous studies. The unconditional $rATE$ is larger (\$17.4, S.E. 4.1) suggesting that hypothetical random allocation of the curricula would yield higher average outcomes compared to the actual assignment. However, the unconditional difference $\Delta$ is insignificant ($\$-3.1$, S.E. 1.8). This suggests that, on average, the selection of versions is not statistically distinguishable from random allocation. 

\begin{figure}[t]
    \centering
    \caption{Effect heterogeneity and decomposition by gender}  \label{fig:dec-female}
    \includegraphics[width=0.8\linewidth]{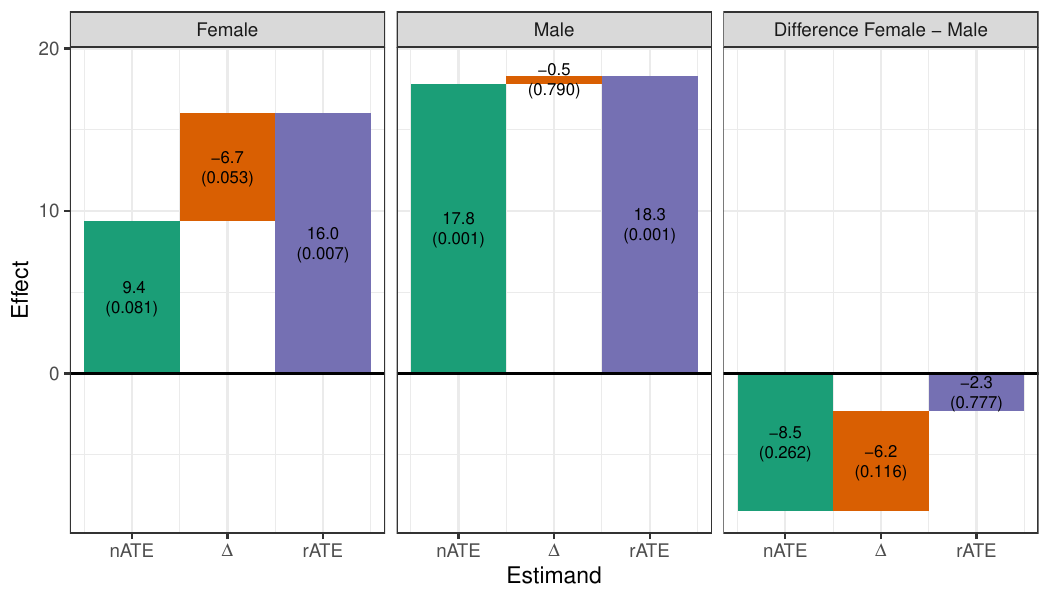}
    
    \subcaption*{\textit{Notes:} The numbers in the bar show the point estimate and the p-value in parentheses.}
\end{figure}

Figure \ref{fig:dec-female} depicts the decomposition of the gender specific effects. We observe that the effect for women with the actual composition of vocational training ($nATE$) is not significant at $\alpha=0.05$, but under the hypothetical treatment composition of the population would show a clear positive effect ($rATE = \$16.0$). The gender gap in effectiveness basically disappears when both groups receive the same hypothetical mix of vocational training. The right part of Figure \ref{fig:dec-female} suggests that 73\% of the gender gap in the effectiveness of JC is due to different training curricula. This means that the worse than average performance of the assignment mechanism seen in the unconditional parameters is mostly driven by women. While the assignment to vocational training for men is as well targeted as random assignment, for women it is even worse. This indicates that there is room for improvement to target vocational training in general and for women in particular. Our results suggest that removing the worse than random targeting of vocational training for women could decrease the gender gap in the effectiveness of access to JC.

\section{Concluding Remarks} \label{sec:conc}
The method proposed in this paper provides a practical way of decomposing effect heterogeneity obtained from analyzing a binary treatment indicator that does not coincide with the effective multi-valued treatment. The approach likely extends to other causal parameters and identification strategies such as continuous effective treatments, selection on unobservables/instrumental variables, or mediation analysis. It would also be interesting to see whether the ideas could be further developed to find the most relevant dimensions of effective treatments for cases with multiple treatment versions instead of requiring the researcher to manually specify them.

The conceptual and empirical results highlight that potential treatment heterogeneity underlying the analyzed binary indicator should be taken more seriously and explicitly discussed in applications, especially when interpreting heterogeneous effects. The decomposition provides one principled way to do this. It requires to observe the effective treatment. Thus, data collection can anticipate the goal of better understanding treatment heterogeneity by recording effective treatment information beyond a binary indicator. Furthermore, the decomposition shows that reducing the analysis to such binary indicators, while facilitating the analysis, comes at the cost of a more intricate interpretation of empirical results.

{\setstretch{1.45} 
\bibliographystyle{apacite}
\renewcommand{\APACrefYearMonthDay}[3]{\APACrefYear{#1}}
\bibliography{references}
}

\begin{appendices}
\counterwithin{figure}{section}
\counterwithin{table}{section}

\huge \noindent \textbf{Appendices}


\normalsize

\appendix		
		
\section{Proof of Theorem \ref{thm_AsyNor}}
\label{app_Theory1}
 
\subsection{Preliminaries}
The proof is structured as follows:	We derive the asymptotically linear representation of the best linear predictor and show its normality using the true covariance matrix. Auxiliary results and estimated variance are provided in Supplementary Appendix \ref{sec_app_auxNEW} and \ref{sec_app_AV1}. We refer with BCCK to \citeA{Belloni2015SomeResults} and with SC to \citeA{Semenova2021DebiasedFunctions}. For the following, we use empirical process notation {\footnotesize\begin{align*}
	E_n[X_i] := \frac{1}{n}\sum_{i=1}^{n}X_i, \quad
	G_n[X_i] :=  \frac{1}{\sqrt{n}}\sum_{i=1}^{n}(X_i - E[X_i]). 
\end{align*}} Recall that $D_i = \sum_{t\neq 0}D_{t,i}$, $D_iY_i = \sum_{t\neq 0}D_{t,i}Y_i$, and $D_{t,i}D_{s,i} = 0$ for $s\neq t$. Assumption \ref{ass:si-v} implies that 
	$E[Y_iD_i|X_i] = \sum_{t\neq 0}e_t(X_i)\mu_t(X_i)$ and
	$E[\mu_t(X_i)D_i|X_i] = \sum_{t\neq 0}e_t(X_i)\mu_t(X_i)$.

\subsection{Machine Learning Bias}
We verify the small bias Assumption 3.5 in SC for our moment functions evaluated at the true $\pi$. Let $u_n = o(1)$ such that with probability of at least $1-u_n$, for all $f \in [K]$, the cross-fitted $\hat{\eta}_f$ belongs to a shrinking neighborhood $\mathcal{H}_n$ around $\eta$. We show that, uniformly over $\mathcal{H}_n$, the moment functions for $nATE$, $rATE$, and $\Delta$ satisfy {\footnotesize \begin{alignat*}{2}
	B_n &=	\sqrt{n} \underset{\hat{\eta} \in \mathcal{H}_n}{\sup}||E[b(Z_i)(\psi(\hat{\eta},\pi) - \psi(\eta,\pi))]|| &&=  o(1)\\
	\Lambda_n &= \underset{\hat{\eta} \in \mathcal{H}_n}{\sup}E[||b(Z_i)(\psi(\hat{\eta},\pi) -  \psi(\eta,\pi))||^2]^{1/2} &&=  o(1).  
\end{alignat*} }
For following expectations are all used omitting prefix ${\sup}_{\hat{\eta} \in \mathcal{H}_n}$ when it does not cause confusion. We will make use of a general decomposition of the AIPW moments: For general binary $D_i$, any $Y_i$, and $\eta = (\mu_i,e_i)$ define {\footnotesize\begin{align*}
	\psi_i(\eta) &= \frac{D_i(Y_i-\mu_i)}{e_i} + \mu_i.
\end{align*}}
Decomposing the function evaluated at two points yields 

\vspace{-12pt}
{\footnotesize\begin{align*}
	\psi_i(\hat{\eta}) - \psi_i(\eta) &= (\hat{\mu}_i - \mu_i)\bigg(1-\frac{D_i}{e_i}\bigg) - (\hat{e}_i- e_i)(Y_i-\mu_i)\frac{D_i}{e_i\hat{e}_i} + (\hat{\mu}_i - \mu_i)(\hat{e}_i - e_i)\frac{D_i}{\hat{e}_ie_i} \\
	&\equiv (a.1) - (a.2) + (a.3). 
\end{align*}}
For the following, we ignore ``control'' part  $\psi^{[0]}_i(W_i,\eta)$ as this is covered by SC and rates are unaffected. Due to cross-fitting, $\hat{e}_t(X_i)$ and $\hat{\mu}_t(X_i)$ only depend on $i$ through $X_i$. Thus, when evaluating expectations with $\hat{\eta}$, we omit conditioning on the cross-fitted fold as in SC. Unconditional convergence then follows from \citeA{Chernozhukov2018}, Lemma 6.1. 
	
\subsubsection{Machine Learning Bias: $nATE$} \label{sec_MLBias_nATE}
	For the $nATE$ the $\psi_i(\eta)$-function has components {\footnotesize\begin{align*}
		\mu_i &= \frac{\sum_{t\neq 0}e_t(X_i)\mu_t(X_i)}{\sum_{t\neq 0}e_t(X_i)}, \
		e_i = \sum_{t\neq 0}e_t(X_i),\
		D_i = \sum_{t\neq 0}D_{t,i},\
		Y_i = \sum_{t}D_{t,i}Y_i(t)
	\end{align*}}
	and equivalently for $\psi(\hat\eta)$. We now verify SC, Assumption 3.5:  For $B_n$, we have that 
	{\footnotesize\begin{align*}
		E[b_i(a.1)] &= E[b_i[\hat{\mu}_i-\mu_i](1-E[D_i|X_i]/e_i)] 
		=0 \\
		E[b_i(a.2)] &= E[b_i[\hat{e}_i-e_i](E[Y_iD_i|X_i] - \mu_i E[D_i|X_i])/(e_i\hat{e}_i)] 
		=0 \\
		||E[b_i(a.3)]|| &\lesssim \sup_{x\in\mathcal{X}}(1-e_0(x))^{-1}E[||b_i||^2]^{1/2}E[(\hat{e}_i-e_i)^2(\hat{\mu}_i-\mu_i)^2]^{1/2}\\
		&\lesssim \sqrt{k}E[(\hat{e}_i-e_i)^{2h_2}]^{\frac{1}{2h_2}} E[(\hat{\mu}_i-\mu_i)^{2h_1}]^{\frac{1}{2h_1}}\\
		&\lesssim \sqrt{k}J^{}s_{0,n,2h_2}\bigg(\sup_{t\neq0} s_{t,n,2h_1} + \sup_{t\neq0,x\in\mathcal{X}}e_t(x)m_{t,n,2h_1}\bigg) \\
		&\lesssim  \sqrt{k}s_{0,n,2h_2}\bigg(Js_{n,2h_1} + m_{n,2h_1}\bigg) 
	\end{align*}} 

 \noindent
	for some $1/h_1 + 1/h_2 = 1$ by H\"older's inequality. The aggregate $\hat{\mu}_i$ rate follows from Assumption A.3 together with expanding

 \noindent
 {\footnotesize\begin{align*}
		\hat{e}_t(x)\hat{\mu}_t(x) - e_t(x)\mu_t(x) &= e_t(x)(\hat{\mu}_t(x)-\mu_t(x)) +  \mu_t(x)(\hat{e}_t(x)-e_t(x))\\ & \ +  (\hat{e}_t(x)-e_t(x))(\hat{\mu}_t(x)-\mu_t(x))
	\end{align*}}
	for all $t$. As potential outcome means are uniformly bounded by A.2, this yields that {\footnotesize\begin{align*}
		E[(\hat{\mu}_i-\mu_i)^{c}]^{1/c} \lesssim J\bigg(\sup_{t\neq0} s_{t,n,c} + \sup_{t\neq0,x\in\mathcal{X}}e_t(x)m_{t,n,c}\bigg)
	\end{align*}}
	for any $c \geq 2$. Overall we have that {\footnotesize\begin{align*}
		B_n^{[nATE]} &= \sqrt{n}\sup_{\hat{\eta}\in \mathcal{H}_n}||E[b_i(\psi_i^{nATE}(\hat{\eta}) - \psi_i^{nATE}(\eta))]|| \\
		&\lesssim \sqrt{nk}s_{0,n,2h_2}\bigg(Js_{n,2h_1} + m_{n,2h_1}\bigg) 
	\end{align*}}
	For $\Lambda_n$ note that: {\footnotesize\begin{align*}
		E[||b_i(\psi(\hat{\eta})-\psi_i(\eta))||^2] \lesssim \xi_k^2 E[(\psi_i(\hat{\eta})-\psi_i(\eta))^2]
	\end{align*}} 
	Decomposing the second term on the right hand side exploiting Assumption A.3 together with independence of the nuisance models and the conditional independence yields:
	{\footnotesize\begin{align*}
		E[(a.1)^2] &= E[(\hat{\mu}_i-\mu_i)^2(1-D_i/e_i)^2] \\
		&= E[(\hat{\mu}_i-\mu_i)^2(1-2 + 1/e_i)] \\
		&\lesssim E[(\hat{\mu}_i-\mu_i)^2] \\
		E[(a.2)^2] &= E[(\hat{e}_i-e_i)^2(Y_i - \mu_i)^2D_i/(e_i^2\hat{e}_i^2)] \\
		&= E[(\hat{e}_i-e_i)^2E[(Y_i - \mu_i)^2|X_i,D_i=1]/(e_i\hat{e}_i^2)]  \\
		&\lesssim E[(\hat{e}_i-e_i)^2] \\
		E[(a.3)^2] &= E[(\hat{\mu}_i-\mu_i)^2(\hat{e}_i - e_i)^2D_i/(e_i^2\hat{e}_i^2)] \\
		&= E[(\hat{\mu}_i-\mu_i)^2(\hat{e}_i - e_i)^2/(e_i\hat{e}_i^2)] \\
		&\lesssim E[(\hat{\mu}_i-\mu_i)^2] 
	\end{align*}}
	where the last inequality uses a constant bound on the control propensities. Here $e_i$ denotes the aggregate treatment propensity with uniformly bounded inverse due to A.3. Thus, using the convergence rate for the aggregate mean $\hat{\mu}_i$ above, we obtain that {\footnotesize\begin{align*}
		\Lambda_n^{[nATE]} 
		&=E[||b_i(\psi(\hat{\eta})-\psi_i(\eta))||^2]^{1/2} \\
		&\lesssim \xi_k \bigg(s_{0,n,2} + \sqrt{J}\bigg(\sup_{t\neq0}s_{t,n,2} + \sup_{t\neq0,x\in\mathcal{X}} m_{t,n,2}e_t(x)\bigg) \bigg)  \\
		&\lesssim \xi_k \bigg(s_{0,n,2} + \sqrt{J}s_{n,2} + J^{-1}m_{n,2} \bigg) 
	\end{align*}}
 \vspace{-36pt}
	\subsubsection{Machine Learning Bias: $rATE$ and $\Delta$} \label{sec_MLBias_rATE}
First note that, conditional on the event $u_n$ {\footnotesize\begin{align*}
	\max_{1\leq i \leq n} \frac{\pi_t}{\hat{e}_t(X_i)} &= \max_{1\leq i \leq n}\frac{e_t(X_i)}{\hat{e}_t(X_i)}\frac{\pi_t}{e_t(X_i)} \\
	&\equiv \chi_{t,n} \\
& \lesssim_P \sup_{x\in\mathcal{X}}\sup_{\hat{e}_t \in E_{t,n}}\frac{e_t(x)}{\hat{e}_t(x)}\sup_{x\in\mathcal{X}}\frac{\pi_t}{e_t(x)} \\
& \lesssim_P \sup_{\hat{e}_t \in E_{t,n},x\in\mathcal{X}}\frac{e_t(x)}{\hat{e}_t(x)} \\
	&\lesssim_P 1
\end{align*}} 	
by A.3 and A.4. This implies that $\sup_{t \neq 0}\chi_{t,n} \lesssim_P 1$. For $rATE$, we use the same decomposition but for each PO term. Omitting the control part $\psi_i^{[0]}(\eta)$, we have: {\footnotesize\begin{align*}
		\psi_i(\hat{\eta},\pi) - \psi_i(\eta,\pi) &= \bigg[\sum_{t\neq 0}\pi_t\bigg]^{-1}\sum_{t\neq 0}\pi_t\bigg[
		(\hat{\mu}_t(X_i) - \mu_t(X_i))\bigg(1-\frac{D_{t,i}}{e_t(X_i)}\bigg) \\ &\ - (\hat{e}_t(X_i)- e_t(X_i))(Y_i-\mu_t(X_i))\frac{D_{t,i}}{e_t(X_i)\hat{e}_t(X_i)} \\ &\ + (\hat{\mu}_t(X_i) - \mu_t(X_i))(\hat{e}_t(X_i) - e_t(X_i))\frac{D_{t,i}}{\hat{e}_t(X_i)e_t(X_i)}\bigg] \\
		&= \bigg[\sum_{t\neq 0}\pi_t\bigg]^{-1}\sum_{t\neq 0}\pi_t\bigg(A_{t,1} - A_{t,2} + A_{t,3}\bigg).
	\end{align*}}
	We now consider $B_{n}^{[rATE]}$. Equivalently to the $nATE$ we have that 
		$E[b_iA_{t,1}] = E[b_iA_{t,2}] = 0$
	for all $t=1,\dots,J$. The remaining term can be bounded as {\footnotesize\begin{align*}
		||E[b_i\sum_{t\neq 0}\pi_tA_{t,3}]|| 
		&= \bigg|\bigg|E[b_i\sum_{t\neq 0}\pi_t(\hat{\mu}_t(X_i) - \mu_t(X_i))(\hat{e}_t(X_i) - e_t(X_i))\frac{D_{t,i}}{\hat{e}_t(X_i)e_t(X_i)}\bigg]\bigg|\bigg| \\
		&\leq_P \sum_{t \neq 0}\pi_t \max_{1\leq i \leq n}e_t(X_i)^{-1}E[||b_i(\hat{\mu}_t(X_i) - \mu_t(X_i))(\hat{e}_t(X_i) - e_t(X_i))||] \\
		&\lesssim_P J\sup_{t\neq 0}\chi_{t,n}\sqrt{k}E[(\hat{\mu}_t(X_i) - \mu_t(X_i))^2(\hat{e}_t(X_i) - e_t(X_i))^2]^{1/2} \\
		&\lesssim  \sqrt{k}Js_{n,2h_1}m_{n,2h_2}
	\end{align*}} 
	by H\"older's inequality with $1/h_1 + 1/h_2 = 1$. Thus, overall we have that {\footnotesize\begin{align*}
		B_{n}^{[rATE]} \lesssim \sqrt{nk}Js_{n,2h_1}m_{n,2h_2}.
	\end{align*}}
	Now consider $\Lambda_n$. First note that {\footnotesize\begin{align*}
		E[(\psi_i&(\hat{\eta},\pi) - \psi(\eta,\pi))^2] = \bigg[\sum_{t\neq 0}\pi_t\bigg]^{-2}\sum_{t\neq 0}\sum_{s\neq 0}\pi_t\pi_sE\bigg[\bigg(A_{t,1} - A_{t,2} + A_{t,3}\bigg)\bigg(A_{s,1} - A_{s,2} + A_{s,3}\bigg)\bigg]
	\end{align*}}
	Now consider the summands for $s=t$: {\footnotesize\begin{align*}
		\pi_t^2E[(A_{t,1} - A_{t,2} + A_{t,3})^2] &\lesssim 4\pi_t^2E[A_{t,1}^2 + A_{t,2}^2 + A_{t,3}^2] 
	\end{align*}}
	Bounding each term separately yields {\footnotesize\begin{align*}
		\pi_t^2E[A_{t,1}^2] &= \pi_t^2E[(\hat{\mu}_t(X_i) - \mu_t(X_i))^2(1-2 + 1/e_t(X_i))] \\
		&\lesssim \pi_t\bigg(\sup_{x\in\mathcal{X}}\bigg|\frac{\pi_t}{e_t(x)}\bigg| + \pi_t\bigg)E[(\hat{\mu}_t(X_i)-\mu_t(X_i))^2] \\
		&\lesssim J^{-1} m_{t,n,2}^2 \\
		\pi_t^2E[A_{t,2}^2] 
		&= \pi_t^2E[(\hat{e}_t(X_i) - e_t(X_i))^2E[(Y_{i}(t)-\mu_t(X_i))^2|X_i]/(e_t(X_i)\hat{e}_t(X_i)^2)] \\
		&\lesssim_P \pi_t^{-1}\sup_{x\in\mathcal{X}}\frac{\pi_t^3}{e_t(x)^3}\chi_{t,n}^2 {E[(\hat{e}_t(X_i)-e_t(X_i))^2]} \\
		&\leq Js_{t,n,2}^2 \\
		\pi_t^2E[A_{t,3}^2] &= \pi_t^2E[(\hat{e}_t(X_i) - e_t(X_i))^2(\hat{\mu}_t(X_i)-\mu_t(X_i))^2/(e_t(X_i)\hat{e}_t(X_i)^2)] \\
		&\leq Js_{t,n,2}^2 \\
	\end{align*}}

 \noindent
	by Assumptions A.3 and A.4. Now consider the summands with $s \neq t$. Recall that $D_{t,i}D_{s,i} = 0$. As a preliminary, note that (conditional on the cross-fitted model): 
 
 \noindent
 {\footnotesize\begin{align*}
		E\bigg[\bigg(1- \frac{D_{t,i}}{e_t(X_i)}\bigg)\bigg(1- \frac{D_{s,i}}{e_s(X_i)}\bigg)\bigg|X_i\bigg] &= -1 \\
		E\bigg[\bigg(1- \frac{D_{t,i}}{e_t(X_i)}\bigg)\frac{D_{s,i}}{e_s(X_i)\hat{e}_s(X_i)}\bigg|X_i\bigg] &= \frac{1}{\hat{e}_s(X_i)} \\
		E\bigg[\bigg(1- \frac{D_{t,i}}{e_t(X_i)}\bigg)D_{s,i}(Y_{i}(s) - \mu_s(X_i))\bigg|X_i\bigg] &= 0
	\end{align*}}
	Thus, summands simplify to {\footnotesize\begin{align*}
		\pi_t\pi_sE\bigg[&-(\hat{\mu}_t(X_i)-\mu_t(X_i))(\hat{\mu}_s(X_i)-\mu_s(X_i)) \\+& (\hat{\mu}_t(X_i)-\mu_t(X_i))(\hat{\mu}_s(X_i)-\mu_s(X_i))\bigg(\frac{\hat{e}_t(X_i)-e_t(X_i)}{\hat{e}_t(X_i)} + \frac{\hat{e}_s(X_i)-e_s(X_i)}{\hat{e}_s(X_i)}\bigg)\bigg] \\
		&\lesssim \bigg[\pi_t\pi_s m_{t,n,2}m_{s,n,2}\bigg] + \bigg[\bigg(\pi_s s_{t,n,2}\chi_{t,n} + \pi_t s_{s,n,2}\chi_{s,n}\bigg) m_{t,n,2h_1}m_{s,n,2h_2}\bigg]
	\end{align*}}
	by repeated application of H\"older's inequality with $1/h_1 + 1/h_2 = 1$. Note that we have $J$ variances and $J(J-1)$ times the covariance term. Thus, we obtain

 {\footnotesize\begin{align*}
		\Lambda_n^{[rATE]} &\lesssim \xi_kE[(\psi_i(\hat{\eta},\pi) - \psi(\eta,\pi))^2]^{1/2} \\
		&\lesssim_p \xi_k\bigg(m_{n,2}^2 + J^2s_{n,2}^2 + J(J-1)( J^{-2}m_{n,2}^2) + J(J-1)(J^{-1}s_{n,2}m_{n,2h_1}m_{n,2h_2})\bigg)^{1/2} \\
		&\lesssim \xi_k (m_{n,2} + Js_{n,2} + \sqrt{Js_{n,2}m_{n,2h_1}m_{n,2h_2}})
	\end{align*}}  
	\vspace{-36pt}

\subsection{Asymptotic Linearization and Normality}
We set $Q = I$ but assume a random design, i.e.~$Q$ is unknown as in BCCK. Derivations are split by $nATE$ and $rATE/\Delta$ and thus $\psi_i(\cdot)$, $r_i$, $\varepsilon_i$, $l_k$, $c_k$, $\beta_0$ are parameter specific. 

\subsubsection{Linearization}
We first expand the $\hat{Q}$-weighted estimator around the best linear predictor. Then, we control the machine learning bias and variation from estimating $\pi$. In the end, we combine these rates with the rates from estimating $\hat{Q}$ to obtain an asymptotic linearization.
\paragraph{ \underline{$rATE/\Delta$:}}
Take a sequence of basis function $b = b_n$ such that $||b|| = 1$. Decompose {\footnotesize\begin{align*}
	\sqrt{n}b'(E_n[b_i\psi_i(\hat{\eta},\hat{\pi})] - \hat{Q}\beta_0) &= \sqrt{n}b'E_n[b_i(\psi_i(\eta,\pi) - b_i'\beta_0)] 
	+ \sqrt{n}b'E_n[b_i(\psi_i(\hat{\eta},\pi)-\psi_i(\eta,\pi))] \\
	&\quad + \sqrt{n}b'E_n[b_i(\psi_i(\hat{\eta},\hat{\pi})-\psi_i(\hat{\eta},\pi))] 
\end{align*}}
The first term will be part of the first order asymptotics used for the normality results later. Now by the derivation in Section \ref{sec_MLBias_rATE} and Chebyshev's inequality we have that
{\footnotesize\begin{align*}
	||\sqrt{n}b'E_n[b_i(\psi_i(\hat{\eta},\pi)-\psi_i(\eta,\pi))]|| \lesssim_P B_n^{[rATE]} + \Lambda_n^{[rATE]} 
\end{align*}} \vspace{-12pt}
The third term can further be decomposed with the multiplicative structure and (H.6): {\footnotesize\begin{align*}
	\sqrt{n}E_n[b_i(\psi_i(\hat{\eta},\hat{\pi})-\psi_i(\hat{\eta},\pi))]  
	&= \sqrt{n}E_n\bigg[b_i\sum_{t \neq 0}\psi_i^{[t]}(\hat{\eta})\bigg(\frac{\hat{\pi}_t}{1-\hat{\pi}_0}-\frac{{\pi}_t}{1-{\pi_0}}\bigg)\bigg] \\
	&= \sum_{t \neq 0}E_n[b_i\psi_i^{[t]}(\hat{\eta})]\sqrt{n}\bigg(\frac{\hat{\pi}_t}{1-\hat{\pi}_0}-\frac{{\pi}_t}{1-{\pi_0}}\bigg) \\
	&= \sum_{t \neq 0}(\hat{\gamma}_t - \gamma_t + \gamma_t)\bigg(\frac{1-\pi_0}{1- \hat{\pi}_0} - 1 + 1\bigg)G_n[a_i^{[t]}] \\
	&= (\hat{\gamma} - \gamma)G_n[a_i]\bigg(\frac{1-\pi_0}{1- \hat{\pi}_0} - 1 + 1\bigg) 
	 + \bigg(\frac{1-\pi_0}{1- \hat{\pi}_0} - 1\bigg)\gamma G_n[a_i] + \gamma G_n[a_i]
\end{align*}}
where $\gamma_t = E[b_i\psi_i^{[t,0]}(\eta)] = E[b_i\tau_t(X_i)]$, $\gamma = (\gamma_1 \dots \gamma_J)$, and $a_i = (a_i^{[1]} \dots a_i^{[J]})'$. 
Now note that, by the iid assumption, Chebyshev's inequality yields {\footnotesize\begin{align*}
	||G_n[a_i]|| \lesssim_P J\sup_{t\neq 0}E[(a_i^{[t]})^2] \lesssim J\sup_{t\neq 0}\pi_t \lesssim 1
\end{align*}} Thus, by (H.5), we obtain that {\footnotesize\begin{align*}
	\sqrt{n}b'E_n[b_i(\psi_i(\hat{\eta},\hat{\pi})-\psi_i(\hat{\eta},\pi))] 
	&= b'\gamma G_n[a_i] + R_{n,\pi}
\end{align*}}
where {\footnotesize\begin{align*}
	||R_{n,\pi}|| &\lesssim_P ||\hat{\gamma} - \gamma||~||G_n[a_i]||~(1 + n^{-1/2}) + n^{-1/2}||\gamma||~||G_n[a_i]|| \\
	&\lesssim_P J\sqrt{k}(n^{-1/2} + m_{n,2} + Js_{n,2}) + n^{-1/2}\sqrt{Jk} \\
	&\lesssim J\sqrt{k}(n^{-1/2} + m_{n,2} + Js_{n,2})
\end{align*}}
as $||\frac{1-\pi_0}{1- \hat{\pi}_0} - 1|| \lesssim_P n^{-1/2}$ as $1-\pi_0$ is bounded away from zero. Recall that $\psi_i(\eta,\pi) - b_i'\beta_0 = \varepsilon_i + r_i$. Now what is left is the remainder due to estimation using $\hat{Q}$. We make use of (H.4) for bounding $||\hat{Q} - I||$. Conditional on the data, observe that {\footnotesize\begin{align*}
	V[b'(\hat{Q}^{-1} - I)G_n[b_i\varepsilon_i]|Z_1,\dots,Z_n] \lesssim b'(\hat{Q}^{-1} - I)\hat{Q}(\hat{Q}^{-1} - I)b 
	\lesssim_P \frac{\xi_k^2 \log k}{n} 
\end{align*}} Moreover, {\footnotesize\begin{align*}
|b'(\hat{Q}^{-1} - I)G_n[b_ir_i]| \lesssim_P \sqrt{\frac{\xi_k^2 \log k}{n} }l_kc_k\sqrt{k} 
\end{align*}} 
as in BCCK, Proof of Lemma 4.1 and using (H.7). For the final term note that $\sup_{z\in\mathcal{Z}}$ $||E[a_ia_i'|Z_i=z]|| \lesssim J^{-1/2}$ as in (H.5). This yields {\footnotesize\begin{align*}
	V[b'(\hat{Q}^{-1} - I)G_n[\gamma a_i]|Z_1,\dots,Z_n] &= b'(\hat{Q}^{-1} - I)\gamma E[a_ia_i'|Z_i]\gamma'(\hat{Q}^{-1} - I)b \\
	&\lesssim_P ||\hat{Q}^{-1}||^2~||\hat{Q}^{-1} - I||^2||~||\gamma||^2J^{-1/2} \\
	&\lesssim \frac{\xi_k^2 \log k}{n}k\sqrt{J}
\end{align*}}
Thus, decomposing and centering the linear predictor yields: {\footnotesize\begin{align*}
		\sqrt{n}b'(\hat{\beta} - \beta_0) &= b'G_n[b_i(\varepsilon_i + r_i) + \gamma a_i] + R_{n,Q} + R_{n,\pi} + R_{n,\eta}
\end{align*}}
where {\footnotesize\begin{align*}
	||R_{n,Q}|| &\lesssim_P \sqrt{\frac{\xi_k^2\log k}{n}}\bigg(1 + k^{1/2}(J^{1/4} + l_kc_k)\bigg) \\
	||R_{n,\pi}|| &\lesssim_P J\sqrt{k}(n^{-1/2} + m_{n,2} + Js_{n,2}) \\
	||R_{n,\eta}|| &\lesssim_P B_n^{[rATE]} + \Lambda_n^{[rATE]} \\
\end{align*}}

\vspace{-30pt}
\paragraph{\underline{$nATE$:}}
For $nATE$ there is no $G_n[a_i]$ term as there are no unconditional probabilities $\pi$ to be estimated. Using the different ML rates, we obtain: {\footnotesize\begin{align*}
	\sqrt{n}b'(\hat{\beta} - \beta_0) &= b'G_n[b_i(\varepsilon_i + r_i)] + R_{n,Q} + R_{n,\eta}  \\
	||R_{n,Q}|| &\lesssim_P \sqrt{\frac{\xi_k^2\log k}{n}}\bigg(1 + k^{1/2}l_kc_k\bigg) \\
	||R_{n,\eta}|| &\lesssim_P B_n^{[nATE]} + \Lambda_n^{[nATE]} \\
\end{align*}}

\noindent
where here $\varepsilon_i = \psi_i^{[nATE]}(\eta) - E[\psi_i^{[nATE]}(\eta)|Z_i]$ and $r_i = E[\psi_i^{[nATE]}(\eta)|Z_i] - b_i'\beta_0$. 

\subsubsection{Asymptotic Normality}
\paragraph{\underline{$rATE/\Delta$:}}
Again let $Q= I$ without loss of generality. Under Assumption C.4 all remainders from the previous subsection are $o_p(1)$ and we are left with leading term {\footnotesize\begin{align*}
	\frac{b'G_n[b_i(\varepsilon_i + r_i) + \gamma a_i]}{||b'\Omega^{1/2}||} 
	= \sum_{i=1}^n \frac{b'[b_i(\varepsilon_i + r_i) + \gamma a_i]}{\sqrt{n}||b'\Omega^{1/2}||}
\end{align*}}
We now verify the Lindeberg condition. First note that the term above has expectation zero. By independence and the binomial formula we obtain. for any $\delta > 0$. {\footnotesize\begin{align*}
	\sum_i^n &E\bigg[\frac{(b'(b_i(\varepsilon_i+r_i) + \gamma a_i))^2}{nb'\Omega b}\mathbbm{1}\bigg(\frac{|b'(b_i(\varepsilon_i+r_i) + \gamma a_i)|}{\sqrt{n}||b'\Omega^{1/2}||} > \delta \bigg)\bigg] \\
	&\leq 4n E\bigg[\frac{(b'b_i(\varepsilon_i+r_i))^2}{nb'\Omega b}\mathbbm{1}\bigg(\frac{|b'b_i(\varepsilon_i+r_i)|}{\sqrt{n}||b'\Omega^{1/2}||} > \frac{\delta}{2} \bigg)\bigg] + 4 n E\bigg[\frac{(b'\gamma a_i)^2}{nb'\Omega b}\mathbbm{1}\bigg(\frac{|b'\gamma a_i|}{\sqrt{n}||b'\Omega^{1/2}||} > \frac{\delta}{2} \bigg)\bigg] \\
	&\equiv (an.1) + (an.2)
\end{align*}}
Now denote {\footnotesize\begin{align*}
	w_{ni} := \frac{b'b_i}{||b'\Omega_1^{1/2}||} \ \Rightarrow \ |w_{ni}| \lesssim \frac{\xi_k}{\sqrt{n}}, \quad nE[|w_{ni}|^2] \lesssim 1
\end{align*}} 
analogously to BCCK Proof of Theorem 4.2 using the conditional moment bound (H.7). 

Now note that, by the eigenvalue assumption C.6, (an.1) is bounded by: {\footnotesize\begin{align*}
	(an.1) &\lesssim \frac{nb'\Omega_1 b}{nb'\Omega b}E\bigg[\frac{(b'b_i(\varepsilon_i+r_i))^2}{nb'\Omega_1 b}\mathbbm{1}\bigg(\frac{|b'b_i(\varepsilon_i+r_i)|}{\sqrt{n}||b'\Omega_1^{1/2}||} > \frac{\delta}{2}\frac{||b'\Omega^{1/2}||}{||b'\Omega_1^{1/2}||} \bigg)\bigg] \\
	&\lesssim 2n E[|w_{ni}|^2\varepsilon_i^2\mathbbm{1}(|\varepsilon_i| + |r_i| > \delta/|w_{ni}|)] + 2n E[|w_{ni}|^2\sup_{z\in\mathcal{Z}}|r(z)|^2\mathbbm{1}(|\varepsilon_i| + |r_i| > \delta/|w_{ni}|)] \\
	&\equiv (an.1.i) + (an.1.ii)
\end{align*}}
Using C.2 $\sup_z|r(z)| \leq l_kc_k$ and (H.9), we obtain that {\footnotesize\begin{align*}
	(an.1.i) &\lesssim nE[|w_{ni}|^2E[\varepsilon_i^2\mathbbm{1}(|\varepsilon_i| > (\delta\sqrt{n}/(2c\xi_k) - l_kc_k))|Z_i]] \\
	& \lesssim \sup_{z\in\mathcal{Z}}E\bigg[\sup_{t\neq 0}\frac{\varepsilon_i(t)^2D_{t,i}}{e_t(X_i)^2}\mathbbm{1}\bigg(\frac{|\varepsilon_i(t)|D_{t,i}}{e_t(X_i)} > \frac{\delta\sqrt{n}}{2\xi_k} - l_kc_k\bigg)\bigg|Z_i=z\bigg] \\
	&\lesssim \sup_{z\in\mathcal{Z}}\sup_{t\neq 0}\sup_{x\in\mathcal{X}}\frac{\pi_t^2}{e_t(x)^2}\pi_t^{-2}E[\sup_{t\neq 0}\varepsilon_i(t)^2D_{t,i}\mathbbm{1}(|\varepsilon_i(t)|D_{t,i} > c_n/J)|Z_i=z]
\end{align*}} 
where $c_n = \bigg(\frac{\delta\sqrt{n}}{2\xi_k} - l_kc_k\bigg)$. Now by the integrated tail formula we have that 

\noindent
{\footnotesize\begin{align*}
	E[&\sup_{t\neq 0}~\varepsilon_i(t)^2D_{t,i}\mathbbm{1}(|\varepsilon_i(t)|D_{t,i} > c_n/J|Z_i=z)] \\
	&= \int_0^{\infty}P(\sup_{t\neq 0} \varepsilon_i(t)^2D_{t,i}\mathbbm{1}(|\varepsilon_i(t)|D_{t,i} > c_n/J) > w|Z_i=z)dw \\
	&\leq \sum_{t\neq 0}P(D_{t,i}=1|Z_i=z) \int_0^{\infty}P(\varepsilon_i(t)^2\mathbbm{1}(|\varepsilon_i(t)| > c_n/J) > w|Z_i=z, D_{t,i} = 1)dw \\
	&\lesssim J \sup_{t\neq 0} \pi_t \int_0^{\infty} P(\varepsilon_i(t)^2\mathbbm{1}(|\varepsilon_i(t)| > c_n/J) > w|Z_i=z, D_{t,i} = 1)dw \\
	&\lesssim \sup_{t\neq 0} E[\varepsilon_i(t)^2\mathbbm{1}(|\varepsilon_i(t)| > c_n/J)|Z_i=z, D_{t,i} = 1]
\end{align*}}
Note that, by Markov's inequality and the moment bound in C.1, we have that {\footnotesize\begin{align*}
	P(|\varepsilon_i(t)| > c_n/J|Z_i=z,D_{t,i}=1) \lesssim (c_n/J)^{-m}
\end{align*}}
H\"older's inequality then yields for the equation above {\footnotesize\begin{align*}
	E[\varepsilon_i(t)^2\mathbbm{1}(|\varepsilon_i(t)| > c_n/J)|Z_i=z, D_{t,i} = 1]
	&\leq \bigg(E[|\varepsilon_i(t)|^{m}|Z_i=z, D_{t,i} = 1](c_n/J)^{-m}\bigg)^{2/m} \\
	&\lesssim \bigg[\bigg(\frac{\delta \sqrt{n}}{2 c \xi_k} - l_kc_k\bigg)\frac{1}{J}\bigg]^{-2}
\end{align*}}
for any $t \neq 0$. Plugging this back into $(an.1.i)$ and using A.3 $\sup_{t\neq 0}\pi_t J \lesssim 1$ yields {\footnotesize\begin{align*}
		(an.1.i) &\lesssim \bigg[\bigg(\frac{\delta \sqrt{n}}{2 c \xi_k} - l_kc_k\bigg)\frac{1}{J}\bigg]^{-2}J^2 
		\lesssim \frac{\xi_k^2J^4}{n}
	\end{align*}}
As $\delta\sqrt{n}/\xi_k - l_kc_k \rightarrow \infty$. For $(an.1.ii)$, equivalently to BCCK Theorem 4.2, {\footnotesize\begin{align*}
	(an.1.ii) \lesssim l_k^2c_k^2 \sup_{z\in\mathcal{Z}}P(|\varepsilon_i| > c\delta\sqrt{n}/\xi_k - l_kc_k|Z_i=z) 
\end{align*}} Analogously to the derivations for $(an.1.i)$ using (H.9) we obtain that {\footnotesize\begin{align*}
P(|\varepsilon_i| > c_n|Z_i=z) &\lesssim \sum_{t \neq 0}P(|\varepsilon_i(t)| > c_n/J|Z_i=z,D_{t,i}=1)P(D_{t,i}=1|Z_i=z) \\
&\lesssim \sup_{t\neq 0} P(|\varepsilon_i(t)| > c_n/J|Z_i=z,D_{t,i}=1) \\
&\lesssim \sup_{t\neq 0} \frac{E[|\varepsilon_i(t)|^{m}|Z_i=z,D_{t,i}=1]}{(c_n/J)^{m}} \\
&\lesssim (c_n/J)^{-m}
\end{align*}}
where the last two steps follow from Markov's inequality and the conditional moment bound. Plugging this back into $(an.1.ii)$ yields 

\noindent
{\footnotesize\begin{align*}
	(an.1.ii) \lesssim \bigg(\frac{(l_kc_k)^{\frac{2}{m}}J}{[\delta\sqrt{n}/\xi_k - l_kc_k]}\bigg)^{m}
\end{align*}}
Now consider $(an.2)$. Note that by (H.5), we have that {\footnotesize\begin{align*}
	(an.2) &\leq \frac{||\gamma||^2}{b'\Omega b}E[||a_i||^2\mathbbm{1}(||a_i|| > (\delta/2)||b'\Omega^{1/2}||/||\gamma||)] \\
	&\lesssim  ||\gamma||^2E[||a_i||^2\mathbbm{1}(||a_i|| > C\delta\sqrt{n/kJ})] \\
	&\lesssim kJ\sum_{t \neq 0}E[(a_i^{[t]})^2\mathbbm{1}(||a_i|| > C\delta\sqrt{n/kJ})] \\
	&\lesssim kJ^2 P(||a_i|| > C\delta\sqrt{n/kJ}) \\
	&\lesssim kJ^2 P(\sup_{t\neq 0}(a_i^{[t]})\sqrt{J} > C\delta\sqrt{n/kJ}) \\
	&\lesssim kJ^3 \sup_{t\neq 0}P(|a_i^{[t]}| > C\delta\sqrt{n/kJ^2}) \\
	&= o(1)
\end{align*}}
if $\sqrt{n/kJ^2} \rightarrow \infty$ as $a_i^{[t]}$ is uniformly bounded for all $t$. Thus using Assumption C.5 yields {\footnotesize\begin{align*}
	(an.1) + (an.2) \lesssim  \frac{\xi_k^2J^4}{n} +  \bigg(\frac{(l_kc_k)^{\frac{2}{m}}J}{[\delta\sqrt{n}/\xi_k - l_kc_k]}\bigg)^{m} = o(1) 
\end{align*}}

\vspace{-13pt}
\paragraph{\underline{$nATE$:}}
There is no $(an.2)$ term for the $nATE$. For $(an.1)$ most derivations follow analogously. However, due to (H.9), improved rates can obtained: {\footnotesize\begin{align*}
	E[\varepsilon_i^2\mathbbm{1}(|\varepsilon_i| > c_n)|Z_i=z] 
	&\lesssim E[\sup_{t\neq 0}\varepsilon_i(t)^2\mathbbm{1}(|\varepsilon_i(t)| >c_n)|Z_i=z] \\
	&\lesssim_P J \sup_{z\in\mathcal{Z}}\sup_{t\neq 0}E[\varepsilon_i(t)^2\mathbbm{1}(|\varepsilon_i(t)| >c_n)|Z_i=z] \\
	&\lesssim Jc_n^{-2}
\end{align*}}
Similarly for $(an.1.ii)$, we have that {\footnotesize\begin{align*}
	P(|\varepsilon_i| > c_n) &\leq J \sup_{t\neq 0} P(|\varepsilon_i(t)| > c_n) \lesssim Jc_n^{-m}
\end{align*}}
Plugging both into $(an.1)$ then with $c_n$ as above yields
{\footnotesize\begin{align*}
	(an.1) \lesssim  J\frac{}{[\delta\sqrt{n}/\xi_k - l_kc_k]^{2}}+ J\frac{l_k^2c_k^2}{[\delta\sqrt{n}/\xi_k - l_kc_k]^{m}} = o(1) 
\end{align*}}
under the assumption B.5. Note that convergence is faster than $(an.1)$ for the $rATE$. Asymptotic normality then follows from the sufficiency of the Lindeberg condition. $\square$
 
\newpage
\section{Supplementary Appendix (not for publication)}
\subsection{Auxiliary results for Appendix A} \label{sec_app_auxNEW}
	\textbf{(H.1): $||\hat{\gamma}_t - \gamma_t||$ bound} \\
	{\footnotesize\begin{align*}
		||\hat{\gamma}_t - \gamma_t|| 
		&\leq ||E_n[b_i(\psi_i^{[t]}(\hat{\eta}) - \psi_i^{[t]}(\eta))]|| - ||E_n[b_i \psi_i^{[t]}(\eta) - E[b_i\psi_i^{[t]}(\eta)]]||
	\end{align*}}
	Using Markov's inequality and Cauchy-Schwarz together with the definition of the moment function yields for the first term {\footnotesize\begin{align*}
		||E_n[b_i(\psi_i^{[t]}(\hat{\eta}) - \psi_i^{[t]}(\eta))]|| 
		&\lesssim_P E[||b_i(\psi_i^{[t]}(\hat{\eta})- \psi_i^{[t]}(\eta))||] \\
		&\lesssim_P E[||b_i||^2]^{1/2}E[(\psi_i^{[t]}(\hat{\eta})- \psi_i^{[t]}(\eta))^2]^{1/2} \\
		&\leq \sqrt{k}E[(\psi_i^{[t]}(\hat{\eta})- \psi_i^{[t]}(\eta))^2]^{1/2} \\
		&\lesssim_P \sqrt{kJ}(m_{t,n,2} + Js_{t,n,2})
	\end{align*}}
	as {\footnotesize\begin{align*}
		E[(\psi_i^{[t]}(\hat{\eta}) - \psi_i^{[t]}(\eta))^2] 
		&\leq 4 E\bigg[(\hat{\mu}_t(X_i)-\mu_t(X_i))^2(1-D_{t,i}/e_t(X_i))^2 + (\hat{e}_t(X_i) - e_t(X_i))^2\varepsilon_i(t)^2\frac{D_{t,i}}{e_t(X_i)^2\hat{e}_t(X_i)^2} \\ &\quad  + (\hat{\mu}_t(X_i)-\mu_t(X_i))^2(\hat{e}_t(X_i) - e_t(X_i))^2\frac{D_{t,i}}{e_t(X_i)^2\hat{e}_t(X_i)^2}\bigg] \\
		&\lesssim_P \sup_{x\in\mathcal{X}}\frac{\pi_t}{e_t(x)}\pi_t^{-1}\bigg(E[(\hat{\mu}_t(X_i) - \mu_t(X_i))^2] + \chi_{t,n}^2\pi_t^{-2}E[(\hat{e}_t(X_i) - e_t(X_i))^2] \bigg) \\
		&\lesssim_P J(m_{t,n,2}^2 + J^2s_{t,n,2}^2)		
	\end{align*}} 
	For the second term we have 

 \vspace{-24pt}
 {\footnotesize\begin{align*}
		||E_n[b_i \psi_i^{[t]}(\eta) - E[b_i\psi_i^{[t]}(\eta)]]||
		&\lesssim_P E[||E_n[b_i \psi_i^{[t]}(\eta)] - E[b_i\psi_i^{[t]}(\eta)]||] \\
		&\leq  E[||E_n[b_i \psi_i^{[t]}(\eta)]||^2]^{1/2} \\
		&= (E[\psi_i^{[t]}(\eta)^2b_i'b_i]/n)^{1/2} \\
		&= (E[E[\psi_i^{[t]}(\eta)^2|X_i]b_i'b_i/n])^{1/2} \\
		&= \bigg(E\bigg[\bigg(\frac{\sigma_t^2(X_i)}{e_t(X_i)} + \mu_t(X_i)^2\bigg)b_i'b_i\bigg]/n\bigg)^{1/2} \\
		&\lesssim \bigg(\bigg(\sup_{x\in\mathcal{X}}\frac{\pi_t}{e_t(x)}\pi_t^{-1} + 1\bigg)E[b_i'b_i]/n\bigg)^{1/2} \\
		&\lesssim \sqrt{\frac{Jk}{n}}
	\end{align*}}
	by conditional independence and bounded second moments. Overall we obtain {\footnotesize\begin{align*}
		||\hat{\gamma}_t - \gamma_t|| \lesssim_P 
		\sqrt{kJ}\bigg(n^{-1/2} + m_{t,n,2} + Js_{t,n,2}\bigg)
	\end{align*}} 
	\textbf{(H.2): $||\gamma_t||$ rate} 
	{\footnotesize\begin{align*}
		||\gamma_t|| &= ||E[b_i\psi_i^{[t]}(\eta)]|| = ||E[b_i\mu_t(X_i)]|| \lesssim \sup_{x\in\mathcal{X}}|\mu_t(x)|E[||b_i||] \lesssim \sqrt{k}
	\end{align*}}
	\textbf{(H.3): $||E_n[b_i(\psi_i(\hat{\eta},\hat{\pi}) - \psi_i(\hat{\eta},{\pi}))]||$ rate} 
	{\footnotesize\begin{align*}
		||E_n[b_i(\psi_i(\hat{\eta},\hat{\pi}) - \psi_i(\hat{\eta},{\pi}))]||
		&= ||\sum_{t \neq 0}E_n[b_i\psi_i^{[t]}(\hat{\eta})]\bigg(\frac{\hat{\pi}_t}{1-\hat{\pi}_0} - \frac{{\pi}_t}{1-{\pi_0}} \bigg) || \\
		&\lesssim_P J\sup_{t\neq 0}(||\gamma_t|| + ||\hat{\gamma}_t - \gamma_t||)|\hat{\pi}_t - \pi_t|(1 + n^{-1/2}) \\
		&\lesssim_P J\sqrt{k}\sqrt{\frac{1}{nJ}} \\
		&= \sqrt{\frac{Jk}{n}}
	\end{align*}}
\textbf{(H.4) Restatement of Lemma 6.2 from BCCK of the \citeA{Rudelson1999RandomPosition} LLN for Matrices} Let $Q_1,\dots,Q_n$ be a sequence of independent symmetric non-negative $k\times k$-matrix valued random variables with $k\geq 2$ such that $Q = E_n[E[Q_i]]$ and $||Q_i|| \leq M$ a.s., then for $\hat{Q} = E_n[Q_i]$ {\footnotesize\begin{align*}
	E[||\hat{Q} - Q||] \lesssim \frac{M \log k}{n} + \sqrt{\frac{M ||Q|| \log k}{n}}.\end{align*}}
	\textbf{(H.5) $\gamma$, $a_i$, and $a_ia_i'$ rates}
	Define $A_n = E_n[a_ia_i']$ and $\hat{A}_n = E_n[\hat{a}_i\hat{a}_i']$.
	Recall the definitions:
		$\gamma = (\gamma_1,\dots,\gamma_J),~\gamma_t = E[b_i\tau_t(X_i)],~a_i = (a_i^{[1]},\dots,a_i^{[J]}).$
	Thus {\footnotesize\begin{align*}
		||\gamma|| &\lesssim \sqrt{J}\sup_{t\neq 0}||\gamma_t|| \lesssim \sqrt{Jk} \\
		||\hat{\gamma} - \gamma || &\lesssim_P \sqrt{J}\sup_{t\neq 0}||\hat{\gamma}_t - \gamma_t|| \lesssim_P \sqrt{kJ^2}(n^{-1/2} + m_{n,2} + Js_{n,2}) \\
		E[||a_i||] &\leq \sqrt{J}\sup_{t\neq 0}E[(a_i^{[t]})^2]^{1/2} 
		\lesssim_P \sqrt{J}\sup_{t\neq 0}\sqrt{\pi_t} \lesssim 1 \\
		E[||\hat{a}_i - a_i||] &\leq \sqrt{J}\sup_{t\neq 0}E[(\hat{a}_i^{[t]} - a_i^{[t]})^2]^{1/2} \lesssim \sqrt{J}\sup_{t\neq 0}\sqrt{\frac{\pi_t}{n}} \lesssim n^{-1/2}
	\end{align*}}
	as $\hat{a}_i^{[t]}-a_i^{[t]} \lesssim_P \pi_t|\hat{\pi}_0 - \pi_0|  + \pi_0|\hat{\pi}_t - \pi_t|$. Equivalently, we have {\footnotesize\begin{align*}
		||A_n|| &\leq E_n[||a_ia_i'||] \lesssim_P E[||a_i||^2] \leq J \sup_{t\neq 0}\pi_t \lesssim 1 \\
		||\hat{A}_n - A_n|| &\lesssim E_n[||(\hat{a_i} - a_i)a_i'||] \lesssim_P  \sqrt{J} \sup_{t\neq 0}\bigg(\pi_t(\hat{\pi}_0 - \pi_0)^2 + (\hat{\pi}_t - \pi_t)^2 \bigg)^{1/2} \lesssim_P n^{-1/2}
	\end{align*}}
	We further have that {\footnotesize\begin{align*}
		\max_{1\leq i \leq n}||a_ia_i'|| 
		&\leq  \sqrt{J}\max_{1\leq i \leq n}||a_ia_i'||_1 \\
		&\lesssim_P \sqrt{J} \max_{1\leq i \leq n} \sum_{t \neq 0}(D_{t,i}(1-\pi_0) - \pi_t + D_{0,i}\pi_t)^2 \\
		&\lesssim_P \sqrt{J} (1 + J^{-2} + J^{-1}) \\
		&\lesssim \sqrt{J}
	\end{align*}}
	and for the expectation {\footnotesize\begin{align*}
		||E[a_ia_i']|| &\leq \sqrt{J}||E[a_ia_i']||_1 \\
		&\leq \sqrt{J}\sup_{t\neq 0}(E[(a_i^{[t]})^2] + \sum_{s\neq t,0}E[a_i^{[t]}a_i^{[s]}]) \\
		&\leq \sqrt{J} \sup_{t\neq 0} (\pi_t  + \sum_{s\neq t,0}\pi_t\pi_s) \\
		&\lesssim J^{-1/2}
	\end{align*}}
	Now note that $a_ia_i'$ are symmetric, non-negative iid matrix valued random variables. Thus using Rudelson's LLN (H.4) 
	and $J = o(n)$ we obtain {\footnotesize\begin{align*}
		E[||E_n[a_ia_i'] - E[a_ia_i']||]
		\lesssim \frac{J^{1/2}\log J}{n} + \sqrt{\frac{J^{1/2}||E[a_ia_i']||\log J}{n}} 
		\lesssim \sqrt{\frac{\log J}{n}}
	\end{align*}} 
		\textbf{(H.6) Linearization of the unconditional weights: }
	{\footnotesize\begin{align*}
		\frac{\hat{\pi}_t}{\sum_{t \neq 0} \hat{\pi}_t } - \frac{\pi_t}{\sum_{t \neq 0} {\pi}_t } 
		&= \frac{\hat{\pi}_t}{1- \hat{\pi}_0 } -  \frac{ {\pi}_t}{1- {\pi}_0 } \\
		&= \frac{\hat{\pi}_t(1-\pi_0) - (1-\pi_0)\pi_t + (1-\pi_0)\pi_t - \pi_t(1-\hat{\pi}_0)}{(1-\pi_0)(1-\hat{\pi}_0)} \\
		&= \frac{1 - \pi_0}{1 - \hat{\pi}_0}\frac{1}{(1 - \pi_0)^2}E_n[([D_{t,i} - \pi_t](1-\pi_0) + [D_{0,i} - \pi_0]\pi_t )]
	\end{align*}}
	Defining $a_i^{[t]} = (1-\pi_0)^{-2}(D_{t,i}(1-\pi_0) + D_{0,i}\pi_t - \pi_t)$ then yields {\footnotesize\begin{align*}
		\sqrt{n}\bigg(\frac{\hat{\pi}_t}{1-\hat{\pi}_0}-\frac{{\pi}_t}{1-{\pi_0}}\bigg)
		&= \frac{1 - \pi_0}{1 - \hat{\pi}_0}G_n[a_i^{[t]}]
	\end{align*}}
 \noindent
	\textbf{(H.7) $rATE$ Conditional mean error variance: }
	Recall that $\varepsilon_i = \psi_i(\eta,\pi) - E[\psi_i(\eta,\pi)|Z_i]$.
	The $rATE$ conditional mean error has finite second conditional moment:
	{\footnotesize\begin{align*}
		E[\varepsilon_i^2|Z_i] &= (1-\pi_0)^{-2}\sum_{t\neq 0}\sum_{t'\neq 0}\pi_t\pi_t'E[(\psi_i^{[t]}(\eta) - E[\psi_i^{[t]}(\eta)|Z_i])(\psi_i^{[t']}(\eta) - E[\psi_i^{[t']}(\eta)|Z_i])|Z_i] \\
		&\lesssim_P \sum_{t \neq 0}\pi_t^2E\bigg[\frac{\sigma_t^2(X_i)}{e_t(X_i)} + (\mu_t(X_i) - E[\mu_t(X_i)|Z_i])^2 \bigg|Z_i\bigg] \\
		&\quad  + \sum_{t \neq 0}\sum_{t' \neq 0}\pi_t\pi_{t'}E[(\mu_t(X_i) - E[\mu_t(X_i)|Z_i])(\mu_{t'}(X_i) - E[\mu_{t'}(X_i)|Z_i])|Z_i] \\
		&\lesssim J\sup_{t\neq 0}\pi_t \sup_{x\in\mathcal{X}}\frac{\pi_t}{e_t(x)} + J^2 \sup_{t\neq 0}\pi_t^2 \\
		&\lesssim 1
	\end{align*}}
	As $\sup_{t,x\in\mathcal{X}} \sigma_t^2(x) + \mu_t(x)$ is uniformly bounded by Assumptions A.2 and B.1/C.1. \\ \noindent
	\textbf{(H.8) Effect of estimating $\pi$ for decomposition term $\Delta = nATE - rATE$:} Note that due to the multiplicative structure of the decomposition we have that, for any $\eta$,
	{\footnotesize\begin{align*}
		\psi_i^{[\Delta]}(\eta,\hat{\pi}) - \psi_i^{[\Delta]}(\eta,{\pi}) 
		&=[\psi_i^{[nATE]}(\eta,\hat{\pi}) - \psi_i^{[rATE]}(\eta,\hat{\pi})] - [\psi_i^{[nATE]}(\eta,{\pi}) - \psi_i^{[rATE]}(\eta,{\pi})]  \\
		&= \sum_{t \neq 0}\psi_i^{[t,0]}(\eta)\bigg[\frac{\pi_t}{\sum_{t \neq 0}\pi_t} - \frac{\hat{\pi}_t}{\sum_{t \neq 0}\hat{\pi}_t}\bigg] \\
		&= - (\psi_i^{[rATE]}(\eta,\hat{\pi}) - \psi_i^{[rATE]}(\eta,{\pi})) 
	\end{align*}}
	Thus an analogous asymptotically linear representation for the estimator of $\Delta$ can be obtained by the $rATE$ results with a sign flip. The leading term for normality and its asymptotic variance will also be identical up to sign. \\ \noindent
	\textbf{(H.9) Error and error tail bounds}
	For the $rATE$, the regression error by definition is a convex combination of centered moment functions {\footnotesize\begin{align*}
		\varepsilon_i = \frac{\sum_{t \neq 0}\pi_t(\psi_i^{[t]}(\eta) - E[\psi_i^{[t]}(\eta)|Z_i])}{\sum_{t \neq 0}\pi_t}
	\end{align*}}
	and thus {\footnotesize\begin{align*}
		|\varepsilon_i| \leq \sup_{t\neq 0}|\psi_i^{[t]}(\eta) - E[\psi_i^{[t]}(\eta)|Z_i]| \lesssim \sup_{t\neq 0}\bigg|\frac{\varepsilon_i(t)D_{t,i}}{e_t(X_i)}\bigg| + 1
	\end{align*}}
	almost surely. 
	For the $nATE$, however, propensity score weights yield {\footnotesize\begin{align*}
		\varepsilon_i = \frac{\sum_{t \neq 0}D_{t,i}\varepsilon_i(t) + e_t(X_i)\mu_t(X_i) -E[D_{t,i}\varepsilon_i(t) + e_t(X_i)\mu_t(X_i)|Z_i]}{\sum_{t \neq 0}e_t(X_i)} 
	\end{align*}}
	and thus 
		$|\varepsilon_i| \lesssim \sup_{t\neq 0}|\varepsilon_i(t)| + 1$ 
	almost surely.

\subsection{Toy example} \label{sec:app-ex}

Consider a setting with a binary heterogeneity variable $X_i \in \{0,1\}$ and three effective treatments $T_i \in \{0,1,2\}$. We impose deterministic potential outcomes that are homogeneous within treatment status, but heterogeneous between treatments:
\begin{center} \footnotesize
	\begin{tabular}{l|ccc} 
		& $Y_i(0)$ & $Y_i(1)$ & $Y_i(2)$ \\ \hline 
$X_i = 0$ & 0 & -1 & 1 \\
$X_i = 1$ & 0 & -1 & 1 
	\end{tabular} 
\end{center}
Both groups defined by $X_i$ have the same potential outcomes under the different treatments. This means there can be no real effect heterogeneity. However, consider now that the probability to receive the effective treatments varies with $X_i$:
\begin{center} \footnotesize
	\begin{tabular}{l|ccc}
	& $P(T_i = 0| X_i)$ & $P(T_i = 1| X_i)$ & $P(T_i = 2| X_i)$ \\ \hline 
	$X_i = 0$ & 0.5 & 1/8 & 3/8 \\
	$X_i = 1$ & 0.5 & 3/8 & 1/8 
\end{tabular}
\end{center}
Collapsing treatments one and two into a binary treatment $D_i = \mathbbm{1}(T_i > 0)$ yields the subgroup conditional average treatment effects ($CATE$):
 {\footnotesize\begin{align*}
CATE(X_i) = 1 - 4 \cdot P(T_i = 1| X_i) =\begin{cases}
		~\ 0.5 &\text{ if } X_i = 0 \\
		-0.5 &\text{ if } X_i = 1.
	\end{cases}
\end{align*}}
Thus, binary aggregation leads us to find a positive effect for one group and a negative effect for another group although the effective treatments actually do not create heterogeneous effect. Everything is driven by groups receiving a different composition.

\subsection{Neyman-orthogonality} \label{app_neyman}
The key insight is that the $nATE$ and $rATE$ scores are Neyman-orthogonal with known probabilities $\pi_t$, $t=1,\dots,J$. We show how the additional estimation error can be incorporated in Appendix \ref{app_Theory1}. Here we consider the Gateaux derivative for $nATE$ and $rATE$ scores with respect to the infinite-dimensional nuisance parameters $\eta = (\mu(x),p(x)) = (\mu_0(x),\dots,\mu_J(x),e_0(x),\dots,e_J(x))'$. As $\pi$ is assumed to be known, we suppress dependence $\psi(\eta,\pi) = \psi(\eta)$ out of convenience for now. Suppressing also the dependencies of the nuisance parameters on $x$, we write the path-wise derivative of the conditional expectation of a score with respect to the vector of nuisance parameters as{\footnotesize\begin{align*}
\partial_{\eta} E[\psi_i(\eta)|X_i=x] = \partial_{r} E[\psi_i(\dots,\mu_t + r(\tilde{\mu}_t - \mu_t),\dots,e_t + r(\tilde{e}_t - e_t),\dots)|X_i=x]|_{r = 0}
\end{align*}}
First, we revisit Neyman-orthogonality of the doubly robust score:
{\footnotesize\begin{align*}
\partial_{r} E&[\psi_i^{[t]}(\eta+r(\tilde{\eta} - \eta)) | X_i = x ]|_{r = 0} \\
&= \partial_{r} E\left[(\mu_t + r(\tilde{\mu}_t - \mu_t)) + \frac{D_{t,i} Y_i}{e_t + r(\tilde{e}_t - e_t)}  - \frac{D_{t,i}(\mu_t + r(\tilde{\mu}_t - \mu_t))}{e_t  + r(\tilde{e}_t - e_t)} \middle| X_i = x \right]\bigg|_{r = 0} \\
&= (\tilde{\mu}_t - \mu_t) - \frac{e_t \mu_t (\tilde{e}_t - e_t)}{e_t^2}  - \frac{e_t^2(\tilde{\mu}_t - \mu_t) - e_t\mu_t (\tilde{e}_t - e_t)}{e_t^2} \\
&=0
\end{align*}}
where we use that $E[D_{t,i} Y_i | X_i = x] = E[D_{t,i} \sum_t D_{t,i} Y_i(t) | X_i = x] = E[D_{t,i} Y_i(t) | X_i = x] = e_t \mu_t$ by the observational rule and Assumption \ref{ass:si-v}. 

$rATE$ score is then a linear combination of doubly robust scores inheriting their Neyman-orthogonality:
{\footnotesize\begin{align*}
\partial_{r} E[\psi_i^{[rATE]}(\eta+r(\tilde{\eta} - \eta)) | X_i = x ]|_{r = 0} 
&=  \sum_{t\neq 0} \frac{\pi_t}{1-\pi_0} \partial_{r} E[\psi_i^{[t]}(\eta+r(\tilde{\eta} - \eta)) | X_i = x ]|_{r = 0} \\
&\quad - \partial_{r} E[\psi_i^{[0]}(\eta+r(\tilde{\eta} - \eta)) | X_i = x ]|_{r = 0} \\
&=0
\end{align*}}
$nATE$ score differs from the standard doubly robust scores but is still orthogonal:
{\footnotesize\begin{align*}
\partial_{r} E&[\psi_i^{[nATE]}(\eta+r(\tilde{\eta} - \eta)) | X_i = x ]|_{r = 0} \\
&=\partial_{r} E\left[\frac{\sum_{t\neq 0} [(\mu_t + r(\tilde{\mu}_t - \mu_t))(e_t + r(\tilde{e}_t - e_t))]}{\sum_{t\neq 0}(e_t + r(\tilde{e}_t - e_t))} + \frac{D_i Y_i}{\sum_{t\neq 0}(e_t + r(\tilde{e}_t - e_t))} \right. \\ 
&\left.\quad\quad\quad\quad - \frac{D_i\sum_{t\neq 0} [(\mu_t + r(\tilde{\mu}_t - \mu_t))(e_t + r(\tilde{e}_t - e_t))]}{[\sum_{t\neq 0}(e_t + r(\tilde{e}_t - e_t))]^2} \middle| X_i = x \right]\Bigg|_{r = 0} \\
&\quad - \partial_{r} E[\psi_i^{[0]}(\eta+r(\tilde{\eta} - \eta)) | X_i = x ]|_{r = 0} \\
&= \frac{\sum_{t\neq 0} [\mu_t (\tilde{e}_t - e_t) + e_t(\tilde{\mu}_t - \mu_t)]\sum_{t\neq 0}e_t}{[\sum_{t\neq 0}e_t]^2} - \frac{\sum_{t\neq 0} \mu_t e_t \sum_{t\neq 0}(\tilde{e}_t - e_t)}{[\sum_{t\neq 0} e_t]^2} \\
&\quad - \frac{\sum_{t\neq 0} e_t \mu_t \sum_{t\neq 0}(\tilde{e}_t - e_t)}{[\sum_{t\neq 0}e_t]^2} - \frac{\sum_{t\neq 0} e_t \sum_{t\neq 0} [\mu_t (\tilde{e}_t - e_t) + e_t(\tilde{\mu}_t - \mu_t)]}{[\sum_{t\neq 0}e_t]^2} \\ 
&\quad + 2 \frac{ \sum_{t\neq 0} \mu_t e_t \sum_{t\neq 0}(\tilde{e}_t - e_t)}{[\sum_{t\neq 0}e_t]^2} - \partial_{r} E[\psi_i^{[0]}(\eta+r(\tilde{\eta} - \eta)) | X_i = x ]|_{r = 0} \\
&=0
\end{align*}}
\noindent

where we use that $E[D_i Y_i | X_i = x ] = E[D_i \sum_{t} D_{t,i} Y_i(t) | X_i = x ] = \sum_{t\neq 0} e_t \mu_t $ by the observational rule and Assumption \ref{ass:si-v}. Consequently, the difference between the $nATE$ and $rATE$ score that forms the $\Delta$ score is Neyman-orthogonal as well:
{\footnotesize\begin{align*}
\partial_{\eta} E[\psi_i^{[nATE]}(\eta) - \psi_i^{[rATE]}(\eta)|X_i=x] = 0
\end{align*}}
\vspace{-48pt}
        \subsection{Estimation of Asymptotic Variance} \label{sec_AsyVar1}
        Let $E_n[X_i] = \frac{1}{n}\sum_{i=1}^{n}X_i$. Define
		{\footnotesize\begin{align}
			\hat{Q} = E_n[b(Z_i)b(Z_i)'], \quad 
			\hat{\Omega} = \hat{Q}^{-1}\hat{\Sigma}\hat{Q}^{-1} \label{eq_AVest1} 
			\end{align}}
        For the $rATE$ we use
		{\footnotesize\begin{align*}
			\hat{\Sigma} &= E_n\bigg[(b(Z_i)e_i + \hat{a}_i - \bar{\hat{a_i}})(b(Z_i)e_i + \hat{a}_i - \bar{\hat{a_i}})'\bigg]
		\end{align*}}
        with $e_i = \psi_i^{[rATE]}(\hat{\eta},\hat{\pi}) - b(Z_i)'\hat{\beta}$, $\hat{\pi}_t = E_n[D_{t,i}]$  and {\footnotesize\begin{align*}
			\hat{a}_i &= \sum_{t \neq 0}\frac{E_n[b(Z_i)(\psi_i^{[t]}(\hat{\eta})-\psi_i^{[0]}(\hat{\eta}))](D_{t,i}(1-\hat{\pi}_0) + D_{0,i}\hat{\pi}_t)}{(1-\hat{\pi}_0)^2} \\
			\hat{\bar{a}}_i &= E_n[\hat{a}_i] 
		\end{align*}}
        For $\Delta$, the $\psi_i^{[rATE]}(\hat{\eta},\hat{\pi})$ has to be replaced by its corresponding score and $\hat{\Sigma}$ changes to
		{\footnotesize\begin{align*}
			\hat{\Sigma} &= E_n\bigg[(b(Z_i)e_i - \hat{a}_i + \bar{\hat{a_i}})(b(Z_i)e_i - \hat{a}_i + \bar{\hat{a_i}})'\bigg].
		\end{align*}}
        For $nATE$ we use $\hat{\Sigma} = E_n[b(Z_i)b(Z_i)'e_i^2]$ as there are no estimated unconditional weights.

\subsection{Asymptotic Variance Estimation Theory} \label{sec_app_AV1}
In the following, we use some Lemmas of BCCK. To do so, we impose the following additional assumption:  $\xi_k^{2m/(m-2)}\log k /n \lesssim 1 $, $\log \xi_k \lesssim \log k$ and Lipschitz constant {\footnotesize\begin{align*}
	\xi_k^L := \sup_{x,x'\in\mathcal{X},x\neq x'} \frac{||b(x) - b(x')||}{||x - x'||}
\end{align*}} 
obeys Condition (A.5) in BCCK, i.e.~$\log \xi_k^L \lesssim \log k$. Moreover, $c_kl_k \lesssim \sqrt{\log k}$ and $\sqrt{\frac{\xi_k^2\log k}{n}}\bigg((nJ)^{1/m}\sqrt{\log k } + \sqrt{k}l_kc_k\bigg) \lesssim \sqrt{\log k}$ as in Theorem 4.6 by BCCK. Moreover, assume $M_{n,1} = o(1)$ for $nATE$ or $M_{n,2} = o(1)$ for $rATE$/$\Delta$ where $M_{n,1}$ and $M_{n,2}$ are defined at the end of the section in equations \eqref{eq_AV_nATE} and \eqref{eq_AV_rATE} respectively. We first provide auxiliary results and then derivations for $rATE/\Delta$. Rates for $nATE$ follow by simplification. 

\subsubsection{Auxiliary Results}
\paragraph{(MA.1) $\psi_i^{[t]}$ bounds}
First note that, for any $t\neq 0$,
{\footnotesize\begin{align*}
	|\psi_i^{[t]}(\eta)| 
	= \bigg|\frac{D_{t,i}\varepsilon_i(t)}{e_t(X_i)} + \mu_t(X_i)\bigg| 
	\leq \sup_{x\in\mathcal{X}}\frac{\pi_t}{e_t(x)}\pi_t^{-1}|D_{t,i}||\varepsilon_i(t)| + |\mu_t(X_i)|
\end{align*}}
Thus {\footnotesize\begin{align*}
	\max_{1\leq i \leq n}|\psi_i^{[t]}(\eta)| &\lesssim_P \sup_{x\in\mathcal{X}}\frac{\pi_t}{e_t(x)}\pi_t^{-1}\max_{1\leq i \leq n}|\varepsilon_i(t)| + \sup_{x\in\mathcal{X}}|\mu_t(x)| 
	\lesssim_P Jn^{1/m} 
\end{align*}}
by the Assumption A.2 and B.1/C.1. Moreover
{\footnotesize\begin{align*}
	|\psi_i^{[t]}(\eta)\psi_i^{[t']}(\eta)| 
	\leq   |\psi_i^{[t]}(\eta)|^2 + |\psi_i^{[t']}(\eta)|^2 
	\leq 2\sup_{t\neq0}|\psi_i^{[t]}(\eta)|^2
\end{align*}}
and similarly
{\footnotesize\begin{align*}
	|\psi_i^{[t]}(\hat{\eta})-\psi_i^{[t]}({\eta})||\psi_i^{[t']}(\hat{\eta})-\psi_i^{[t']}({\eta})|	&\leq 2\sup_{t\neq0}|\psi_i^{[t]}(\hat{\eta})-\psi_i^{[t]}({\eta})|^2
\end{align*}}
Now consider the product of the moment functions for any $\eta$. 
For the square note that {\footnotesize\begin{align*}
	|\psi_i^{[t]}(\eta)|^2 
	&=  \bigg|\frac{D_{t,i}\varepsilon_i(t)}{e_t(X_i)} + \mu_t(X_i)\bigg|^2 \\
	&= \frac{D_{t,i}\varepsilon_i(t)^2}{e_t(X_i)^2} + 2\mu_t(X_i)\frac{D_{t,i}\varepsilon_i(t)}{e_t(X_i)} + \mu_t(X_i)^2 \\
	&\leq \sup_{x\in\mathcal{X}}\bigg|\frac{\pi_t}{e_t(x)}\bigg|^2 \pi_t^{-2}\varepsilon_i(t)^2D_{t,i} + 2\sup_{x\in\mathcal{X}}\mu_t(x)\sup_{x\in\mathcal{X}}\bigg|\frac{\pi_t}{e_t(x)}\bigg|\pi_t^{-1}|\varepsilon_i(t)|D_{t,i} + \sup_{x\in\mathcal{X}}\mu_t(x)^2 \\
	&\lesssim \pi_t^{-2}\varepsilon_i(t)^2D_{t,i} + \pi_t^{-1}|\varepsilon_i(t)|D_{t,i} + 1
\end{align*}}
Looking at the max then yields {\footnotesize\begin{align*}
	\sup_{t\neq 0}&\max_{1\leq i \leq n}|\psi_i^{[t]}(\eta)|^2 \\
	&\lesssim_P \sup_{t\neq 0} \pi_t^{-2}\max_{1\leq i \leq n}|\varepsilon_i(t)|^2\max_{1\leq i \leq n}D_{t,i} + \sup_{t\neq 0}\pi_t^{-1}\max_{1\leq i \leq n}|\varepsilon_i(t)|\max_{1\leq i \leq n}D_{t,i} + 1 \\
	&\lesssim_P J^2n^{2/m} + Jn^{1/m} + 1 \\
	&\lesssim J^2n^{2/m}
\end{align*}}
Equivalently we obtain 
{\footnotesize\begin{align*}
	\sup_{t\neq 0}\max_{1\leq i \leq n}|\psi_i^{[t]}(\hat{\eta})| 
	&\leq \sup_{\hat{\eta} \in \mathcal{H}_n}\sup_{t\neq 0}\max_{1\leq i \leq n}|\psi_i^{[t]}(\hat{\eta})| \\
	&\lesssim_P \sup_{t\neq 0}\chi_{t,n} \pi_t^{-1} \max_{1\leq i \leq n}|\varepsilon_i(t)|\max_{1\leq i \leq n}D_{t,i} \\
	&\lesssim_P J n^{1/m} \\
	\sup_{t\neq 0}\max_{1\leq i \leq n}|\psi_i^{[t]}(\hat{\eta})|^2 
	&\leq \sup_{\hat{\eta} \in \mathcal{H}_n}\sup_{t\neq 0}\max_{1\leq i \leq n}|\psi_i^{[t]}(\hat{\eta})|^2 \\
	&\lesssim_P \sup_{t\neq 0}\chi_{t,n} \pi_t^{-2} \max_{1\leq i \leq n}|\varepsilon_i(t)|^2\max_{1\leq i \leq n}D_{t,i} \\
	&\lesssim_P J^2 n^{2/m}
\end{align*}}
\vspace{-48pt}
\paragraph{(MA.2) $\kappa_{n,1}$ and $\kappa_{n,2}$ rates}
{\footnotesize\begin{align*}
	E[\max_{1\leq i \leq n}|\psi_i(\hat{\eta},\pi)-\psi_i({\eta},\pi)|] 
	&\lesssim_P \sum_{t\neq 0}\pi_t E[\max_{1\leq i \leq n}|\psi_i^{[t]}(\hat{\eta})-\psi_i^{[t]}({\eta})|] \\
	&\lesssim_P J \sup_{t\neq0}\pi_t E[\max_{1\leq i \leq n}|\psi_i^{[t]}(\hat{\eta})-\psi_i^{[t]}({\eta})|] \\
	&\leq \sup_{t\neq 0}E[\max_{1\leq i \leq n}|\psi_i^{[t]}(\hat{\eta})-\psi_i^{[t]}({\eta})|] \\
	&\leq \kappa_{n,1}
\end{align*}}
\vspace{-36pt}
{\footnotesize\begin{align*}
	E[\max_{1\leq i \leq n}|\psi_i(\hat{\eta},\pi)-\psi_i({\eta},\pi)|^2] 
	&\lesssim_P \sum_{t\neq 0}\sum_{t'\neq 0}\pi_t\pi_{t'} E[\max_{1\leq i \leq n}|\psi_i^{[t]}(\hat{\eta})-\psi_i^{[t]}({\eta})||\psi_i^{[t']}(\hat{\eta})-\psi_i^{[t']}({\eta})|] \\
	&\lesssim_P J \sup_{t\neq0}\pi_t E[\max_{1\leq i \leq n}|\psi_i^{[t]}(\hat{\eta})-\psi_i^{[t]}({\eta})|^2]\\
	&\leq \sup_{t\neq 0} E[\max_{1\leq i \leq n}|\psi_i^{[t]}(\hat{\eta})-\psi_i^{[t]}({\eta})|^2] \\
	&\leq \kappa_{n,2}
\end{align*}}
\vspace{-36pt}
\paragraph{(MA.3) $v_i$ decomposition}
For arbitrary $\eta$ and $\pi$ define 
	$\hat{\beta}(\eta,\pi) = \hat{Q}^{-1}E_n[b_i\psi_i(\eta,\pi)]$
and thus $\hat{\beta} = \hat{\beta}(\hat{\eta},\hat{\pi})$. Rewriting the residual using estimated nuisances then yields
{\footnotesize\begin{align*}
	v_i = \psi_i(\hat{\eta},\hat{\pi}) - b_i'\hat{\beta}(\hat{\eta},\hat{\pi}) 
	= \psi_i(\hat{\eta},{\pi}) - b_i'\hat{\beta}(\hat{\eta},{\pi})  + \psi_i(\hat{\eta},\hat{\pi}) - \psi_i(\hat{\eta},{\pi})   + b_i'(\hat{\beta}(\hat{\eta},{\pi}) - \hat{\beta}(\hat{\eta},\hat{\pi}))
\end{align*}}
\vspace{-48pt}

\paragraph{(MA.4) Probability rates}
For $t\neq 0$, estimators of unconditional weights obey
{\footnotesize\begin{align*}
	\frac{\hat{\pi}_t}{1-\hat{\pi}_0} - \frac{{\pi}_t}{1-{\pi}_0} 
	&= \frac{\hat{\pi}_t(1-\pi_0) - \pi_t(1-\hat{\pi}_0)}{(1-\hat{\pi}_0)(1-{\pi}_0)} \\
	&\lesssim_P \bigg(\pi_t|\hat{\pi}_0 - \pi_0| + \pi_0|\hat{\pi}_t - \pi_t|\bigg)(1+ O_p(|\hat{\pi}_0 - \pi_0|)) \\
	&\lesssim_P \pi_tn^{-1/2} + (n/\pi_t)^{-1/2} \\
	&\lesssim {(Jn)}^{-1/2}
\end{align*}}
by Chebyshev's inequality as $E[\hat{\pi}_t -\pi_t] = 0$ and {\footnotesize\begin{align*}
	E[(\hat{\pi}_t -\pi_t)^2] &= \frac{V[D_{it}]}{n} 
	= \pi_t(1-\pi_t)/n
	\lesssim \pi_t/n
\end{align*}}
with $|\hat{\pi}_0 - \pi_0| \lesssim_P n^{-1/2}$ as control propensities are bounded away from $0$ and $1$.

\paragraph{(MA.5) Maximal impact of nuisances on predictions}
{\footnotesize\begin{align*}
	\max_{1\leq j \leq n}|b_j'(\hat{\beta}(\hat{\eta},\hat{\pi}) - \hat{\beta}(\hat{\eta},\pi))| 
	&=	\max_{1\leq j \leq n}|b_j'\hat{Q}^{-1}E_n[b_i(\psi_i(\hat{\eta},\hat{\pi}) - \psi_i(\hat{\eta},\pi))]| \\
	&\lesssim_P \sup_{z\in\mathcal{Z}}||b(z)||\ ||\hat{Q}^{-1}||\ ||E_n[b_i(\psi_i(\hat{\eta},\hat{\pi}) - \psi_i(\hat{\eta},\pi))]|| \\
	&\lesssim_P \xi_k \sqrt{\frac{Jk}{n}}
\end{align*}}
where the second to last line comes from H.1 - H.3. Moreover, {\footnotesize\begin{align*}
	\max_{1\leq j \leq n}|b_j'(\hat{\beta}(\hat{\eta},{\pi}) - \hat{\beta}({\eta},\pi))| 
	&=	\max_{1\leq j \leq n}|b_j'\hat{Q}^{-1}E_n[b_i(\psi_i(\hat{\eta},{\pi}) - \psi_i({\eta},\pi))]| \\
	&\lesssim_P \xi_k ||\hat{Q}^{-1}||\ ||E_n[b_i(\psi_i(\hat{\eta},{\pi}) - \psi_i({\eta},\pi))]|| \\
	&\lesssim_P {\xi_k}n^{-1/2} (B_n^{[rATE]} + \Lambda_n^{[rATE]}) \\
	&\lesssim_P {\xi_k}n^{-1/2}
\end{align*}}
\vspace{-36pt}
due to Markov's inequality. The deviation from the best linear predictor follows from BCCK, Theorem 4.3, under the assumptions stated in the beginning of this section: {\footnotesize\begin{align*}
	\max_{1\leq j \leq n}|b_j'(\hat{\beta}(\eta,\pi) - \beta_0)| \lesssim_P \xi_k\sqrt{\frac{\log(k)}{n}}
\end{align*}}
\vspace{-36pt}
\paragraph{(MA.6) Error term tail bounds}
C.1 implies the following tail bound for $rATE$ {\footnotesize\begin{align*}
	\max_{1\leq i \leq n}|\varepsilon_i| 
	&= 	\max_{1\leq i \leq n}|\psi_i^{[rATE]}(\eta,\pi) - E[\psi_i^{[rATE]}(\eta,\pi)|Z_i]| \\
	&\lesssim_P  \max_{1\leq i \leq n}\sum_{t \neq 0}\frac{\pi_t D_{t,i}|\varepsilon_i(t)|}{e_t(X_i)} \\
	&\lesssim_P \max_{1\leq i \leq n}\sup_{t\neq 0}\sup_{x\in\mathcal{X}}\frac{\pi_t}{e_t(x)}|\varepsilon_i(t)|\sum_{t\neq 0} D_{t,i} \\
	&\lesssim_P \max_{1\leq i \leq n}\sup_{t\neq 0}|\varepsilon_i(t)|
\end{align*}}
and equivalently by B.1 for the $nATE$ with $\sup_{x\in\mathcal{X}}\frac{\pi_t}{e_t(x)}$ replaced by $1$. Note that {\footnotesize\begin{align*}
	E[\max_{1\leq i \leq n}\sup_{t\neq 0}|\varepsilon_i(t)|] 
	&= \int_{-\infty}^{(nJ)^{1/m}}P(\max_{1\leq i \leq n}\sup_{t\neq 0}|\varepsilon_i(t)| > w)dw  + \int_{(nJ)^{1/m}}^{\infty}P(\max_{1\leq i \leq n}\sup_{t\neq 0}|\varepsilon_i(t)| > w)dw \\ \end{align*} \begin{align*}
	&\leq (Jn)^{1/m} + \int_{(nJ)^{1/m}}^{\infty}n \sum_{t \neq 0}P(|\varepsilon_i(t)| > w)dw \\
	&\leq (Jn)^{1/m} + (Jn)^{1/m}\int_{(nJ)^{1/m}}^{\infty}w^{1-1/m}P(|\varepsilon_i(t)| > w)dw \\
	&\lesssim (Jn)^{1/m}(1+o(1)) \\
	&\lesssim (Jn)^{1/m}
\end{align*}}
where the second term is convergent due to the conditional moment bound B.1/C.1 for $\varepsilon_i(t)$. Overall, this implies that $\max_{1\leq i \leq n}|\varepsilon_i|\lesssim_P (Jn)^{1/m}$ by Markov's inequality. 

\subsubsection{Definitions and Decomposition}
Define ${\Sigma} = E[(b_i(\varepsilon_i + r_i) - \gamma a_i)(b_i(\varepsilon_i + r_i) - \gamma a_i)']$, 
	${\Sigma}_n = E_n[(b_i(\varepsilon_i + r_i) - \gamma a_i)(b_i(\varepsilon_i + r_i) - \gamma a_i)']$, and 
	$\hat{\Sigma}_n = E_n[(b_iv_i - \hat{\gamma} \hat{a}_i)(b_iv_i - \hat{\gamma} \hat{a}_i)']$ 
where $v_i = \psi_i(\hat{\eta},\hat{\pi}) - b_i'\hat{\beta}$ and $\hat{a}_i$ obtained by replacing true $\pi$ with estimates $\hat{\pi}$: {\footnotesize\begin{align*}
	\hat{a}_i = (\hat{a}_i^{[1]},\dots,\hat{a}_i^{[J]}), \quad 
	\hat{a}_i^{[t]} = \frac{D_{t,i}(1-\hat{\pi}_0) + D_{0,i}\hat{\pi}_t - \hat{\pi}_t}{(1-\hat{\pi}_0)^2}
\end{align*}}
In the following we proof that $||\hat{\Sigma}_n-\Sigma|| = o_p(1)$. The remaining rates for convergence of $\hat{\Omega} = \hat{Q}^{-1}\hat{\Sigma}_n\hat{Q}^{-1}$ to $\Omega$ follows from BCCK, Proof of Theorem 4.6. We start with decomposing 
	$||\hat{\Sigma}_n-\Sigma|| \leq ||\hat{\Sigma}_n-\Sigma_n|| + ||{\Sigma}_n-\Sigma||$.
\subsubsection{Bounds for $||\hat{\Sigma}_n - \Sigma_n||$}
We bound the three components separately of the following decomposition: 
{\footnotesize\begin{align*}
	||\hat{\Sigma}_n-\Sigma_n|| \leq ||E_n[b_ib_i'(v_i^2 - (\varepsilon_i + r_i)^2)]|| 
	 + ||E_n[\hat{\gamma}\hat{a}_i\hat{a}_i'\hat{\gamma}' - \gamma a_i a_i' \gamma']|| 
	 +  ||E_n[b_i(v_i(\hat{\gamma}\hat{a}_i)' - (\varepsilon_i+r_i)(\gamma a_i)) ]||  
\end{align*}} 
\vspace{-36pt}
\paragraph{Part 1}
{\footnotesize\begin{align*}
	||&E_n[b_ib_i'(v_i^2 - (\varepsilon_i + r_i)^2)]|| \\
	&\leq ||E_n[b_ib_i'((\psi_i(\hat{\eta},{\pi}) - b_i'\hat{\beta}(\hat{\eta},{\pi})  + \psi_i(\hat{\eta},\hat{\pi}) - \psi_i(\hat{\eta},{\pi})   + b_i'(\hat{\beta}(\hat{\eta},{\pi}) - \hat{\beta}(\hat{\eta},\hat{\pi})))^2 - (\varepsilon_i + r_i)^2)]|| \\
	&\leq ||E_n[b_ib_i'(\psi_i(\hat{\eta},{\pi}) - b_i'\hat{\beta}(\hat{\eta},{\pi}) - (\varepsilon_i + r_i)^2)]|| 
	+ ||E_n[b_ib_i'(\psi_i(\hat{\eta},\hat{\pi}) - \psi_i(\hat{\eta},{\pi}))^2]|| \\
	&\quad + ||E_n[b_ib_i'(b_i'(\hat{\beta}(\hat{\eta},{\pi}) - \hat{\beta}(\hat{\eta},\hat{\pi})))^2]|| \\
	&\equiv (v.1) + (v.2) + (v.3)
\end{align*}}
For $(v.1)$ we can use the proof of Theorem 3.3 in SC for the $nATE$ and (MA.6) to obtain {\footnotesize\begin{align*}
	(v.1) \lesssim_P \bigg((Jn)^{1/m} + \sup_{z\in\mathcal{Z}}||r(z)||\bigg)\kappa_{n,1} + \kappa_{n,2}
\end{align*}}
where, for $nATE$, $rATE$, and $\Delta$, tail are allowed to differ. For the second term we have
{\footnotesize\begin{align*}
	(v.2) &= \bigg|\bigg|E_n\bigg[b_ib_i'\bigg(\sum_{t \neq 0}\psi_i^{[t]}(\hat{\eta})\bigg(\frac{\hat{\pi}_t}{1-\hat{\pi}_0} - \frac{{\pi}_t}{1-{\pi}_0}\bigg)\bigg)^2\bigg]\bigg|\bigg| \\
	&= \bigg|\bigg|\sum_{t \neq 0}\sum_{t'\neq 0}E_n\bigg[b_ib_i'\psi_i^{[t]}(\hat{\eta})\psi_i^{[t']}(\hat{\eta})\bigg(\frac{\hat{\pi}_t}{1-\hat{\pi}_0} - \frac{{\pi}_t}{1-{\pi}_0}\bigg)\bigg(\frac{\hat{\pi}_{t'}}{1-\hat{\pi}_0} - \frac{{\pi}_{t'}}{1-{\pi}_0}\bigg)\bigg)\bigg]\bigg|\bigg| \\
	&\lesssim_P J^2 \sup_{t,t'\neq 0}|\psi_i^{[t]}(\hat{\eta})\psi_i^{[t']}(\hat{\eta})||E_n[b_ib_i']||\frac{\sqrt{\pi_t\pi_t'}}{n} \\
	&\lesssim_P J^2 \sup_{t\neq 0}\pi_t^{-2}\max_{1\leq i \leq n}|\varepsilon_i(t)|^2 ||\hat{Q}|| \frac{\pi_t}{n} \\
	&\lesssim_P J^3n^{-(1-2/m)}
\end{align*}}
For the third term we exploit (H.3) to obtain
{\footnotesize\begin{align*}
	(v.3) &= ||E_n[b_ib_i(b_i'\hat{Q}^{-1}E_n[b_i'(\psi(\hat{\eta},\hat{\pi}) - \psi_i(\hat{\eta},\pi))])^2]|| \\
	&\lesssim_P \max_{1\leq j \leq n} |b_j'\hat{Q}^{-1}E_n[b_i'(\psi(\hat{\eta},\hat{\pi}) - \psi_i(\hat{\eta},\pi))]|^2||\hat{Q}|| \\
	&\lesssim_P \sup_{z\in\mathcal{Z}}||b(z)||^2||\hat{Q}^{-1}||^2||E_n[b_i'(\psi(\hat{\eta},\hat{\pi}) - \psi_i(\hat{\eta},\pi))]||^2||\hat{Q}|| \\
	&\lesssim_P \xi_k^2 \bigg(\sqrt{\frac{kJ}{n}}\bigg)^2 \\
	&= \frac{\xi_k^2kJ}{n}
\end{align*}}
\vspace{-36pt}
\paragraph{Part 2}
 Let $A_n = E_n[a_ia_i']$ and $\hat{A}_n = E_n[\hat{a}_i\hat{a}_i']$. We use (H.5) to bound the second term
{\footnotesize\begin{align*}
	||E_n[\hat{\gamma}\hat{a}_i\hat{a}_i'\hat{\gamma}' - \gamma a_i a_i' \gamma']|| 
	&= ||\hat{\gamma}\hat{A}_n\hat{\gamma}' - \gamma A_n  \gamma' || \\
	&\leq ||\hat{\gamma}-\gamma||^2||\hat{A}_n - A_n|| + ||\hat{\gamma} - \gamma||^2||A_n|| + ||\hat{\gamma}-\gamma||\ ||\hat{A}_n - A_n||\ ||\gamma|| \\ &\quad \ + ||\hat{\gamma}-\gamma||\ ||A_n||\ ||\gamma|| + ||\hat{A}_n - A_n||\ ||\gamma||^2 \\
	&\lesssim_P ||\hat{\gamma}-\gamma||\ ||A_n||\ ||\gamma|| + ||\hat{A}_n - A_n||\ ||\gamma||^2 \\
	&\lesssim_P \sqrt{kJ^2}(n^{-1/2} + m_{n,2} + Js_{n,2})\sqrt{kJ} + n^{-1/2}kJ \\
	&\lesssim \sqrt{kJ^2}(n^{-1/2} + m_{n,2} + Js_{n,2})
\end{align*}} \vspace{-24pt}
The covariance term is bounded by {\footnotesize\begin{align*}
	||E_n[b_i(v_i(\hat{\gamma}\hat{a}_i)' - (\varepsilon_i+r_i)(\gamma a_i)) ]||    
	&\leq ||E_n[b_i(v_i-(\varepsilon_i + r_i))(\gamma a_i)'] || + ||E_n[b_i(\varepsilon_i + r_i)(\hat{\gamma}\hat{a}_i - \gamma a_i)']||\\ &\quad + ||E_n[b_i(v_i-(\varepsilon_i + r_i))(\hat{\gamma}\hat{a}_i - \gamma a_i)']|| \\
	&\equiv (c.1) + (c.2) + (c.3)
\end{align*}}
Now we use (H.5) and (MA.5) to bound $(c.1)$
{\footnotesize\begin{align*}
	||E_n[b_i(v_i - (\varepsilon + r_i))(\gamma a_i)']|| 
	&\lesssim_P \xi_k\max_{1\leq i \leq n}|b_i'(\hat{\beta}(\hat{\eta},\hat{\pi}) - \beta_0)|E_n[||\gamma a_i||]\\
	&\quad + ||E_n[b_i(\psi_i(\hat{\eta},\hat{\pi}) - \psi_i(\hat{\eta},{\pi}))(\gamma a_i)']|| + ||E_n[b_i(\psi_i(\hat{\eta},{\pi}) - \psi_i({\eta},{\pi}))(\gamma a_i)']|| \\
	&\equiv (c.1.1) + (c.1.2) + (c.1.3)
\end{align*}}
with decomposing the BLP error using the rates in (H.5) and (MA.5)
{\footnotesize\begin{align*}
	(c.1.1)	&\leq\xi_k J ||\gamma||E_n[||a_i||] \max_{1\leq i \leq n}|b_i'(\hat{\beta}(\hat{\eta},\hat{\pi}) - \beta_0)| \\
	&\lesssim_P \xi_k \sqrt{kJ}\bigg(\xi_k\sqrt{Jk/n} + \xi_k n^{-1/2} + \xi_k\sqrt{\log(k)/n}\bigg) \\
	&\lesssim \frac{\xi_k^2kJ}{\sqrt{n}}	
\end{align*}}
For $(c.1.2)$ note that {\footnotesize\begin{align*}
	(c.1.2) &= ||\sum_{t\neq 0}\bigg(\frac{\hat{\pi}_t}{1-\hat{\pi}_0} - \frac{{\pi}_t}{1-{\pi}_0}\bigg)E_n[b_i\psi_i^{[t]}(\hat{\eta})a_i^{[t]}]\gamma_t'|| \\
	&\lesssim_P J \sup_{t\neq 0}\sqrt{\frac{\pi_t}{n}}\max_{1\leq i \leq n}|\psi_i^{[t]}(\hat{\eta})|E_n[||b_ia_i^{[t]}||]||\gamma_t|| \\
	&\lesssim_P J^{3/2}n^{-(\frac{1}{2} - \frac{1}{m})}k
\end{align*}}
by (MA.1) and (H.5) in conjunction with Markov's inequality. For $(c.1.3)$, we have that {\footnotesize\begin{align*}
	(c.1.3) &= ||\sum_{t \neq 0}\frac{\pi_t}{1-\pi_0}E_n[b_i(\psi_i^{[t]}(\hat{\eta})-\psi_i^{[t]}({\eta}))a_i^{[t]}]\gamma_t'||\\
	&\leq J \sup_{t\neq 0}\pi_t \max_{1\leq j \leq n}|\psi_j^{[t]}(\hat{\eta})-\psi_j^{[t]}({\eta})||a_i^{[t]}|E_n[||b_i||]||\gamma_t|| \\
	&\lesssim_P \kappa_{1,n}k
\end{align*}}
also by (H.5) and Markov's inequality.
For $(c.2)$ note that {\footnotesize\begin{align*}
	(c.2) &\leq \xi_k \max_{1\leq j \leq n}|\varepsilon_j + r_j|E_n[||\hat{\gamma}\hat{a}_i - \gamma a_i||] \\
	&\lesssim_P \xi_k ((Jn)^{1/m} + \sup_{z\in\mathcal{Z}}||r(z)||)E_n[||\hat{\gamma}\hat{a}_i - \gamma a_i||] \\
	&\lesssim_P \xi_k ((Jn)^{1/m} + \sup_{z\in\mathcal{Z}}||r(z)||)\sqrt{kJ^2}(n^{-1/2} + m_{n,2} + Js_{n,2})
\end{align*}}
as (H.5) implies {\footnotesize\begin{align*}
	E_n[||\hat{\gamma}\hat{a}_i - \gamma a_i||] 
	&= E_n[||\hat{\gamma} - \gamma||\ ||a_i||] + E_n[||{\gamma}||\ ||\hat{a}_i - a_i||] +  E_n[||\hat{\gamma} - \gamma||\ ||\hat{a}_i - a_i||] \\
	&\lesssim_P ||\hat{\gamma} - \gamma||E_n[||a_i||] + ||\gamma||E_n[||\hat{a}_i - a_i||] \\
	&\lesssim_P \sqrt{kJ^2}(n^{-1/2} + m_{n,2} + Js_{n,2}) + \sqrt{Jk/n} \\
	&\lesssim  \sqrt{kJ^2}(n^{-1/2} + m_{n,2} + Js_{n,2})
\end{align*}}
Note that $(c.3)$ is at most of rate $(c.1) + (c.2)$ which completes the covariance. 

\subsubsection{Bounds for $||\Sigma_n - \Sigma||$}
Now consider the remaining difference {\footnotesize\begin{align*}
	||\Sigma_n - \Sigma || &\leq ||E_n[b_ib_i'(\varepsilon_i + r_i)] -E[b_ib_i'(\varepsilon_i + r_i)]|| + ||E_n[\gamma a_ia_i'\gamma'] - E[\gamma a_ia_i'\gamma']|| \\ &\quad + ||E_n[b_i(\varepsilon_i+r_i)(\gamma a_i)' + \gamma a_i (\varepsilon_i + r_i)b_i'] - E[b_i(\varepsilon_i+r_i)(\gamma a_i)' + \gamma a_i (\varepsilon_i + r_i)b_i']|| \\
	&= (d.1) + (d.2) + (d.3)
\end{align*}}
$(d.1)$ is bounded in BCCK, Proof of Theorem 4.6, with adapted tail rates using (MA.6) {\footnotesize\begin{align*}
	(d.1) \lesssim_P \bigg((Jn)^{1/m} + \sup_{z\in\mathcal{Z}}||r(z)||\bigg)\sqrt{\frac{\xi_k^2 \log k}{n}}
\end{align*}}
For $(d.2)$ note that {\footnotesize\begin{align*}
	(d.2) &\leq ||\gamma||^2||E_n[a_ia_i'] - E[a_ia_i']|| \\
	&\lesssim_P kJ\sqrt{\frac{\log J}{n}}
\end{align*}}
due to (H.5). For $(d.3)$ first note that {\footnotesize\begin{align*}
	\max_{1\leq i \leq n}|a_i'\gamma'b_i| \lesssim_P \xi_k ||\gamma || \max_{1\leq i \leq n}||a_i|| \lesssim_P \xi_k \sqrt{k}J \\
	E_n[||b_ia_i'\gamma'||] \leq \xi_k ||\gamma||E_n[||a_i||] \lesssim_P \xi_k \sqrt{kJ}
\end{align*}}
due to (H.5). Now we use the Symmetrization Lemma to bound $(d.3)$. Let $w_i$ for $i=1,\dots,n$ denote independent Rademacher random variables independent of the data. Denote $E_{w}[\cdot]$ the expectation operator with respect to the measure of $w$. We have that {\footnotesize\begin{align*}
	(d.3) &\lesssim E\bigg[E_w\bigg[||E_n[w_i(\varepsilon_i+r_i)(b_ia_i\gamma' + \gamma a_i b_i')]||\bigg]\bigg] \\
	&\lesssim \sqrt{\frac{\log k}{n}}E\bigg[||E_n[(\varepsilon_i+r_i)^2(b_ia_i'\gamma'b_ia_i'\gamma' + b_ia_i'\gamma'\gamma a_i b_i' + \gamma a_ib_i'b_ia_i'\gamma' + \gamma a_i b_i'\gamma a_i b_i')]||^{1/2}\bigg] \\
	&\lesssim_P \sqrt{\frac{\log k}{n}}\max_{1\leq i \leq n}|\varepsilon_i + r_i|\bigg(  2 \max_{1\leq i \leq n}|a_i\gamma b_i|^{1/2}E[E_n[||b_ia_i'\gamma'||]]^{1/2} \\
	&\quad + \max_{1\leq i \leq n}|a_i'\gamma'\gamma a_i|^{1/2}E[E_n[||b_ib_i'||]]^{1/2} + \max_{1\leq i \leq n}|b_i'b_i|^{1/2}E[E_n[||\gamma a_i a_i'\gamma'||]]^{1/2}\bigg) \\
	&\lesssim_P \sqrt{\frac{\log k}{n}}\max_{1\leq i \leq n}|\varepsilon_i + r_i|\bigg((\xi_k\sqrt{k}J)^{1/2}(\xi_k\sqrt{kJ})^{1/2} + ||\gamma||\max_{1\leq i \leq n}||a_i||\sqrt{k} + \xi_k ||\gamma||\bigg) \\
	&\lesssim \sqrt{\frac{\log k}{n}}\bigg((nJ)^{1/m} + \sup_{z\in\mathcal{Z}}||r(z)||\bigg)\bigg(J^{3/4}\xi_k\sqrt{k} + kJ\bigg) 
\end{align*}}
where the second line is due to Khinchin's inequality as $(b_ia_i\gamma' + \gamma a_i b_i')$ are iid symmetric. The remaining steps follow from (H.5) and (MA.6) and repeated Markov's inequality. 

\subsubsection{Full variance}
Collecting all the rates, we obtain that {\footnotesize\begin{align}
	||\hat{\Sigma}_n - \Sigma|| &\lesssim_P  
	\bigg((Jn)^{1/m} + \sup_{z\in\mathcal{Z}}||r(z)||\bigg)\kappa_{n,1} + \kappa_{n,2} 
	+ J^3n^{-(1-2/m)} 
	+ \frac{\xi_k^2kJ}{n}  + \frac{\xi_k^2kJ}{\sqrt{n}} 
	+ J^{3/2}n^{-(\frac{1}{2} - \frac{1}{m})}k \notag \\ 
	&
	+ \kappa_{1,n}k + \xi_k ((Jn)^{1/m} + \sup_{z\in\mathcal{Z}}||r(z)||)\sqrt{kJ^2}(n^{-1/2} + m_{n,2} + Js_{n,2}) + kJ\sqrt{\frac{\log J}{n}} \notag \\
	& + \bigg((Jn)^{1/m} + \sup_{z\in\mathcal{Z}}||r(z)||\bigg)\sqrt{\frac{\xi_k^2 \log k}{n}} 
	+ \sqrt{\frac{\log k}{n}}\bigg((nJ)^{1/m} + \sup_{z\in\mathcal{Z}}||r(z)||\bigg)\bigg(J^{3/4}\xi_k\sqrt{k} + kJ\bigg) \notag \\
	&\equiv M_{n,2} \label{eq_AV_rATE}
\end{align}}
$nATE$ derivations are analogous, but without $\gamma a_i$ terms. Thus the solution simplifies to {\footnotesize\begin{align}
	||\hat{\Sigma}_n - \Sigma|| &\lesssim_P 	\bigg((Jn)^{1/m} + \sup_{z\in\mathcal{Z}}||r(z)||\bigg)\bigg(\kappa_{n,1} + \sqrt{\frac{\xi_k^2 \log k}{n}}\bigg) + \kappa_{n,2} = M_{n,1} \label{eq_AV_nATE}
\end{align}}
Thus $M_{n,1} = o(1)$ for $nATE$ or $M_{n,2} = o(1)$ for $rATE$/$\Delta$ matches Assumption A.V.

\subsection{Supplementary Material for Section \ref{sec_MC1}} \label{app_MCdesign}
We simulate $n$ observations of $(Y_i,X_i,T_i)$. We create the following two designs for the $k$-dimensional vector $X_i$ for two designs: A) $X_{i,j} \sim \mathcal{U}[-1,1]$ for $j=1,\dots,p$ and B) like A) but with nonlinear transformations 
		$X_{i,2} = \sqrt{1/3}~\text{sign}(X_{i,1})$, 
		$X_{i,3} = \sqrt{2/3}\sin(\pi X_{i,1})$,
		$X_{i,4} = \sqrt{5/12}(X_{i,1}^2 - 1)$,
		$X_{i,5} = \sqrt{1/5}(4X_{i,1}\mathbbm{1}(X_{i,1} > 0) - 1)$.
The parameterizations for design B) assure that we have an equal signal to noise ratio in both designs, i.e.~all predictors have $E[X_{i,j}] = 0$ and $V[X_{i,j}]=1/3$. For the potential outcomes let
   $u_i \sim \mathcal{N}(0,1)$, $Y_i(t) = u_i$ for $t \neq 1$, and $Y_i(1) = \tau + u_i$. Treatment probabilities $P(T_i=t|X_i) = e_t(X_i)$ for $t=0,1\dots,J$ (with $t=0$ denoting control) are generated as {\footnotesize\begin{align*}
	{e}_t(x) &= \frac{\exp(\sum_{l}x_l\beta_{t,l})}{1+ \sum_{j\neq 0}\exp(\sum_{l}x_l\beta_{j,l})} 
\end{align*}}
Design A sets $\beta_{2,1} = \beta_{3,1} = 1$ and $\beta_{t,l} = 0$ else.
For design $B$, we set $\beta_{0,l} = 0$ for all $l$, $\beta_{1,1} = 1$ and $\beta_{1,l} = 0$ for $l > 1$, and $\beta_{2,1} = \beta_{2,2} = \beta_{2,3} = \beta_{2,4} =  \beta_{2,5} = 1$, $\beta_{2,l} = 0$ for $l > 5$. 
Thus, conditional treatment effects are given by $\tau_1(x) = \tau = 10$ and $\tau_t(x) = 0$ for all $t \neq 1$. This implies the following conditional decomposition terms: 
{\footnotesize\begin{align*}
	E[rATE(X_i)|X_{i,1} = x_1] &= \tau\bigg[\frac{\pi_1}{1-\pi_0}\bigg] \\
	E[nATE(X_i)|X_{i,1} = x_1] &= \tau\bigg[\frac{e_1(x_1)}{1-e_0(x_1)}\bigg] \\
	E[\Delta(X_i)|X_{i,1} = x_1] &= \tau\bigg[\frac{e_1(x_1)}{1-e_0(x_1)} - \frac{\pi_1}{1-\pi_0}\bigg]
\end{align*}} 
Note that $E[X_{i,1}] = 0$ and $V[X_{i,1}] = 1/3$. Thus the best linear approximation of $E[\Delta(X_i)|X_{i,1}]$ has population parameters $(\alpha,\beta)$ with {\footnotesize\begin{align*}
	\alpha &= \tau E\bigg[\frac{e_1(X_{i,1})}{1-e_0(X_{i,1})} - \frac{\pi_1}{1-\pi_0}\bigg] -  \beta E[X_{i,1}] =  \tau E\bigg[\frac{e_1(X_{i,1})}{1-e_0(X_{i,1})} - \frac{\pi_1}{1-\pi_0}\bigg] \\
	\beta &= \frac{\tau}{V[X_{i,1}]}E\bigg[\frac{e_1(X_{i,1})}{1-e_0(X_{i,1})}X_{i,1} - \frac{\pi_1}{1-\pi_0}X_{i,1}\bigg] =3\tau  E\bigg[\frac{e_1(X_{i,1})}{1-e_0(X_{i,1})}X_{i,1}\bigg] 
\end{align*}}
and equivalently for the $rATE$ and $nATE$. Evaluating the expectations for both designs A and B yields the following true decomposition parameter values:

\begin{table}[!h]	\centering \caption{Monte Carlo: Decomposition Estimands}
	\begin{subtable}{0.45\textwidth} \footnotesize	\centering \caption{Design A}
		\begin{tabular}{l|ccc}\hline \hline \\[-1ex]
			& $rATE$ & $nATE$ & $\Delta$ \\ \hline \\[-1ex]
			$\alpha$& 5.127& 5.000 &-.127 \\
			$\beta$ &0.000 &2.383&  2.383 \\ \hline\hline
		\end{tabular}
	\end{subtable} 	
\begin{subtable}{0.45\textwidth}\footnotesize		\centering \caption{Design B}
	\begin{tabular}{l|ccc}\hline \hline \\[-1ex]
		& $rATE$ & $nATE$ & $\Delta$ \\ \hline \\[-1ex]
		$\alpha$&   5.679 &4.710 &-.969  \\  
		$\beta$	&   0.000 &5.749  &5.749	  \\ \hline\hline
	\end{tabular}
\end{subtable} 
\begin{justify}\footnotesize
	The tables contain the best linear predictor parameters for all decomposition terms, i.e.~the regression estimands from equation $\psi^{[parameter]}(W_i,\eta) = \alpha + \beta X_{i,1} + \varepsilon_i$ for $parameter$ equal to $nATE$, $rATE$, or $\Delta$. $\alpha$ and $\beta$ do not depend on $k$ or $n$. 
\end{justify}
\end{table}

\subsection{Supplementary Material for Section \ref{sec:app-sc1}} \label{app:app-sc1}

\begin{figure}[h]
    \centering
    \caption{Distribution of smoking intensities along heterogeneity variables}\label{fig:smoke-dis}
    \includegraphics[width=0.9\linewidth]{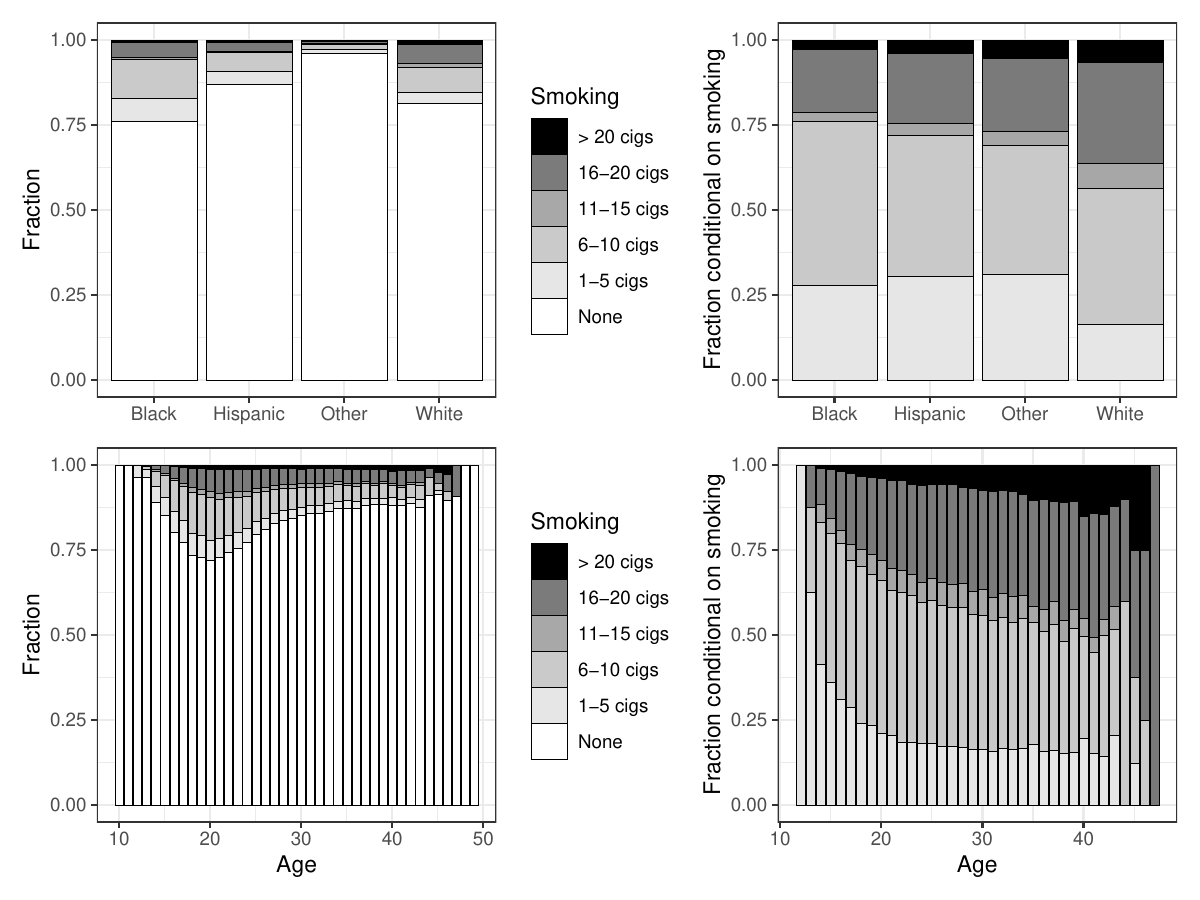}
\end{figure}

\begin{figure}[h]
    \centering
    \caption{Average potential outcomes of smoking intensities}\label{fig:apo-smoke}
    \includegraphics[width=0.6\linewidth]{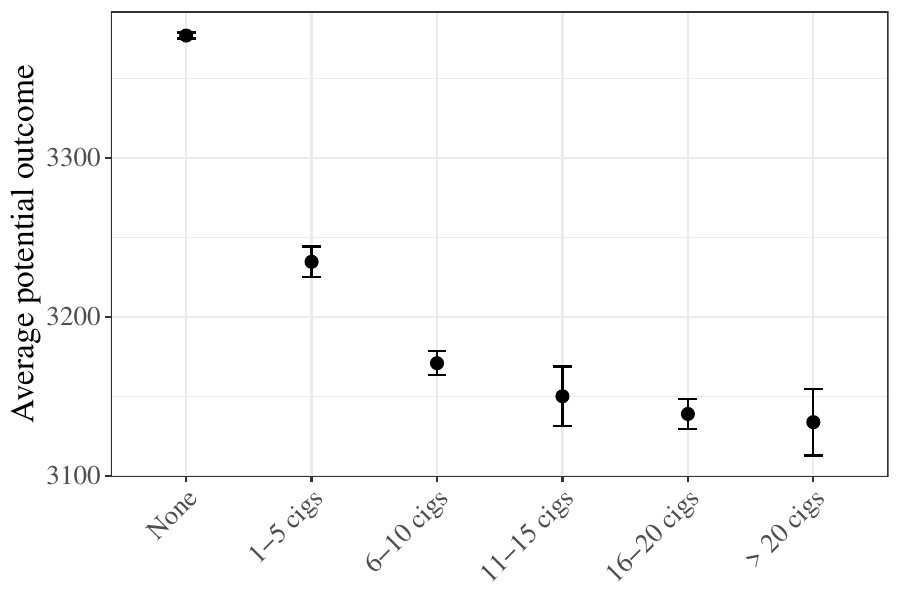}
    \subcaption*{\textit{Note:} Average potential outcomes estimated with Double Machine Learning using an ensemble of Ridge, Lasso and Random Forest regression. Point estimates and 95\%-confidence interval.}
\end{figure}

\begin{figure}[h]
    \centering
    \caption{Re-scaled propensity scores}\label{fig:resclaed-smoke}
    \includegraphics[width=0.75\linewidth]{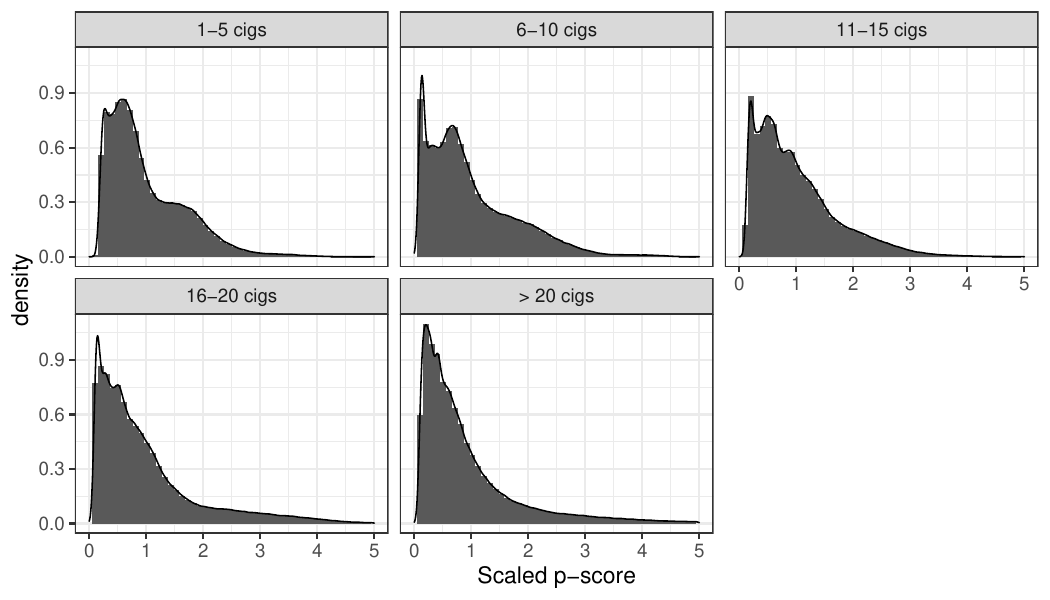}
\end{figure}

The distribution of smoking intensities is shown in Figure \ref{fig:smoke-dis} and Table \ref{tab:smoke-dist}. The majority of mothers do not smoke during pregnancy ranging from 76\% for Black mothers to 96\% in the category ``Other''. However, the right panel of Figure \ref{fig:smoke-dis} shows that conditional on smoking white mothers and older mothers smoke more heavily. 
Figure \ref{fig:apo-smoke} replicates the solid line of Figure 1 in \citeA{Cattaneo2010EfficientIgnorability} with Double Machine Learning as a byproduct. Our results are very similar and show that average potential outcomes become smaller the higher the intensity of smoking.
Figure \ref{fig:resclaed-smoke} contains the re-scaled propensity scores $e_t(x)/\pi_t$ for all treatment versions.

\begin{table}[h]
\caption{Distribution of smoking intensities by ethnicity (in percent)}
\label{tab:smoke-dist}
\centering \footnotesize	
\begin{tabular}{lccccc}
  \hline
 & Black & Hispanic & Other & White & All \\ 
  \hline
$>$ 20 cigs & 0.7 & 0.5 & 0.2 & 1.2 & 1.1 \\ 
  16-20 cigs & 4.4 & 2.7 & 0.9 & 5.5 & 5.1 \\ 
  11-15 cigs & 0.7 & 0.5 & 0.2 & 1.3 & 1.2 \\ 
  6-10 cigs & 11.5 & 5.4 & 1.5 & 7.4 & 7.8 \\ 
  1-5 cigs & 6.7 & 4.0 & 1.2 & 3.0 & 3.6 \\ 
  None & 76.1 & 87.0 & 96.0 & 81.6 & 81.2 \\ 
   \hline
\end{tabular}
\end{table}


\subsection{Supplementary Material for Section \ref{sec:app-sc2}} \label{app:app-sc2}

\begin{figure}[h]
    \centering
    \caption{Distribution of treatment versions by gender}\label{fig:version-dis}
    \includegraphics[width=0.7\linewidth]{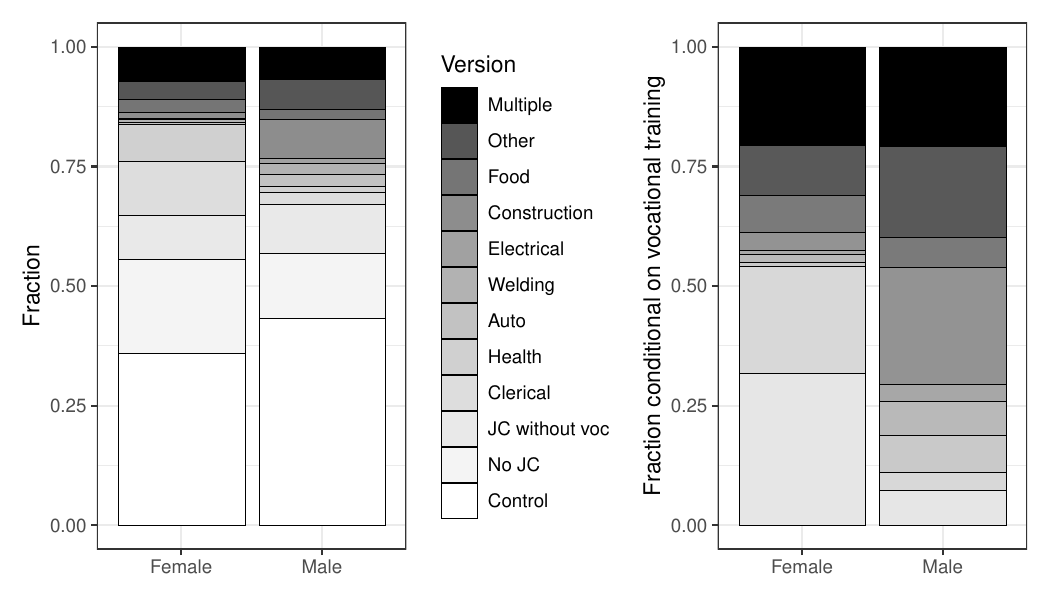}
\end{figure}

The distribution of versions is shown in Figure \ref{fig:version-dis} and Table \ref{tab:version-dist}. We observe that women are overrepresented in clerical, health and food training, while men are more likely to be observed in automechanics, welding, electrical and construction training.

\begin{table}[h] 
\caption{Share of observations in treatment versions (in percent)}
\label{tab:version-dist}
\centering \footnotesize	
\begin{tabular}{lccc}
  \hline
 & Female & Male & All \\ 
  \hline
Control & 36.1 & 43.3 & 40.2 \\ 
  No JC & 19.6 & 13.5 & 16.1 \\ 
  JC without voc & 9.3 & 10.3 & 9.8 \\ 
  Clerical & 11.1 & 2.4 & 6.1 \\ 
  Health & 7.8 & 1.3 & 4.1 \\ 
  Auto & 0.3 & 2.5 & 1.6 \\ 
  Welding & 0.6 & 2.4 & 1.6 \\ 
  Electrical & 0.3 & 1.1 & 0.8 \\ 
  Construction & 1.4 & 8.0 & 5.2 \\ 
  Food & 2.7 & 2.1 & 2.4 \\ 
  Other & 3.6 & 6.2 & 5.1 \\ 
  Multiple & 7.2 & 6.8 & 7.0 \\ 
   \hline
\end{tabular}
\end{table}

\begin{figure}[h]
    \centering
    \caption{Average potential outcomes of treatment versions}\label{fig:apo-jc}
    \includegraphics[width=0.5\linewidth]{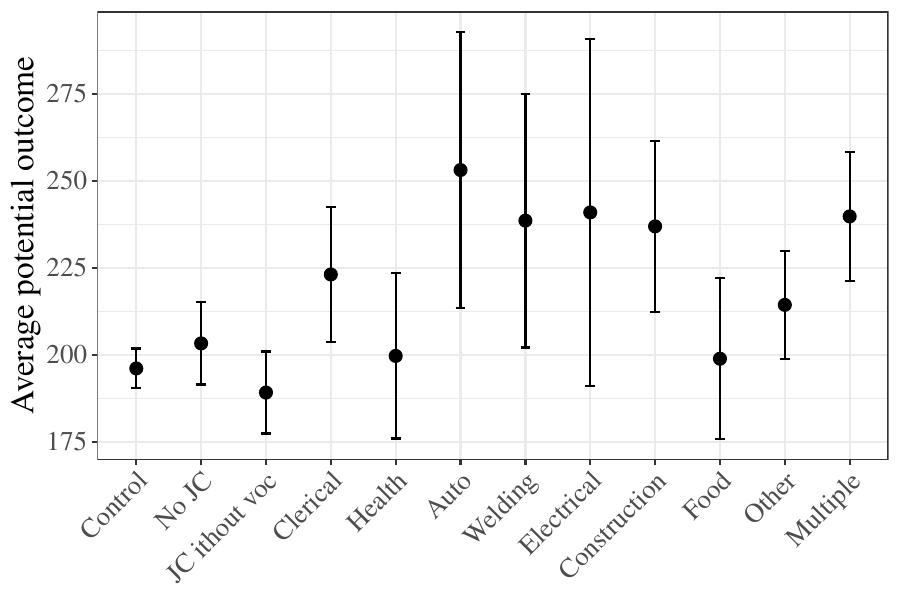}
    \subcaption*{\textit{Note:} Average potential outcomes estimated with Double Machine Learning using an ensemble of Ridge, Lasso and Random Forest regression. Point estimates and 95\%-confidence interval.}
\end{figure}

\begin{figure}[h]
    \centering
    \caption{Re-scaled propensity scores}\label{fig:resclaed-jc}
    \includegraphics[width=\linewidth]{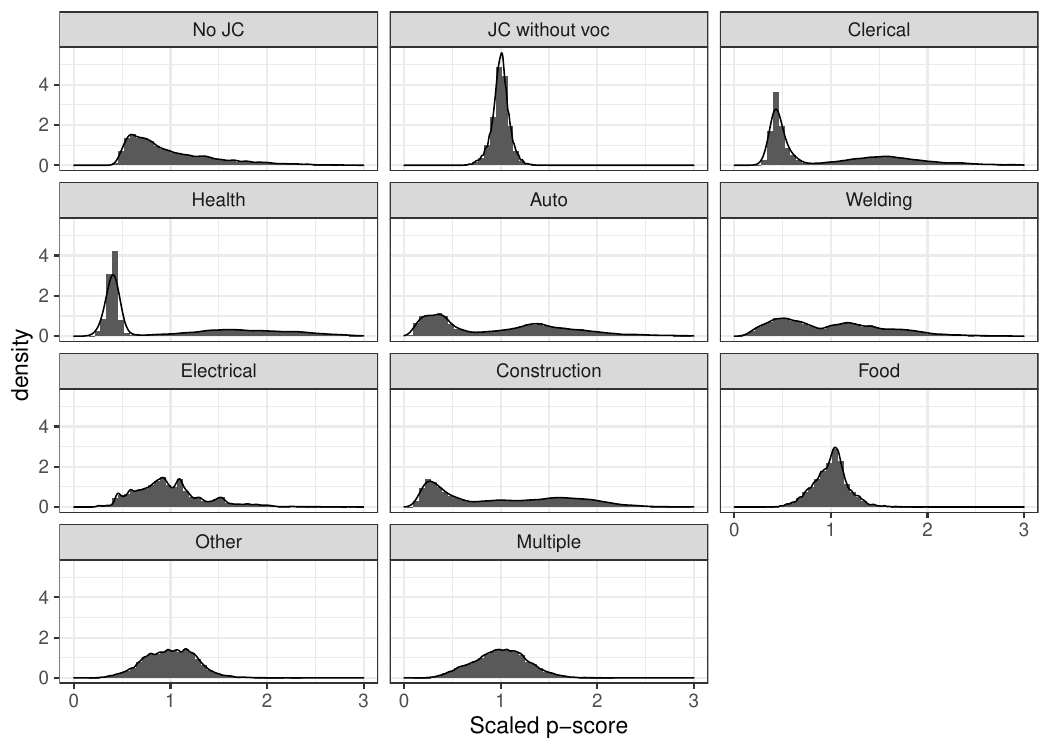}
\end{figure}

As a byproduct of the decomposition estimation, we create the AIPW scores for every treatment version. This allows us to inspect their often noisily estimated average potential outcomes in Figure \ref{fig:apo-jc}. We observe a clear pattern. The point estimates of the predominantly male trainings are all larger than the predominantly female ones. Figure \ref{fig:resclaed-jc} contains the re-scaled propensity scores $e_t(x)/\pi_t$ for all treatment versions. 

\end{appendices}

\end{document}